\documentclass[a4paper,11pt]{article}
\pdfoutput=1 % if your are submitting a pdflatex (i.e. if you have
\usepackage{jheppub} % for details on the use of the package, please
\usepackage[T1]{fontenc} % if needed
\usepackage{mathtools}
\usepackage{blkarray}
\usepackage{graphicx}
\usepackage{amsfonts}
\usepackage{soul}
\usepackage{amssymb}
\usepackage{amsmath}
\usepackage{caption}
\usepackage{subfigure}
\usepackage{comment}
\usepackage{tensor}
\usepackage{cases}
\usepackage[utf8]{inputenc}
\usepackage{cite}
\usepackage{amsmath,amssymb,mathtools}
\usepackage[numbers]{natbib}
\providecommand{\abs}[1]{\lvert#1\rvert}
\providecommand{\bd}[1]{\boldsymbol{#1}}
\providecommand{\ro}[1]{\mathrm{#1}}
\providecommand{\ca}[1]{\mathcal{#1}}
\providecommand{\Mp}{M_{\mathrm{Pl}}}

\title{\boldmath Ultraviolet Sensitivity of Peccei--Quinn Inflation}

\author[a,b,c]{Davide Dal Cin}
\author[a,b,c,d]{and Takeshi Kobayashi}

\affiliation[a]{SISSA, International School for Advanced Studies, \\ Via Bonomea 265, 34136 Trieste, Italy }
\affiliation[b]{INFN, Sezione di Trieste,\\ Via Valerio 2, 34127 Trieste, Italy}
\affiliation[c]{IFPU, Istitute for Fundamental Physics of the Universe,\\ Via Beirut 2, 34014 Trieste, Italy}

\affiliation[d]{Kobayashi-Maskawa Institute for the Origin of Particles and the Universe,\\ Nagoya University, Nagoya 464-8602, Japan}

\emailAdd{ddalcin@sissa.it}
\emailAdd{takeshi.kobayashi@sissa.it}

\abstract{The radial direction of the Peccei--Quinn field can drive cosmic inflation, given a non-minimal coupling to gravity. This scenario has been considered to simultaneously explain inflation, the strong $CP$ problem, and dark matter. We argue that Peccei--Quinn inflation is extremely sensitive to higher-dimensional operators. Further combining with the discussion on the axion quality required for solving the strong $CP$ problem, we examine the validity of this scenario. We also show that after Peccei--Quinn inflation, resonant amplifications of the field fluctuation is inevitably triggered.}

\begin{document} 
\maketitle
\flushbottom

\section{Introduction}
\label{sec:intro}

The Peccei--Quinn (PQ) mechanism \citep{Peccei:1977hh}, based on a global $\ro{U}(1)_{\ro{PQ}}$ symmetry, provides a solution to the strong $CP$ problem by promoting the QCD $\theta$ angle to a dynamical axion field~\citep{Weinberg:1977ma,Wilczek:1977pj}.
The axion can be interpreted as the phase of a complex scalar---the PQ field. 
When the PQ field settles down to the minimum of a Mexican hat potential, the $\ro{U}(1)_{\ro{PQ}}$ symmetry is spontaneously broken and the axion plays the role of a pseudo Nambu--Goldstone boson.
Axions also provide a viable candidate for the dark matter of our universe~\citep{Preskill:1982cy,Abbott:1982af,Dine:1982ah}.

Cosmology requires an additional scalar field for driving cosmic inflation, to explain the observed flatness and homogeneity of the universe, as well as the origin of the primordial curvature perturbation~\citep{Starobinsky:1979ty,Starobinsky:1980te,Sato:1980yn,Guth:1980zm,Mukhanov:1981xt,Linde:1981mu}. 
Here, the radial component of the PQ field can also serve as the inflaton, with the help of a non-minimal coupling to gravity~\citep{Fairbairn:2014zta,Kearney:2016vqw,Ballesteros:2016xej,Boucenna:2017fna,Hamaguchi:2021mmt}.
This provides an intriguing possibility that the strong $CP$ problem, dark matter, and inflation are all explained by the PQ field. (Even more problems can be solved with some extensions~\citep{Ballesteros:2016xej}.)
The PQ inflation scenario also predicts the value of the tensor-to-scalar ratio as $r \approx 0.004$, which can be tested with upcoming cosmic microwave background (CMB) experiments such as~\citep{LiteBIRD:2022cnt}.

The crucial ingredient of the PQ inflation scenario is a coupling between the PQ field and the Ricci scalar. This dimension-four interaction, which is expected to exist from the point of view of an effective field theory in curved space~\citep{Parker:2009uva}, flattens the potential to realize a slow-roll~\citep{Salopek,Kaiser:1994vs,Okada:2010jf,Linde:2011nh}. 
In this sense it has many features in common with the model of~\citep{Bezrukov:2007ep} where the Higgs field with a non-minimal gravitational coupling drives inflation. 
Higgs inflation requires a very large gravitational coupling
in order to be consistent with both the measurement of the Higgs self-coupling, and the curvature perturbation amplitude inferred from analyses of the CMB.
It has been pointed out that the large gravitational coupling lowers the cutoff scale of the effective field theory, making the predictions of the model unreliable~\citep{Barbon:2009ya,Burgess:2009ea,Hertzberg:2010dc,Burgess:2014lza} (though a counterargument has been presented in~\citep{Bezrukov:2010jz}).
On the other hand for the PQ field, its self-coupling is not constrained from experiments and thus is effectively a free parameter of the model. 
It can be chosen to allow for a small gravitational coupling, such 
that the cutoff scale is larger than the relevant energy scales during and after inflation.
This has been considered as another virtue of PQ inflation.\footnote{It was pointed out in \citep{Hamaguchi:2021mmt} that the gravitational coupling also helps evade the axion quality problem.}

Here we note that a small gravitational coupling, on the other hand, requires a large field excursion for the PQ field.
This is easily understood by noting that, with a power-law potential $V \propto \phi^n$ and in the absence of a non-minimal coupling to gravity, slow-roll inflation requires super-Planckian field ranges.
The necessity of large field excursions renders the model sensitive to higher-dimensional operators, which in turn depends on the details of the ultraviolet completion of the theory, bringing us back to the situation similar to Higgs inflation.\footnote{The precise statement is that the PQ field excursion in the Jordan frame increases with a decreasing non-minimal coupling. The excursion of the canonically normalized field in the Einstein frame is larger than $10^{18}\, \ro{GeV}$, independently of the value of the non-minimal coupling.}
The objective of this paper is to sharpen this message by evaluating the effects of higher-dimensional operators on PQ inflation. By calculating their impact on curvature perturbations and the duration of inflation, we derive constraints on Planck-suppressed operators.
(See also \citep{Jinno:2019und} which performed a similar study for a Higgs-like real inflaton.) 

The ultraviolet sensitivity of PQ inflation is reminiscent of the so-called axion quality problem~\citep{Kamionkowski:1992mf,Holman:1992us,Kallosh:1995hi}, which is based on the observation that $\ro{U}(1)_{\ro{PQ}}$-breaking 
higher-dimensional operators can spoil the axion as a solution to the strong $CP$ problem by displacing the axion field from the $CP$-conserving vacuum.
While this vacuum displacement is an effect sourced by higher-dimensional operators at low energies where the PQ field is localized at the potential minimum, the effects on PQ inflation is at the inflationary scale where the PQ field is largely displaced from its minimum. Hence the two effects induced by higher-dimensional operators have different nature, and we find that the resulting constraints are complementary to each other, excluding a wide range of operators when combined.

We also analyze the PQ field dynamics after inflation. 
It has been claimed in \citep{Fairbairn:2014zta,Boucenna:2017fna} that after PQ inflation ends, axion dark matter is produced due to the misalignment of the nearly homogeneous axion field from the vacuum, and also that the isocurvature fluctuation of the axions is suppressed by the large PQ field displacement during inflation~\citep{Linde:1990yj,Linde:1991km}. 
On the other hand in the specific PQ(-like) inflation model  of~\citep{Ballesteros:2016xej}, the oscillation of the inflaton about its origin induces a resonant amplification of the field fluctuation, which leads to a restoration of the PQ symmetry.
(See \citep{Tkachev:1995md,Kasuya:1996ns,Kasuya:1998td,Tkachev:1998dc,Harigaya:2015hha,Kobayashi:2016qld,Co:2020dya,Ballesteros:2021bee} for related works).
The subsequent symmetry breaking thus yields a highly inhomogeneous axion field as well as axionic strings, whose decay contributes to the axion production.
Here, one may expect that $\ro{U}(1)_{\ro{PQ}}$-breaking higher-dimensional operators should suppress resonant effects, since they can source an angular momentum to the PQ field and prevent the field from oscillating violently along its radial direction.
However, we show that with higher-dimensional operators allowed for a consistent PQ inflation, a resonant amplification of the PQ field fluctuation is inevitably triggered.
Our finding thus implies that axion production after PQ inflation does not proceed as in the vanilla vacuum misalignment scenario.

The plan of this paper is as follows. In Section \ref{Section2} we review the basic properties of the PQ inflation model.
In Section \ref{Section3} we derive constraints on higher-dimensional 
operators by numerically solving the PQ field dynamics during inflation.
We then analytically derive an approximate expression for the constraints in Section \ref{Section4}.
In our calculations we treat PQ inflation as an effectively single-field model; we justify this treatment in Section \ref{Section5}.
In Section \ref{Section6} we study the PQ field dynamics after inflation and show that a resonant amplification of field fluctuations is unavoidable.
In Section \ref{Section7} we compare the constraint on higher-dimensional operators from PQ inflation with that from axion quality arguments.
We then conclude in Section \ref{Section8}. 
In the main text we focus on higher-dimensional operators that are suppressed by the present-day Planck scale, however in Appendix~\ref{sec:Mp_eff} we study operators suppressed by the effective Planck scale during inflation.
In Appendix \ref{AppA} we present a detailed analysis of the axion contribution to the curvature perturbation based on the $\delta N$~formalism.
In Appendix~\ref{app:iso} we consider a hypothetical situation where axion dark matter is produced from a vacuum misalignment after PQ inflation, and evaluate the axion isocurvature perturbation.

\section{Review of Peccei--Quinn inflation}\label{Section2}

We consider a PQ field $\Phi$ that is coupled to the Ricci scalar,
\begin{equation}
 S = \int d^4 x \sqrt{-g}
\left[
\left( \frac{M^2}{2} + \xi \Phi \Phi^* \right) R
- g^{\mu \nu} \partial_\mu \Phi \partial_\nu \Phi^*
- V (\Phi) 
\right],
\end{equation}
with $V$ being a Mexican hat potential,
\begin{equation}
 V (\Phi) = \frac{\lambda}{6} \left( |{\Phi}|^2 - \frac{f^2}{2}  \right)^2,
 \label{eq:Pqpotential}
\end{equation}
and $f$ is the axion decay constant.
The self-coupling constant~$\lambda$ and gravitational coupling~$\xi$ are assumed to be non-negative.
At the symmetry-breaking vacuum, i.e. $\abs{\Phi} = f / \sqrt{2}$, the mass scale $M$ is related to the reduced Planck mass,  $M_{\ro{Pl}} \approx 2.4 \times 10^{18}\, \ro{GeV}$, by 
\begin{equation}
    M^2_{\ro{Pl}}=M^2+\xi f^2 .
\end{equation}
We assume 
\begin{equation}
\xi f^2\ll M^2_{\ro{Pl}},
\label{eq:xi-f}
\end{equation}
so that $ M_{\ro{Pl}}\simeq M$. 
Rewriting $\Phi$ in terms of its radial and phase (axion) directions as
\begin{equation}
 \Phi = \frac{\varphi}{\sqrt{2}}e^{i \theta },
\label{eq:2.4}
\end{equation}
where $\varphi \geq 0$, 
then the above action becomes
\begin{equation}
 S = \int d^4 x \sqrt{-g}
\left[
\frac{M^2 + \xi \varphi^2}{2} R
- \frac{1}{2} g^{\mu \nu } \partial_\mu \varphi \partial_\nu \varphi
- \frac{1}{2} \varphi^2 g^{\mu \nu } \partial_\mu \theta \partial_\nu
\theta  - V
\right],
\end{equation}
with 
\begin{equation}
 V = \frac{\lambda }{4!} \left( \varphi^2 - f^2 \right)^2.
\label{eq:2.5}
\end{equation}
Performing a conformal transformation 
$\tilde{g}_{\mu \nu} = \Omega^2 g_{\mu \nu }$ with conformal factor
\begin{equation}
 \Omega^2 = 1 + \xi \frac{\varphi^2}{M^2_{\ro{Pl}}},
\end{equation}
one goes to the Einstein frame where the action takes the form
\begin{equation}
 S = \int d^4 x \sqrt{-\tilde{g}}
\left[
\frac{M^2_{\ro{Pl}}}{2} \tilde{R}
- \frac{1}{2}
\tilde{g}_{\mu \nu} \partial_\mu \chi \partial_\nu \chi
-\frac{1}{2} \frac{\varphi^2}{\Omega^2}\tilde{g}_{\mu \nu }
\partial_\mu \theta \partial_\nu \theta
- U
\right].
\label{eq:S-einstein}
\end{equation}
Here $U = V / \Omega^4$, and 
$\chi$ is an almost canonically normalized field defined as
\begin{equation}
d \chi = I \, d\varphi,
\quad
I = 
\frac{\sqrt{1 +  \xi (1 + 6 \xi) \frac{\varphi^2}{\Mp^2} }}{1 + \xi \frac{\varphi^2}{\Mp^2} }.
\label{eq:dchi-dvarphi}
\end{equation}

Hereafter we carry out the analyses in the Einstein frame, and consider
a flat  Friedmann–Robertson–Walker background, $\tilde{g}_{\mu \nu } dx^\nu dx^\mu  = -dt^2 + a(t)^2 d \bd{x}^2$. 
The $\ro{U}(1)_{\ro{PQ}}$ symmetry allows the axion to stay fixed to its initial position. We treat the model as effectively single-field and introduce the slow-roll parameters in terms of the canonical radial field as
\begin{equation}
 \epsilon = \frac{M^2_{\ro{Pl}}}{2} \left(\frac{1}{U} \frac{\partial U}{\partial \chi }\right)^2,
\quad
 \eta = \frac{M^2_{\ro{Pl}}}{U} \frac{\partial^2 U}{\partial \chi^2 }.
\end{equation}
When these are both smaller than unity, the radial field drives slow-roll inflation with its dynamics described by
\begin{equation}
3 H \dot{\chi} \simeq - \frac{\partial U}{\partial \chi},
\quad
3 M^2_{\ro{Pl}} H^2 \simeq U.
\label{eq:slow-roll}
\end{equation}
Here an overdot denotes a derivative in terms of the physical time~$t$,
and $H = \dot{a} / a$ is the Hubble rate.
The scalar power spectrum amplitude, scalar spectral index, and the tensor-to-scalar ratio at some wave number of interest~$k_*$ 
are written respectively as
\begin{equation}
 A_s \simeq \frac{U_*}{24\pi^2 M^4_{\ro{Pl}} \epsilon_*} , 
\quad
 n_s -1 \simeq 2\eta_*-6\epsilon_* , 
 \quad 
r \simeq 16\epsilon_*.
\label{spectralindex2}
\end{equation}
The subscript~$*$ denotes that the quantities are evaluated when the mode~$k_*$ exits the Hubble horizon. 

Ignoring the decay constant $f$ in the potential~(\ref{eq:2.5}) and using $V = \lambda \varphi^4 / 4!$, the slow-roll parameters are obtained as
\begin{equation}
 \epsilon = \frac{ 8 \frac{M^2_{\ro{Pl}}}{\varphi^2} }{1 + \xi (1 + 6 \xi) \frac{\varphi^2}{M^2_{\ro{Pl}}} }, 
\quad
 \eta = \frac{12 \frac{M^2_{\ro{Pl}} }{\varphi^2} + 4 \xi (1 + 12 \xi) - 8 \xi^2 (1 + 6 \xi) \frac{\varphi^2}{M^2_{\ro{Pl}}} }{ \left\{ 1 + \xi (1 + 6 \xi) \frac{\varphi^2}{M^2_{\ro{Pl}}} \right\}^2 }.
 \label{eq:epsiloeta}
\end{equation}
The amplitudes of the two parameters become unity for similar field values. We thus roughly estimate the field value when inflation ends by solving $\epsilon = 1$, giving,
\begin{equation}
\left( \frac{\varphi_{\ro{end}}}{M_{\ro{Pl}}} \right)^2 \sim 
\frac{16}{1 + \sqrt{(1+8\xi) (1+24\xi) }}
\simeq
 \begin{dcases}
 8 & \mathrm{for}\, \, \,  \xi \ll 10^{-1}, \\
  \frac{2}{\sqrt{3} \xi }
  & \mathrm{for}\, \, \,  \xi \gg 10^{-1}.
 \end{dcases}
\label{eq:phiendPQ}
\end{equation}
Hence as long as the decay constant satisfies both $f^2 \ll \Mp^2$ and (\ref{eq:xi-f}), it follows that $f^2 \ll \varphi_{\ro{end}}^2$. 
This justifies our neglecting of $f$ during inflation.

In particular, if the radial field value is large enough such that 
\begin{equation}
\xi \frac{\varphi^2}{M^2_{\ro{Pl}}} \gg 1, 
\label{eq:large-field}
\end{equation}
then the slow-roll parameters become
\begin{equation}
 \epsilon \simeq \frac{8 M^4_{\ro{Pl}}}{\xi (1 + 6 \xi ) \varphi^4}, 
\quad
\eta \simeq - \frac{8 M^2_{\ro{Pl}}}{(1 + 6 \xi) \varphi^2}.
\label{eq:2.13}
\end{equation}
These satisfy a hierarchical relation of $\epsilon \ll \abs{\eta} \ll 1$,
and using (\ref{spectralindex2}) we obtain
\begin{equation}
 A_s \simeq  \frac{\lambda   (1+6\xi )\varphi_*^4}{4608 \pi^2 \xi M^4_{\ro{Pl}} },
\quad
 n_s - 1 \simeq -\frac{16 M^2_{\ro{Pl}}}{(1+6\xi) \varphi_*^2}, 
\quad
  r \simeq \frac{128 M^4_{\ro{Pl}}}{ \xi (1+6 \xi) \varphi_*^4 } .
\label{eq:2.16}
\end{equation}
The slow-roll approximations~(\ref{eq:slow-roll}) in the large-field regime~(\ref{eq:large-field}) take the forms,
\begin{gather}
 3 H \dot{\varphi} \simeq -\frac{\lambda }{6 \xi^2 (1 + 6 \xi )}
\frac{M^4_{\ro{Pl}}}{\varphi },
\label{eq:varphi-SL}
\\
 H^2 \simeq \frac{\lambda M^2_{\ro{Pl}}}{72 \xi^2}.
\label{eq:FR-SL}
\end{gather}
These can be used to obtain the number of $e$-folds between the exit of the $k_*$~mode and the end of inflation as
\begin{equation}
  N_* = \int_{t_*}^{t_{\ro{end}}} dt \, H  \simeq 
\frac{1+6 \xi }{8 M^2_{\ro{Pl}}} \left(\varphi_*^2-\varphi^2_{\ro{end}} \right),
  \label{eq:numberofefoldspq}
\end{equation}
where we assumed the field condition~(\ref{eq:large-field}) to (at least marginally) hold until the end of inflation. Supposing $\varphi_*^2 \gg \varphi_{\ro{end}}^2$, then (\ref{eq:2.16}) is rewritten in terms of $N_*$ as,
\begin{equation}
 A_s \simeq \frac{\lambda N_*^2}{72 \pi^2 \xi (1+6 \xi)},
\quad
n_s - 1 \simeq - \frac{2}{N_*},
\quad
r\simeq \frac{2 (1+6 \xi)}{\xi N_*^2}.
\label{eq:ns_r_N}
\end{equation}

\begin{figure}[!ht]
    \centering
    \subfigure[Self-coupling]{
    \includegraphics[width=0.42\textwidth]{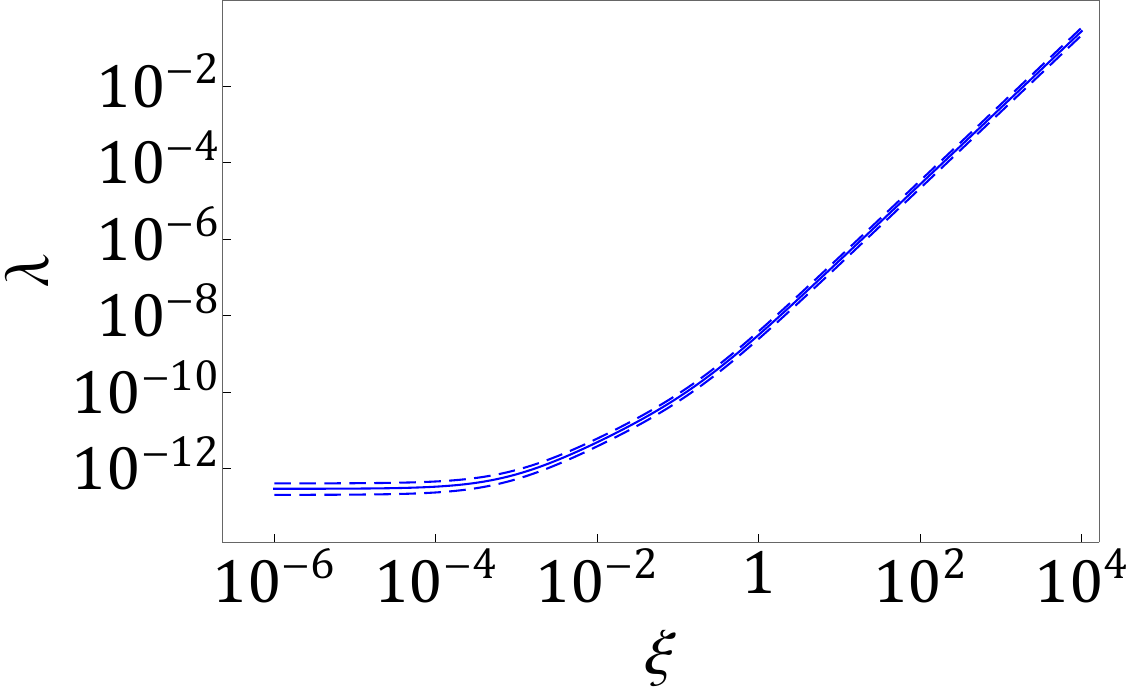}
    \label{fig:lambda}
    }
    \subfigure[Field values]{
    \includegraphics[width=0.42\textwidth]{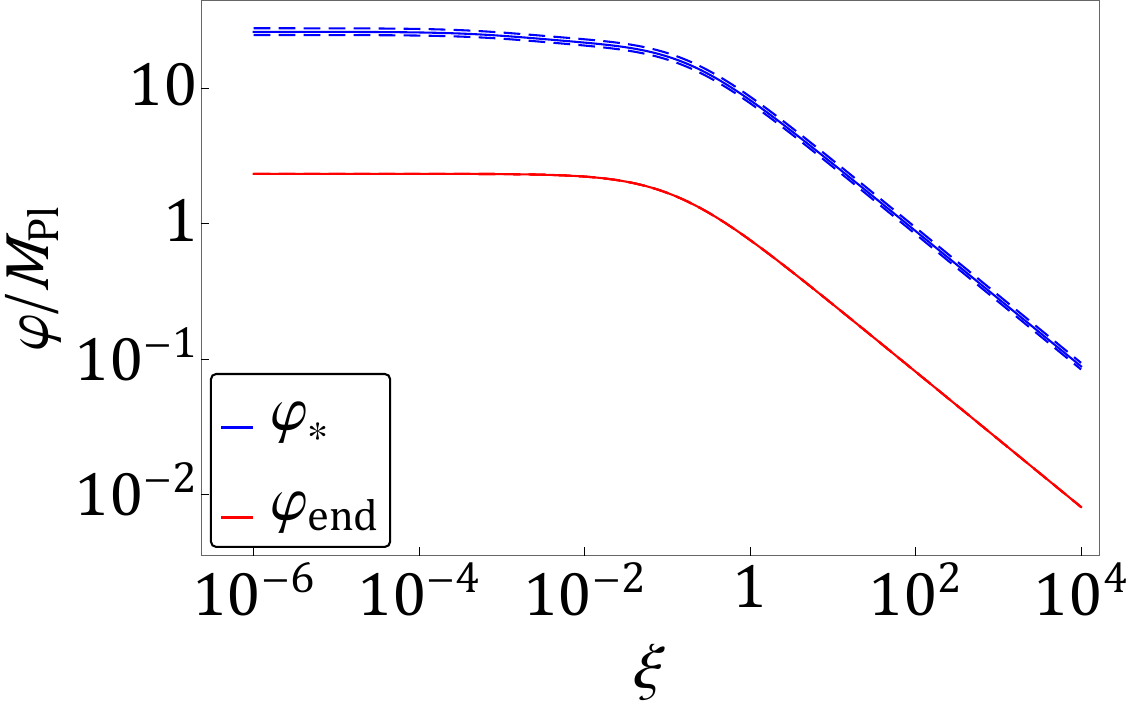}
    \label{fig:varphi}
    }
    \subfigure[Number of $e$-folds]{
    \includegraphics[width=0.42\textwidth]{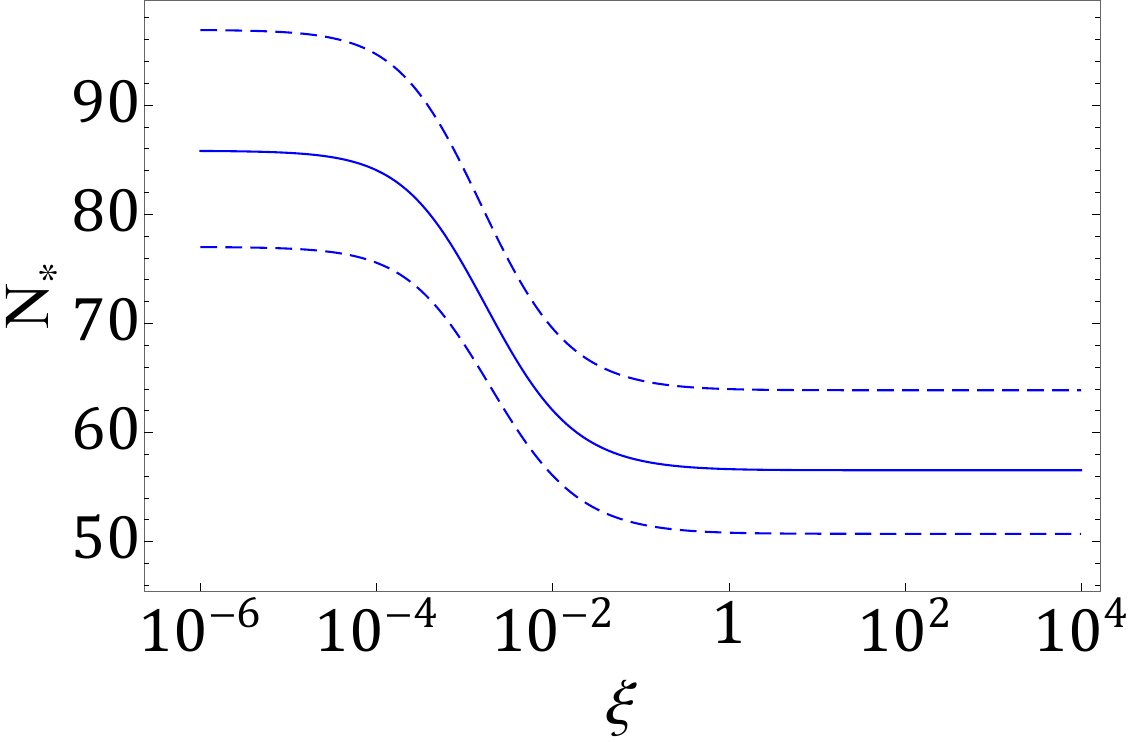}
    \label{fig:N_star}
    }
    \subfigure[Tensor-to-scalar ratio]{
    \includegraphics[width=0.42\textwidth]{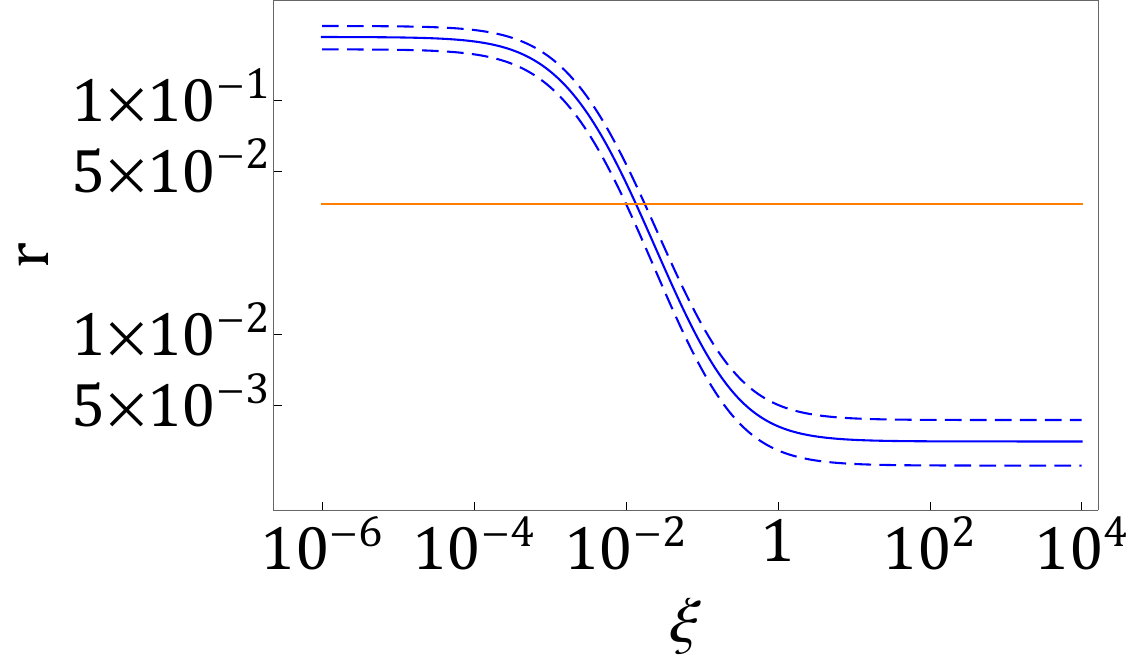}
    \label{fig:r}
    }
    \caption{Inflationary predictions and model parameters of PQ inflation, as functions of the non-minimal gravitational coupling. 
Between the dashed lines the scalar spectral index takes values within the Planck $68\%$~C.L. region $n_s = 0.9649 \pm 0.0042$, with the solid line corresponding to the central value. 
In Fig.~\ref{fig:r} the orange line shows the BICEP/Keck upper limit for~$r$.}
    \label{number1}
\end{figure}

We have also numerically computed the inflationary predictions and model parameters of PQ inflation, which are shown in Fig.~\ref{number1} as functions of $\xi$.
Here $\lambda$ and $\varphi_*$ are fixed using the expressions (\ref{spectralindex2}), such that the scalar power spectrum amplitude takes the best-fit value $A_s=2.1\times 10^{-9} $, and the spectral index lies within 
the $68\%$ confidence region $n_s = 0.9649 \pm 0.0042$ from the Planck constraints at the pivot scale $k_*=0.05 \, \mathrm{Mpc}^{-1} $~\citep{Planck:2018jri}.
The initial field velocity $\dot{\varphi}_*$ is fixed
from the slow-roll expression~(\ref{eq:varphi-SL}).
The equation of motion of $\varphi$ and the Friedmann equation are then numerically solved until the end of inflation, namely, when $-\dot{H}/H^2 = 1$. The axion field~$\theta$ is fixed to a constant in this computation. 
We show in the plots the values\footnote{The values of $\lambda$ and $\xi$ should be considered as those at the inflation scale. Throughout this paper we ignore the renormalization group running of the couplings during inflation, by considering the logarithmic corrections to be small.} 
of the self-coupling~$\lambda$, 
the radial field values at horizon exit~$\varphi_*$ and at the end of inflation~$\varphi_{\ro{end}}$,
$e$-folding number~$N_*$ from when the pivot scale exits the horizon until the end of inflation, 
and tensor-to-scalar ratio~$r$ at the pivot scale.
The blue solid and dashed lines indicate values derived from, respectively, the best-fit and $1\sigma$ uncertainty for~$n_s$.
(These are barely distinguishable in the log plots for $\varphi_*$ and $\lambda$.)
The red line in Fig.~\ref{fig:varphi} shows the field at the end of inflation~$\varphi_{\ro{end}}$,
which also barely depends on the detailed value of~$n_s$.
In Fig.~\ref{fig:r} we also show the observational upper limit
$r<0.036$ ($95\%$ C.L., BICEP/Keck~\citep{BICEP:2021xfz})
by the orange line.

For $\xi \lesssim 10^{-4}$, the model reduces to a simple $\varphi^4$~inflation which is strongly disfavoured by the Planck data~\citep{Planck:2018jri}. The exclusion is explicitly seen in the plots as $r$ exceeding its upper limit,
as well as $N_*$ far exceeding~$60$.
At $\xi \gtrsim 1$, the values of $N_*$ and $r$ approach those of Higgs inflation which invokes $\lambda \sim 10^{-1}$ and $\xi \sim 10^{4}$~\citep{Bezrukov:2007ep}.
Here the central value of the number of $e$-folds becomes $N_* \approx 56.5$. 
Moreover the tensor-to-scalar ratio approaches $r \approx 0.004$, which can be tested in upcoming experiments such as~\citep{LiteBIRD:2022cnt}.
The upper limit on~$r$ requires $\xi \gtrsim 10^{-2}$. 

With the scale of inflation for this model, and assuming that inflation is followed by a phase of inflaton oscillation, and then by radiation domination, the number of inflationary $e$-folds from the horizon exit of the pivot scale $k_*=0.05 \, \mathrm{Mpc}^{-1} $ should satisfy $N_* \lesssim 56$; here the upper limit corresponds to the case of instantaneous reheating. 
One sees in Fig.~\ref{fig:N_star} that if $n_s$ takes a value that is $+1 \sigma$ away from the best fit, the upper limit on~$N_*$ is violated for any value of~$\xi$.

One can hope to avoid a breakdown of the effective field theory by choosing a set of sufficiently small couplings $\lambda$ and $\xi$, as shown in Fig.~\ref{fig:lambda}.
However Fig.~\ref{fig:varphi} indicates that 
for $\xi \lesssim 10^2$, the field value~$\varphi_* $ exceeds $ \Mp$.
It should also be noted that, independently of~$\xi$, the excursion of the canonical field~(\ref{eq:dchi-dvarphi}) obeys\footnote{This can also be understood from the Lyth bound~\citep{Lyth:1996im} for a canonical inflaton,
$\abs{d \chi / d N} \simeq \, \sqrt{r/8} \, \Mp$,
combined with the values of $r$ shown in Fig.~\ref{fig:r}.} 
$\Delta \chi > \Mp$.
The large field excursions would make the inflationary predictions sensitive to operators suppressed by~$\Mp$. 
We look into this in detail in the next section.

\section{Impact of higher-dimensional operators}\label{Section3}

We now study how the naive picture of PQ inflation described above
is affected by higher-dimensional operators, and show that such 
operators need to be strongly suppressed during inflation.

\subsection{Higher-dimensional operators} \label{Higher-dim}

We include an operator of dimension $2 m + n$ in the Jordan frame such that the potential (\ref{eq:Pqpotential}) is modified to
\begin{equation}
 V (\Phi) = \frac{\lambda}{6} \left( |{\Phi}|^2 - \frac{f^2}{2}  \right)^2
+ \Lambda-\left(
g \frac{{\abs{\Phi}}^{2m } \Phi^n}{M^{2m+n-4}_{\ro{Pl}}}
+ \ro{h.c.}
\right),
\label{eq:HDO}
\end{equation}
with $g$ being a dimensionless and complex constant.
We used $\Mp$ as the suppression scale of the operator, by considering that some new physics must intervene at or below~$\Mp$. One may thus expect $\abs{g} \gtrsim 1$, however we will later show that PQ inflation generically requires a much smaller~$\abs{g}$.
Operators that are suppressed instead by the effective Planck scale $ (\Mp^2 + \xi \varphi^2)^{1/2} $ are studied in Appendix~\ref{sec:Mp_eff}.
See also \citep{Barbon:2009ya,Burgess:2009ea,Hertzberg:2010dc,Bezrukov:2010jz,Burgess:2014lza} 
for discussions on the suppression scales of higher-dimensional operators in the context of Higgs inflation.

The parameter~$\Lambda$ has mass dimension four and its value is chosen such that the vacuum energy at the potential minima vanishes.
At leading order in $g$, its amplitude is of
\begin{equation}
\abs{\Lambda} \sim 2|g|M^4_{\ro{Pl}}\left(\frac{f}{\sqrt{2}M_{\ro{Pl}}}\right)^{2 m + n} .
\label{eq:Lambda}
\end{equation}
We also write the dimension of the operator as
\begin{equation}
l = 2 m + n ,
\end{equation}
and the coupling as $g = |g| e^{i \delta}$ with a real phase~$\delta$. 
Rewriting the complex scalar as (\ref{eq:2.4}), the above potential becomes
\begin{equation}
 V = \frac{\lambda }{4!} \left( \varphi^2 - f^2 \right)^2
+ \Lambda
- 2 \abs{g} M^4_{\ro{Pl}}
\left( \frac{\varphi}{\sqrt{2} M_{\ro{Pl}}} \right)^{l }
\cos( n\theta + \delta) .
\label{fullpotential}
\end{equation}
If $n=0$ the axion remains massless.
On the other hand if $n\neq0$, the $\ro{U}(1)_{\ro{PQ}}$ symmetry is broken and the axion acquires a mass;
the presence of such operators is also suggested by the breaking of global symmetries due to quantum gravity   \citep{Abbott,Banks:2010zn,Harlow:2018tng}.
In the following we will consider both PQ-breaking and preserving operators.

The homogeneous equations of motion and the Friedmann equation in the presence of the higher-dimensional operator are 
\begin{multline}
 0 = \ddot{\chi} + 3 H \dot{\chi}
+ \frac{1}{\Omega^4 
\sqrt{ \Omega^2 + 6 \xi^2 \frac{\varphi^2}{M^2_{\ro{Pl}}} }} 
\Biggl[
\left(1 + \xi \frac{f^2}{M^2_{\ro{Pl}}}\right)
\frac{\lambda}{6} (\varphi^2  -f^2) \varphi - \frac{4 \xi \varphi \Lambda}{M^2_{\ro{Pl}}}
- \Omega^2 \varphi \dot{\theta}^2
\\
- \left\{ l \Omega^2  - 4 \xi \frac{\varphi^2}{M^2_{\ro{Pl}}} \right\}
\frac{\Omega^2 \varphi  m_\theta^2}{n^2}
\cos(n \theta +\delta )
\Biggr],
\label{eq:chi-EoM}
\end{multline}
\begin{equation}
 0 = \ddot{\theta} + 3 H \dot{\theta}
+ \frac{2 \dot{\theta} \dot{\varphi}}{\Omega^2 \varphi }
+ \frac{m_\theta^2}{n} \sin (n \theta +\delta ),
\label{eq:theta-EoM}
\end{equation}
\begin{equation}
 3 M^2_{\ro{Pl}} H^2 = \frac{\dot{\chi}^2}{2}
+ \frac{\varphi^2 \dot{\theta}^2}{2 \Omega^2}
+ \frac{1}{\Omega^4}
\left\{
\frac{\lambda }{4!}(\varphi^2 - f^2)^2 + \Lambda
- \frac{\Omega^2 \varphi^2 m_\theta^2}{n^2}
\cos (n \theta +\delta ) 
\right\}.
\label{eq:Friedmann}
\end{equation}
Here $m_\theta$ represents the mass of the angular direction around its potential minima, which is defined as 
\begin{equation}
 m_\theta^2 = \frac{1}{ \varphi^2 \Omega^2 }
\left. \frac{\partial^2 V}{ \partial\theta^2} \right|_{n \theta + \delta=0} =
\frac{n^2 \abs{g} M^2_{\ro{Pl}} }{\Omega^2}
\left( \frac{\varphi}{\sqrt{2}M_{\ro{Pl}}} \right)^{l -2 }.
\label{massoftheta}
\end{equation}
The mass induced by higher-dimensional operators depends on the inflaton field value. In particular when the inflaton reaches its potential minimum, the axion mass becomes $m^2_\theta \simeq n^2 \abs{g} M^2_{\ro{Pl}} \left( f/\sqrt{2} M_{\ro{Pl}}  \right)^{l -2 }$.

\subsection{Constraints from number of $e$-folds}\label{Section:3.2}

Higher-dimensional operators change the tilt of the inflaton potential, which in turn affects the number of inflationary $e$-folds $N$. 
We compute $N$ by numerically solving the set of equations
(\ref{eq:chi-EoM}), (\ref{eq:theta-EoM}), and (\ref{eq:Friedmann}),
from the time when the pivot scale~$k_*$ exits the horizon until the end of inflation at $-\dot{H}/H^2 = 1$.
The scalar power spectrum amplitude and spectral index are fixed to the central values $A_s=2.1\times 10^{-9} $ and $n_s = 0.965$;
we impose this condition by evaluating the observables using the slow-roll expressions (\ref{spectralindex2}) with $V$ given by (\ref{fullpotential}), and 
choosing the values of $\lambda$ and $\varphi_*$ accordingly.\footnote{We will show that higher-dimensional operators significantly affect
$N$ for fixed values of $A_s$ and $n_s$. This is equivalent to saying that for a fixed~$N$, the values of $A_s$ and $n_s$
are highly sensitive to the operators.}

We also use the slow-roll approximations to fix the field velocities at the horizon exit of the pivot scale, namely, 
(\ref{eq:slow-roll}) for the radial field and 
\begin{equation}
3H\dot\theta \simeq - \frac{m^2_\theta}{n} \sin (n\theta+\delta)
\label{eq:axion_slow-roll}
\end{equation}
for the axion.
The solution (\ref{eq:axion_slow-roll}) assumes $m_{\theta}^2 \ll H^2$. In particular for $n = 0$, the right-hand side vanishes and thus the axion is fixed to a constant value during inflation.
On the other hand if $m_{\theta}^2 \gtrsim H^2$, the axion does not follow (\ref{eq:axion_slow-roll}), but quickly becomes stabilized at one of its potential minima. Hence instead of treating $\theta_*$ as a free parameter, we write it as
\begin{equation}
n\theta_*+\delta=\left(n\theta_{i}+\delta\right)\exp{\left(-\frac{m^2_{\theta*}}{H^2_*}\right)},
    \label{eq:smothingtheta}
\end{equation}
in terms of an `initial' angle~$\theta_{i}$, which represents the axion field value a few $e$-foldings before the pivot scale exits the horizon. If $m_{\theta*}^2 \ll H_*^2$ then $\theta_* \simeq \theta_i$, while if $m_{\theta*}^2 \gg H_*^2$ then $n\theta_*+\delta$ is effectively zero independently of the value of $\theta_i$.
We also note that for $m_{\theta*}^2 \gg H_*^2$, 
the parametrization (\ref{eq:smothingtheta}) combined with 
(\ref{eq:axion_slow-roll}) yields $\dot{\theta}_* \simeq 0$, corresponding to an axion at rest in a minimum.
Hence (\ref{eq:axion_slow-roll}) and (\ref{eq:smothingtheta}) allow us to 
systematically analyze the range of possibilities arising from higher-dimensional operators.

\begin{figure}[t!]
\centering
    \includegraphics[width=0.6\textwidth]{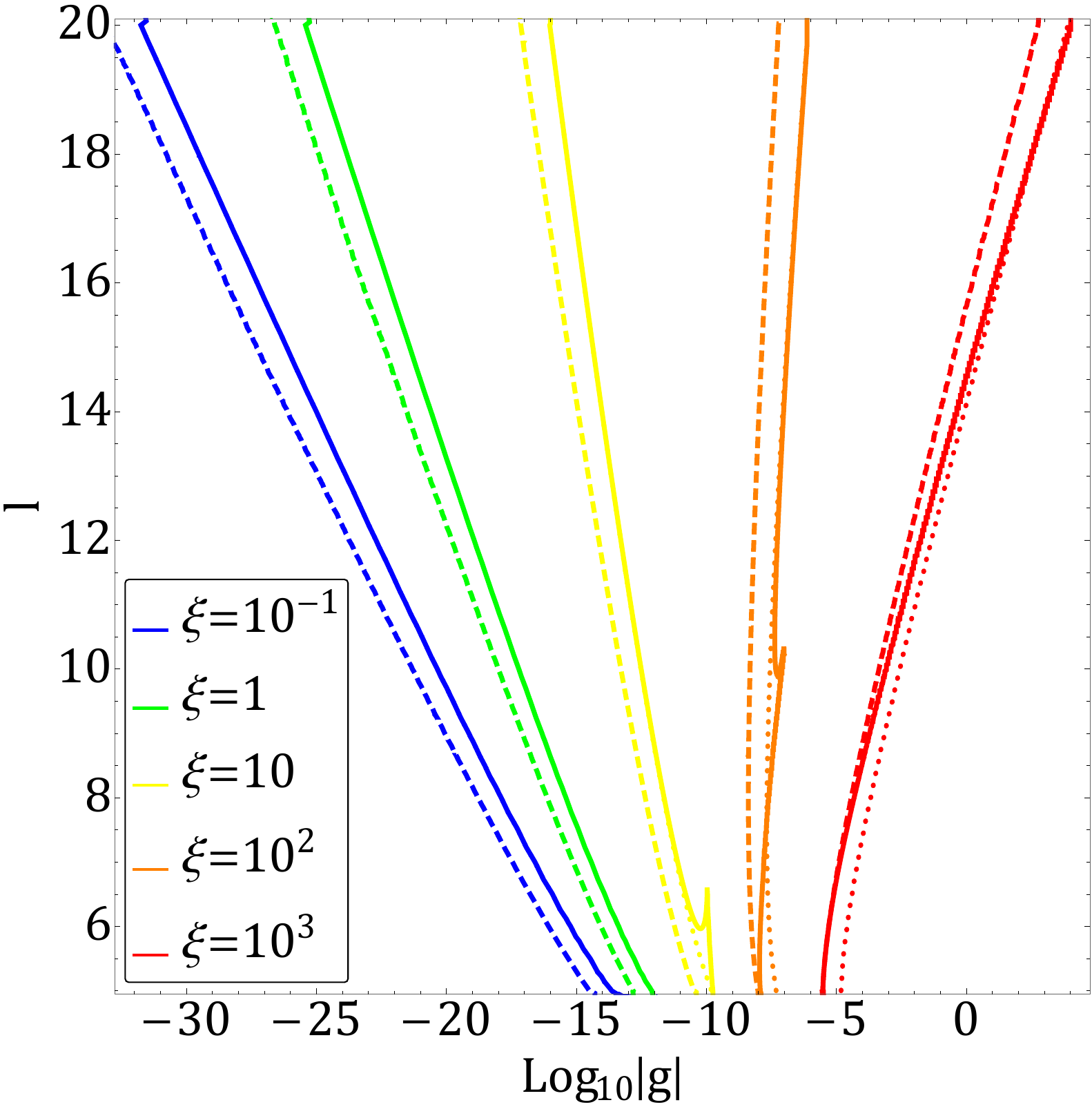}
    \captionof{figure}{Constraints on the coupling constant $|g|$ and dimension $l$ of higher-dimensional operators in PQ inflation, from the requirement that the scalar power spectrum amplitude and spectral index match with observations, and the Planck pivot scale exits the horizon $50$ to $60$ $e$-folds before inflation ends. The allowed regions are on the left of the curves. The colors denote different values for the non-minimal coupling $\xi$. The other parameters are taken as 
$n = 1$ and $\theta_i + \delta = \pi - 0.5$ (solid),
$n = 0$ and $\delta = \pi - 0.5$ (dotted),
$n = 1$ and $\theta_i + \delta = 0.5$ (dashed),
$n = 0$ and $\delta = 0.5$ (dashed);
the last two cases overlap and thus are shown by the same dashed lines. 
The dotted lines overlap with the solid in most part of the plot.
The requirement of very small values for~$\abs{g}$ manifests the extreme sensitivity of PQ inflation to Planck-suppressed higher-dimensional operators.}  
    \label{fig:newquality}
\end{figure}

Since the inflationary dynamics is insensitive to the precise values of  $f$ and $\Lambda$, we are thus left with five free parameters: 
$|g|$, $l$, $n$, $n \theta_i + \delta$, and $\xi$. 
In Fig.~\ref{fig:newquality} we show constraints on higher-dimensional operators in the $|g|$ - $l$ plane. 
(We performed computations also for fractional~$l$.)
As a rough guide, we require the number of $e$-folds from when the pivot scale~$k_*$ exits the horizon until the end of inflation,
to lie within the range\footnote{Note that $N_* \leq 60$ yields a conservative constraint, as the actual upper limit is smaller; see discussions at the end of Section~\ref{Section2}.} $50 \leq N_* \leq 60$.
The regions on the right sides of the lines are excluded since there the number of $e$-folds~$N_*$ is either larger than $60$ or smaller than $50$, and/or $A_s$ and $n_s$ cannot simultaneously take the observed values. 
In each line the non-minimal coupling is taken as $\xi = 10^{-1}$ (blue), 1 (green), 10 (yellow), $10^2$ (orange), $10^3$ (red). 
Each of these are further classified by whether the $\ro{U}(1)_{\ro{PQ}}$ is broken or conserved, and whether $\cos (n \theta_i + \delta)$ is positive (if $n \neq 0$ this corresponds to starting close to a minimum of the axion potential) or negative (close to a hilltop);
we took the combinations of
$n = 1$ and $\theta_i + \delta = \pi - 0.5$ (solid),
$n = 0$ and $\delta = \pi - 0.5$ (dotted),
$n = 1$ and $\theta_i + \delta = 0.5$ (dashed),
$n = 0$ and $\delta = 0.5$ (dashed).
The last two cases are both shown by the same dashed lines since they overlap in the plot. Moreover, the dotted lines overlap with the solid in most part of the plot.
 
As one increases $|g|$, the number of $e$-folds basically increases (decreases) from the base value $N_* \approx 56.5$ for positive (negative) values of $\cos (n \theta_i + \delta)$, which can be understood from the potential (\ref{fullpotential}) being flattened (steepened) by the higher-dimensional operator.
In the plot, the dashed lines for cases with $n \theta_i + \delta = 0.5$
show where $N_* = 60$; on the left of these lines are the regions where $56.5 \leq N_* < 60$.
The dotted lines for $n = 0$ and $\delta = \pi - 0.5$
show where $N_* = 50$, with the regions on the left giving $50 < N_* \leq 56.5$.
On the other hand, the solid lines for $n = 1$ and $\theta_i + \delta = \pi - 0.5$ consist of multiple constraints; we look into this case in detail in the following subsection.
(However the reader interested primarily in the general behavior of the constraints may go straight to Section~\ref{Section4} upon first reading.)

In our computations we only included one higher-dimensional operator as shown in (\ref{eq:HDO}).
In the presence of a tower of operators, a consistent PQ inflation can be achieved if all operators satisfy the constraints of Fig.~\ref{fig:newquality};
here it should be noted that even if operators with different dimensions cancel each other at some particular value of~$\varphi$, as the field rolls the operators will become non-negligible.
The constraints thus indicate that even operators with very high dimensions need to be strongly suppressed.
We remark that there is the possibility that multiple higher-dimensional operators conspire to realize a completely different but flat potential 
at large field values. However we do not pursue such directions since in this paper we are interested in providing discussions that do not rely on the details of the ultraviolet completion.

\subsection{Constraints around the hilltop}
\label{sec:no-go}

If the higher-dimensional operator breaks $\ro{U}(1)_{\ro{PQ}}$ and gives a sufficiently large axion mass, then even if the axion is initially placed near a potential hilltop (i.e. $\cos(n \theta_i + \delta) < 0$), it would roll down to the vicinity of a minimum by the time the pivot scale exits the horizon (i.e. $\cos (n \theta_* + \delta) > 0$).
This is the reason why in Fig.~\ref{fig:newquality} some of the solid lines ($n = 1$ and $\theta_i + \delta = \pi - 0.5$) connect the dotted lines ($n = 0$ and $\delta = \pi - 0.5$, thus $\cos \delta < 0$)
to the dashed lines ($n \theta_i + \delta = 0.5$, thus $\cos (n \theta_* + \delta) > 0$).

\begin{figure}[t!]
\centering
    \subfigure[$\xi = 10$, $\theta_i+\delta=0.5$]{
    \includegraphics[width=0.4\textwidth]{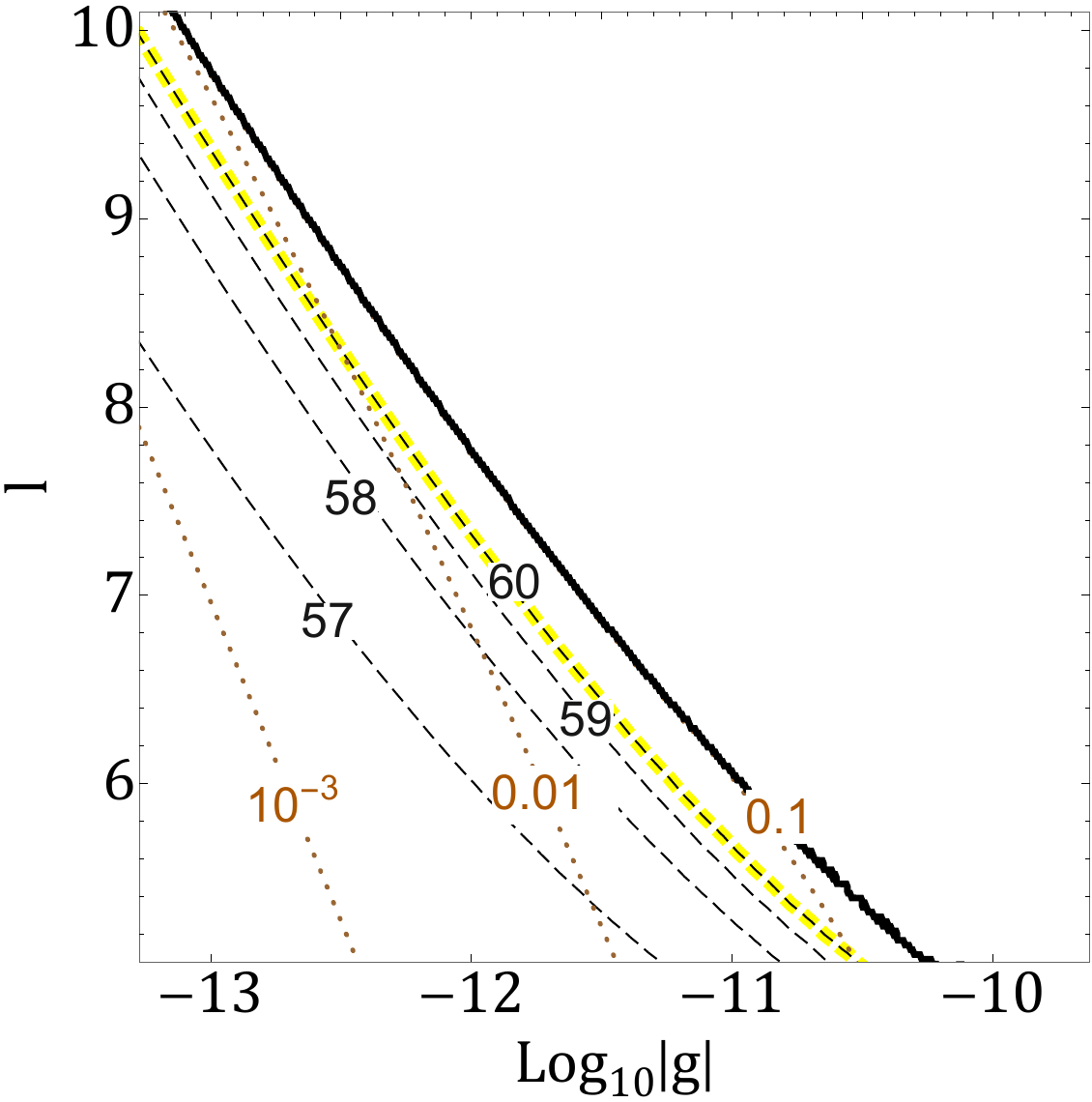}
    \label{fig:3_xi-10pos}
    }
    \subfigure[$\xi = 10$, $\theta_i+\delta=\pi-0.5$]{
    \includegraphics[width=0.4\textwidth]{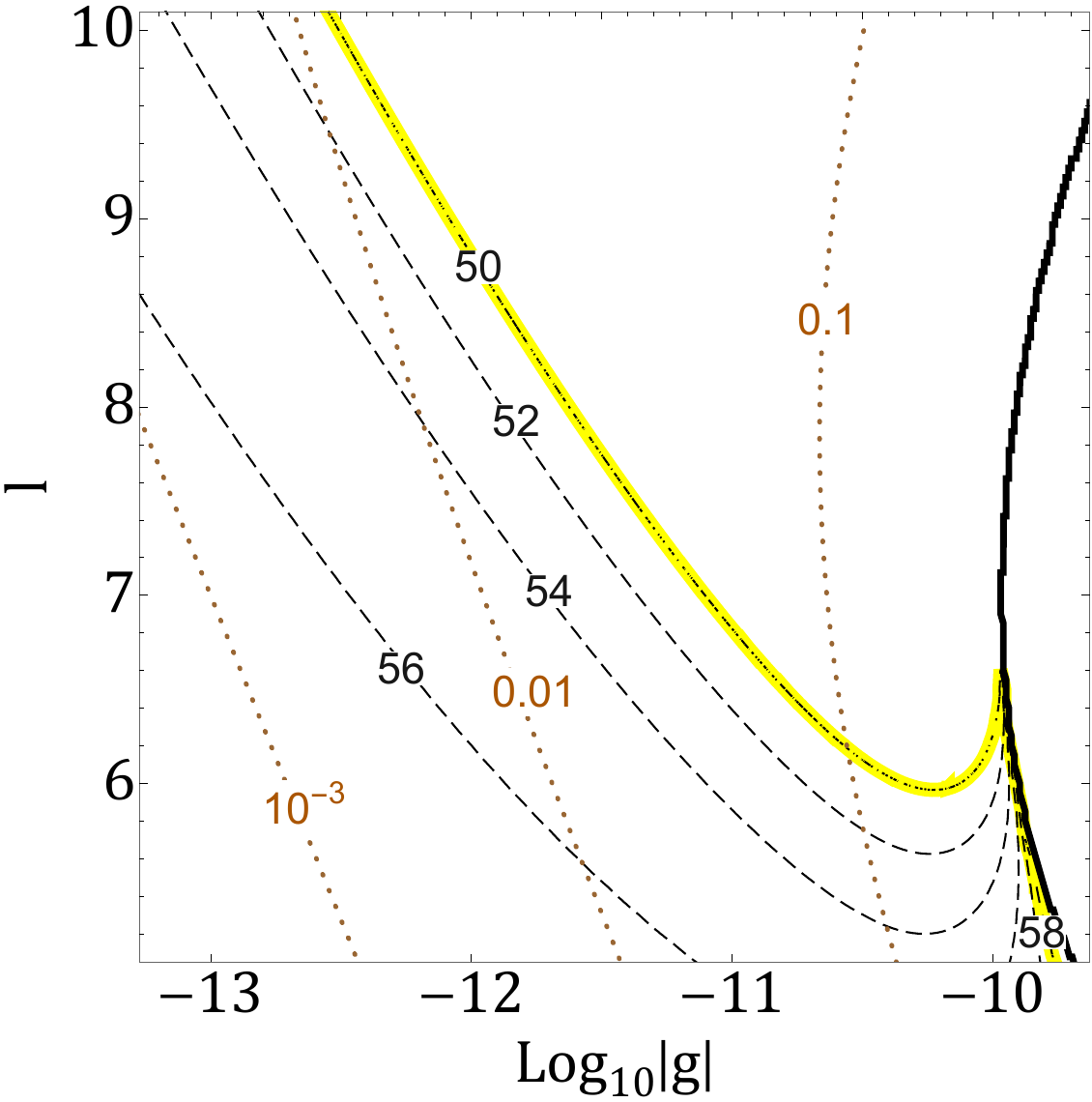}
    \label{fig:3_xi-10}
    }
    \subfigure[$\xi = 10^2$, $\theta_i+\delta=0.5$]{
    \includegraphics[width=0.4\textwidth]{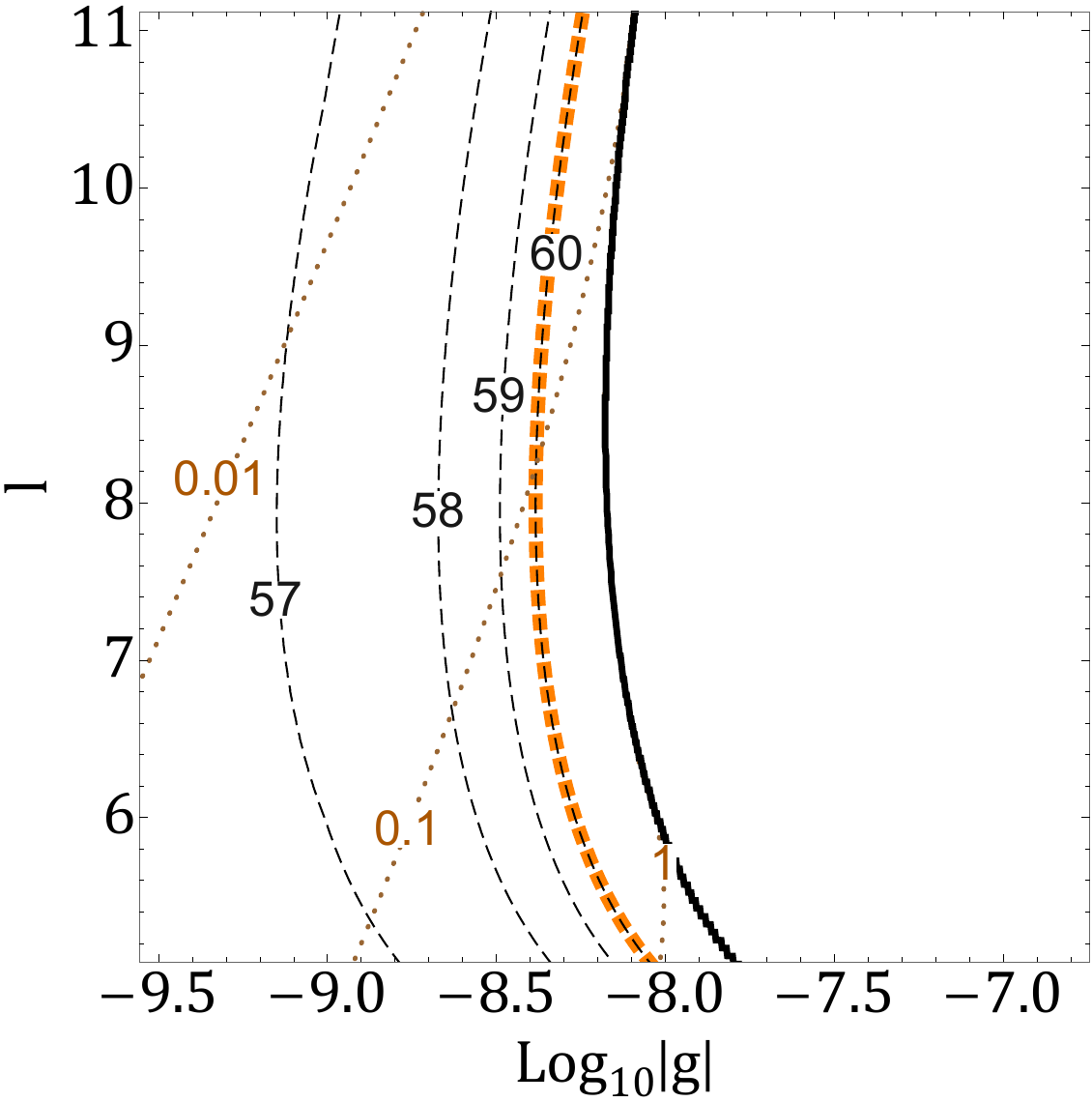}
    \label{fig:3_xi-100pos}
    }
    \subfigure[$\xi = 10^2$, $\theta_i+\delta=\pi-0.5$]{
    \includegraphics[width=0.4\textwidth]{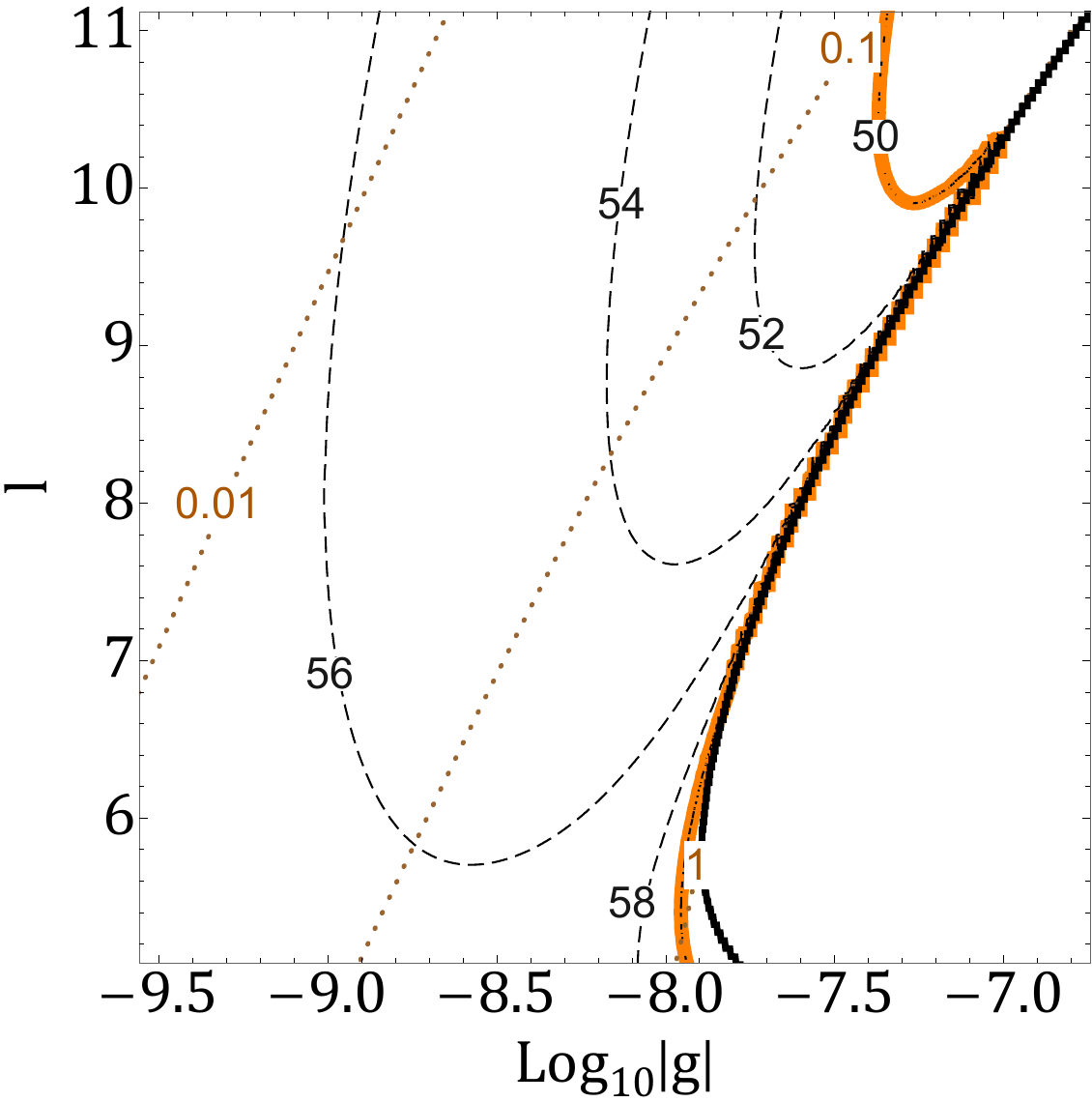}
    \label{fig:3_xi-100}
    }
    \captionof{figure}{Detailed view of Fig.~\ref{fig:newquality}, zooming in on regions where different constraints meet. 
The black dashed contours show values of~$N_*$, and the brown dotted show $m_{\theta *}^2 / H_*^2$. 
All the results are for $n = 1$, while $\xi$ and $\theta_i + \delta$ are varied in each panel.
The combined constraints are shown with the same color scheme (yellow/orange) and line patterns (solid/dashed) as in Fig.~\ref{fig:newquality}.
On the right of the black solid lines, $A_s$ and $n_s$ cannot simultaneously take the observed values.
}  
    \label{fig:newqualityxi100}
\end{figure}

On the solid lines with $\xi \lesssim 1$ (blue and green) the axion mass is sufficiently small such that the axion barely rolls,\footnote{
We later explicitly show in (\ref{eq:4.14}) that the maximum allowed value of $m^2_\theta/H^2$ increases with $\xi$. 
} 
hence the lines overlap with the dotted and show where $N_* = 50$.
On the solid lines with $\xi \gtrsim 10$ (yellow, orange, and red), the axion can roll away from the hilltop region, and as a consequence the lines exhibit kinks. (The red line has a kink at $l \approx 22$, which is not seen in the displayed area.) 
For these lines, the segments above the kinks are set by $N_* = 50$, while the segments right below the kinks arise from not being able to produce the observed values for $A_s$ and $n_s$, and further below the kinks are set by $N_* = 60$.

In Fig.~\ref{fig:newqualityxi100} we zoom into the regions where the various constraints meet, for cases with $n = 1$.
The combined constraints are shown with the same color scheme and line patterns as in Fig.~\ref{fig:newquality}:
$\xi = 10$ (yellow), $10^2$ (orange),
with
$\theta_i + \delta = 0.5$ (dashed), $\pi - 0.5$ (solid).
Here we further show the contour lines for the values of $N_*$ (black dashed) and $m_{\theta *}^2 / H_*^2$ (brown dotted).
On the right of the black solid lines are the regions where there is no combination of ($\lambda$, $\varphi_*$) that simultaneously yield 
$A_s=2.1\times 10^{-9} $ and $n_s = 0.965$.
(The black solid lines are not smooth because of the limited numerical resolution.)
In Figs.~\ref{fig:3_xi-10pos} and \ref{fig:3_xi-100pos} where $\theta_i + \delta = 0.5$, as $\abs{g}$ increases
the $e$-folding number simply increases
from its base value~$N_* = 56.5$. 
On the other hand 
in Figs.~\ref{fig:3_xi-10} and \ref{fig:3_xi-100}
where $\theta_i + \delta = \pi - 0.5$, the $e$-folding number decreases when $m_{\theta *}^2 /H_*^2 \lesssim 0.1$, while it instead increases when $m_{\theta *}^2 / H_*^2 \gtrsim 0.1$ due to the axion rolling down to the vicinity of a minimum.
Moreover, the contour lines of $N_*$ hit the region where the higher-dimensional operator prevents $A_s$  and $n_s$ from simultaneously taking the observed values. One sees that the kink in the combined constraint corresponds to where the $N_* = 50$ contour hits this no-go region.
The constraint for $\xi = 10^3$ has a similar structure, except for that the kink appears beyond the parameter range displayed in Fig.~\ref{fig:newquality}.

\begin{figure}[t!]
\centering
    \subfigure[$n = 0$, $\delta=0.5$]{
\includegraphics[width=0.42\textwidth]{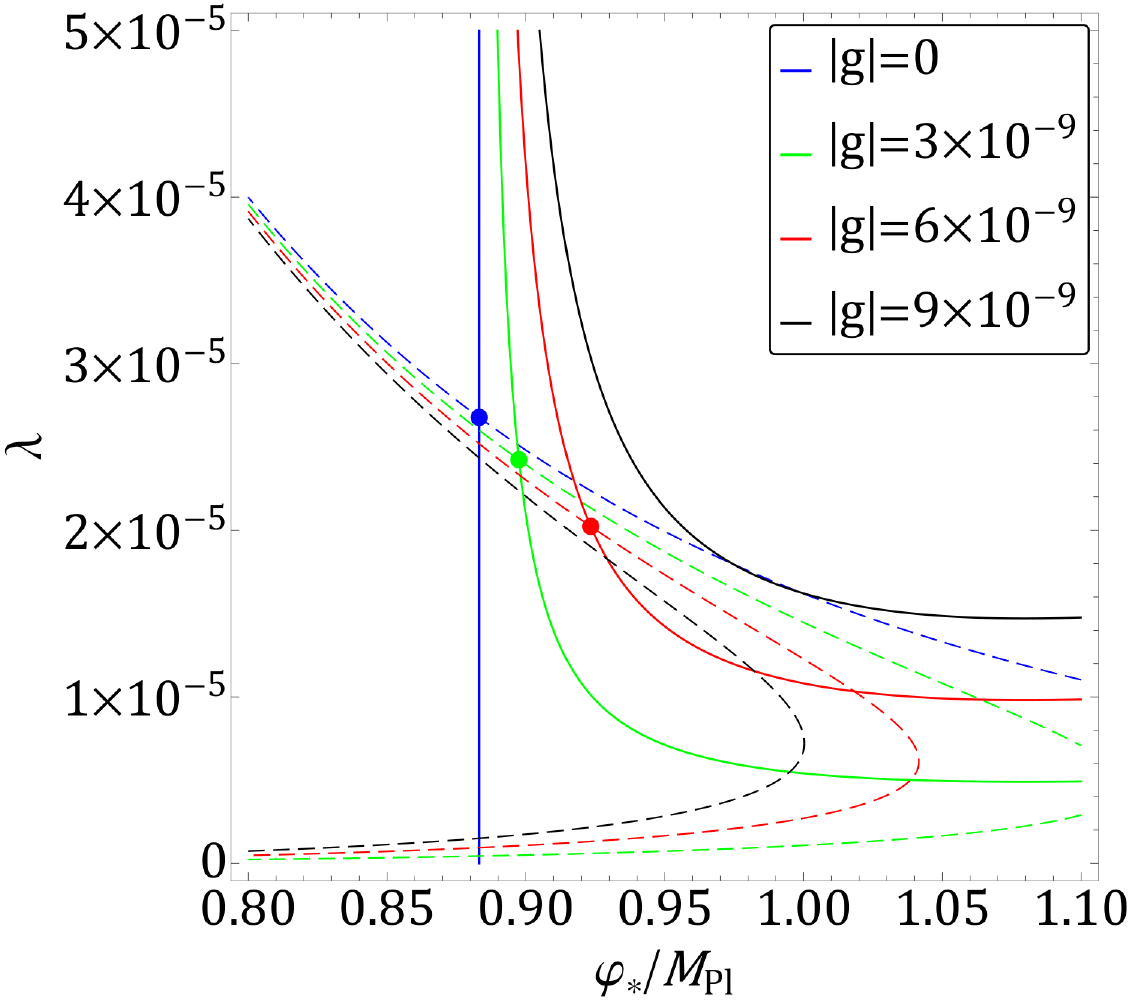}
\label{fig:n0_0.5}
    }
    \subfigure[$n = 0$, $\delta=\pi-0.5 $]{
\includegraphics[width=0.42\textwidth]{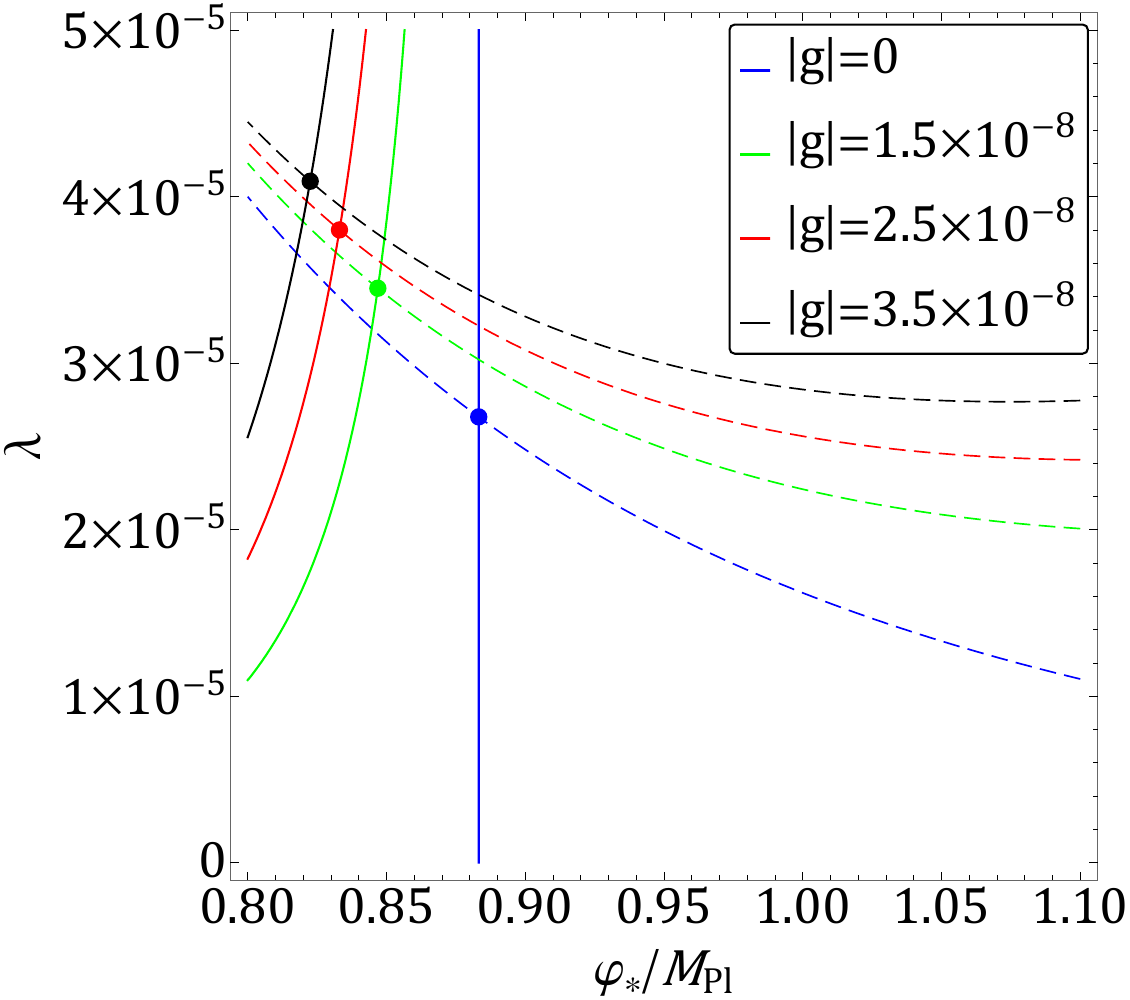}
\label{fig:n0_pi-0.5}
    }
    \subfigure[$n = 1$, $\theta_i+\delta=0.5$]{
\includegraphics[width=0.42\textwidth]{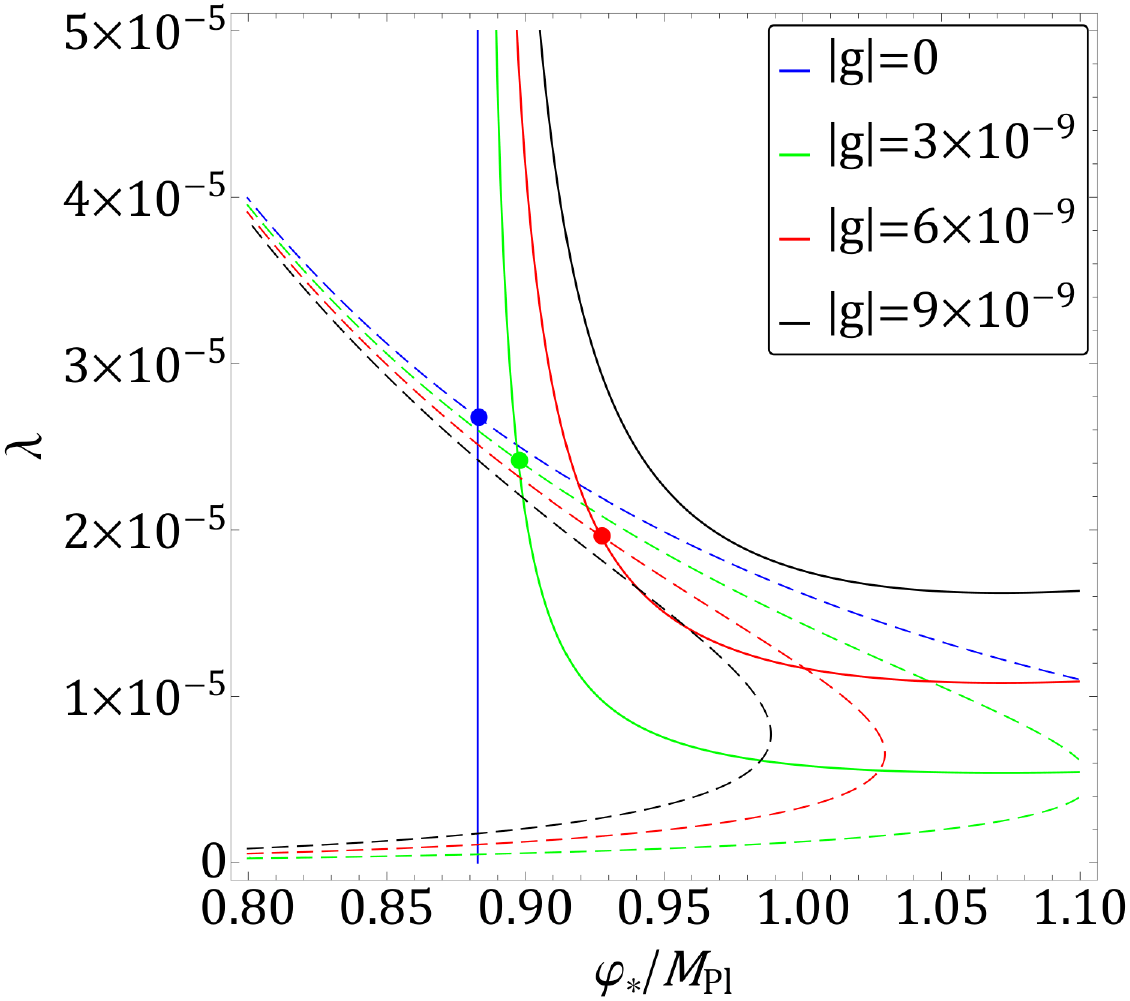}
\label{fig:n1_0.5}
    }
    \subfigure[$n = 1$, $\theta_i+\delta=\pi - 0.5$]{
\includegraphics[width=0.42\textwidth]{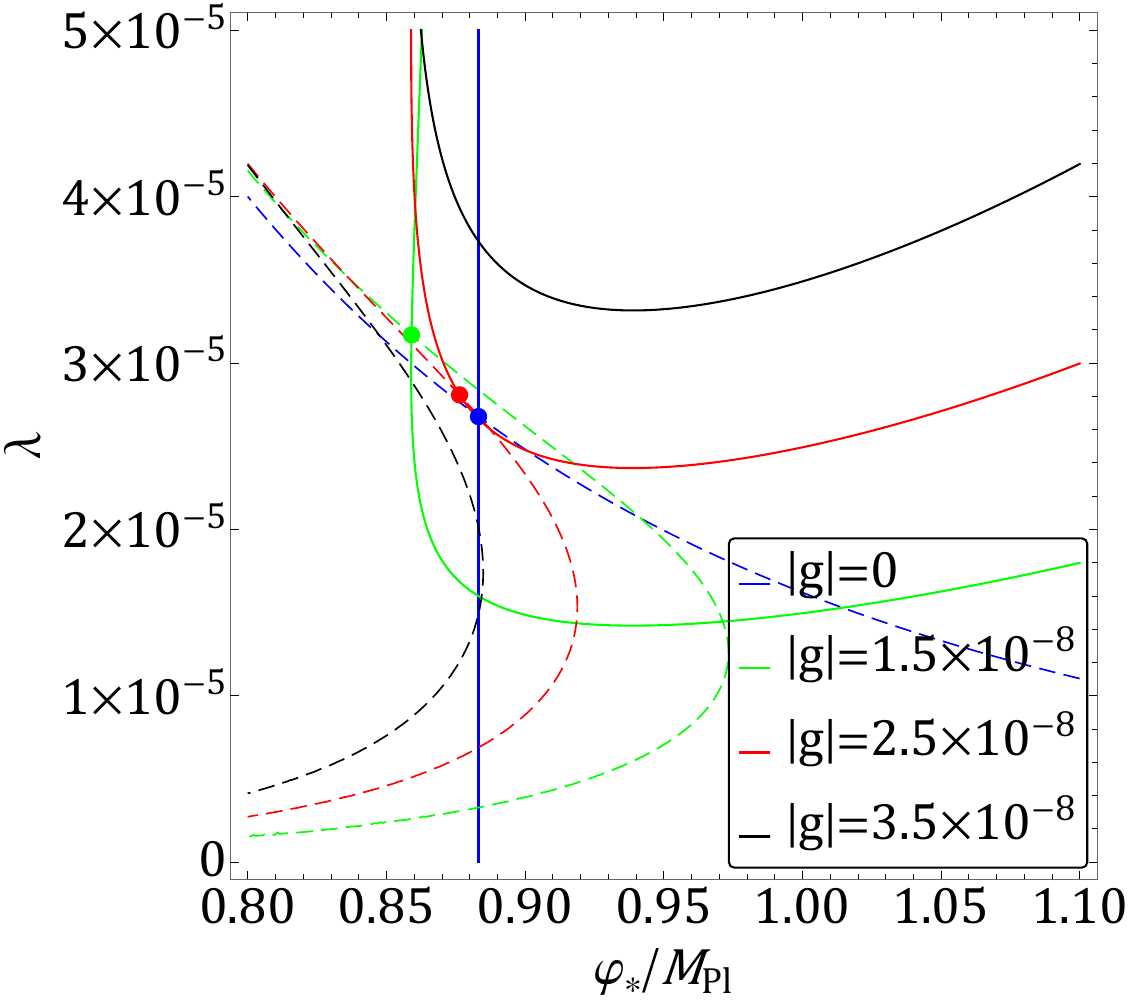}
\label{fig:n1_pi-0.5}
    }
    \captionof{figure}{Combinations of $\varphi_*$ and $\lambda$ that satisfy $ A_s = U_* / 24\pi^2 M^4_{\ro{Pl}} \epsilon_* $ (dashed lines)
and $ n_s -1 = 2\eta_*-6\epsilon_* $ (solid lines). 
The observed values $A_s=2.1\times10^{-9}$ and $n_s=0.965$ are simultaneously realized at the intersections of the dashed and solid lines, which are indicated by the dots.
Different colors correspond to different values of~$\abs{g}$.
Values of $n$ and $ n \theta_i + \delta $ are varied in the four panels.
The other parameters are fixed to $ l = 8$ and $\xi = 10^2$. 
}  
    \label{fig:nosolution}
\end{figure}

In order to see the emergence of the no-go regions, in Fig.~\ref{fig:nosolution} we show the combinations of $\varphi_*$ and $\lambda$ that are solutions to 
$ A_s = U_* / 24\pi^2 M^4_{\ro{Pl}} \epsilon_* $ (shown as dashed lines)
and $ n_s -1 = 2\eta_*-6\epsilon_* $ (solid lines). 
The observed values of $A_s$ and $n_s$ are simultaneously realized at the intersection of the dashed and solid lines, whose position is indicated by a dot. 
Lines and dots with different colors correspond to different values of~$\abs{g}$.
Moreover, $n$ and $ n \theta_i + \delta $ are varied in the four panels, while the rest of the parameters are fixed to $ l = 8$ and $\xi = 10^2$. 
Looking at Fig.~\ref{fig:n1_pi-0.5}
in which $n = 1$ and $\theta_i + \delta = \pi - 0.5$,
as $\abs{g}$ increases the intersection eventually disappears
at around $\abs{g} = 2.5 \times 10^{-8}$ and one enters the no-go region; this is also seen directly in Fig.~\ref{fig:3_xi-100}. 
The disappearance of the intersection is also seen in Fig.~\ref{fig:n1_0.5} in which $n = 1$ and $\theta_i + \delta = 0.5$,
however this happens for values of~$\abs{g}$ that give $N_* > 60$;
hence the no-go region appears in Fig.~\ref{fig:3_xi-100pos} on the right side of the orange line.\footnote{Values of $\varphi_*$ and $\lambda$ are derived in (\ref{eq:zeroth}) and (\ref{varphilinear})
up to linear order in~$g$, however we note that the no-go region emerges where the $g$-expansion breaks down.}

One sees in Fig.~\ref{fig:n1_pi-0.5} that as $\abs{g}$ increases from zero, the value of $\varphi_*$ at the intersection point first decreases (as is the case in Fig.~\ref{fig:n0_pi-0.5} for $\delta = \pi - 0.5$), and then turns to increase (as in Figs.~\ref{fig:n0_0.5} and \ref{fig:n1_0.5} for $n \theta_i + \delta = 0.5$).
This non-monotonic behavior is due to $m_{\theta *}^2 /H_*^2$ approaching unity and forcing the axion to roll down to the vicinity of a minimum, as was shown in Fig.~\ref{fig:3_xi-100}. 
Finally, we should comment on the second branch of intersections seen
in Figs.~\ref{fig:n0_0.5}, \ref{fig:n1_0.5}, and \ref{fig:n1_pi-0.5}.
As $\abs{g}$ increases the solutions in the two branches approach each other, and merge right before one enters the no-go region.
We have ignored this second branch, since it appears at values of $\varphi_*$ larger than those in the first branch, and hence tends to yield $N_*$ that is too large.

\section{Analytic arguments}\label{Section4}

We now analytically derive an approximate expression for the constraints in Fig.~\ref{fig:newquality}, by evaluating the modulation of the number of $e$-folds by higher-dimensional operators. 
For this purpose we can ignore the time evolution of the axion and fix it to a constant value at $\theta = \theta_*$, as we will later verify. 
For the equation of motion of the radial field~(\ref{eq:chi-EoM}) and the Friedmann equation~(\ref{eq:Friedmann}), we assume them to be approximated respectively by the slow-roll expressions (\ref{eq:varphi-SL}) and (\ref{eq:FR-SL}). The necessary conditions for this can be obtained by comparing the approximations and their derivatives with the full equations, which gives
\begin{equation}
 \frac{M^2_{\ro{Pl}}}{\xi \varphi^2}, \, 
\frac{m_\theta^2\varphi^2}{n^2 H^2 M^2_{\ro{Pl}}},  \, 
\frac{\xi f^2}{M^2_{\ro{Pl}}}, \, 
\frac{\xi \abs{\Lambda}}{\lambda \varphi^2 M^2_{\ro{Pl}}}
\ll 1.
\label{eq:SLcond}
\end{equation}
Upon deriving these conditions, 
we assumed $l$
to be larger than four but at most of order ten, 
and cosine factors to be of order unity.
The smallness of the first term corresponds to the large-field condition (\ref{eq:large-field}), and the third term is equivalent to the condition~(\ref{eq:xi-f}).
Given that $\Lambda$ is chosen as (\ref{eq:Lambda}), then the forth term being small follows from the smallness of the other quantities.
The smallness of the second term is imposed by the requirement that the second line of (\ref{eq:chi-EoM}) is negligible;
let us name this parameter as, 
\begin{equation}
    \kappa= \frac{m_\theta^2 \varphi^2 }{n^2 H^2 M^2_{\ro{Pl}}}\simeq 
\frac{72 |g| \xi}{\lambda}
\left(\frac{\varphi}{\sqrt{2}M_{\ro{Pl}}}\right)^{l-2}.
\label{kappa}
\end{equation}
Here upon moving to the far right-hand side, we used (\ref{eq:FR-SL}) and the definition of the axion mass~(\ref{massoftheta}).

In order to evaluate the effect of the higher-dimensional operator on the number of $e$-folds, let us go beyond the leading-order approximations
(\ref{eq:varphi-SL}) and (\ref{eq:FR-SL}) by including terms that 
explicitly depend on $g$.
Under the slow-roll conditions~(\ref{eq:SLcond}),
one can check that the most relevant $g$-dependent term
in the full equations (\ref{eq:chi-EoM}) and (\ref{eq:Friedmann})
is the second line of (\ref{eq:chi-EoM}).  
Hence instead of (\ref{eq:varphi-SL}), let us use 
\begin{equation}
  3 H \dot{\varphi} \simeq 
- \frac{\lambda }{6 \xi^2 (1 + 6 \xi )}
\frac{M^4_{\ro{Pl}}}{\varphi }
\Biggl\{ 1
- 
\frac{6 (l-4) \xi \abs{g} }{\lambda }
\left( \frac{\varphi}{\sqrt{2} M_{\ro{Pl}}}\right)^{l-2 }
\cos(n \theta_* +\delta)
\Biggr\},
\label{eq:varphi-SL_NL}
\end{equation}
which combined with (\ref{eq:FR-SL}) gives 
\begin{equation}
 dN = \frac{H}{\dot{\varphi}} d \varphi
\simeq - \frac{1 + 6 \xi }{4 M^2_{\ro{Pl}}} \varphi \, d \varphi \Biggl\{ 1 +  
\frac{6 (l-4) \xi \abs{g} }{\lambda }
\left( \frac{\varphi}{\sqrt{2}M_{\ro{Pl}}}\right)^{l-2 }
\cos(n \theta_* +\delta )
\Biggr\}.
\label{eq:3.4}
\end{equation}
Here we note that the time evolution of $\theta$ affects (\ref{eq:chi-EoM}) and (\ref{eq:Friedmann}) through terms proportional to either $\dot{\theta}^2$ or $m_\theta^2 \dot{\theta} \sin (n \theta + \delta)$. Given that the axion velocity is sourced by the higher-dimensional operator, we expect $\dot{\theta} \propto m_\theta^2 \propto g$. Hence the evolution of $\theta$ affects the $e$-folding number at quadratic order in~$g$, and this justifies setting $\theta$ to a constant.
  
Supposing the condition (\ref{eq:SLcond}) to hold throughout inflation and integrating (\ref{eq:3.4}) gives
\begin{equation}
  N_* \simeq
(1 + 6 \xi)
\Biggl\{ \frac{\varphi_*^2}{8M^2_{\ro{Pl}}}  
+\frac{3(l-4)}{l}
\frac{\xi \abs{g}}{\lambda }
 \left(\frac{\varphi_*}{\sqrt{2}M_{\ro{Pl}}}\right)^{l} 
\cos{(n\theta_*+\delta)}
\Biggr\},
\label{eq:starting}
\end{equation}
where we have neglected terms that depend on the field value at the end of inflation.
The first term in curly brackets is the leading contribution which was 
derived in (\ref{eq:numberofefoldspq}). 
The second term represents the correction from the higher-dimensional operator; this can either increase or decrease $N_*$ depending on the sign of $\cos (n \theta_* + \delta)$, as discussed in Section~\ref{Section:3.2}.  

We further fix $\varphi_*$ and $\lambda$ in terms of the scalar power spectrum amplitude and spectral index, as was done in the numerical study in the previous section. 
Expanding the slow-roll results for the observables (\ref{spectralindex2}) up to linear order in~$g$ gives 
\begin{align}
    &  A_s \simeq \frac{\lambda \left(1+6 \xi \right) \varphi_*^4}{4608 \pi^2 \xi M_{\ro{Pl}}^4} + \frac{ \left(l-4\right)\left(1+6\xi  \right) \abs{g}}{ 96 \pi^2 } \left(\frac{\varphi_*}{\sqrt{2} M_{\ro{Pl}}} \right)^{l+2} \cos\left( n \theta_* + \delta \right),
    \label{eq:3.21}
\\
    & n_s-1\simeq -\frac{16 M^2_{\ro{Pl}}}{(1+6\xi)\varphi_*^2} - \frac{24 \left(l-4\right)^2 \xi \abs{g}}{ (1+6\xi) \lambda}\left(\frac{\varphi_*}{\sqrt{2} M_{\ro{Pl}}}\right)^{l-4} \cos \left(n \theta_* +\delta \right).
    \label{eq:3.20}
\end{align}
Here we obtained each term in the right-hand sides at leading order in~$\Mp^2 / \xi \varphi_*^2$, and neglected terms containing $f$ and $\Lambda$.
These equations can be solved for $\varphi_*$ and $\lambda$. 
Let us expand the quantities in powers of $g$,
\begin{equation}
 \varphi_* = \varphi_{*0}+\varphi_{*1} + \cdots, 
\quad 
 \lambda = \lambda_0 + \lambda_1 + \cdots,
    \label{eq:expasnionphil}
\end{equation}
where the numbers in the subscript represent orders of~$g$.
Then at the zeroth order we get
\begin{equation}
\varphi_{*0}\simeq \frac{4 M_{\ro{Pl}}}{\sqrt{(1-n_s) (1+6\xi)}}, 
\quad 
\lambda_0 \simeq 
18 \pi^2 A_s (1-n_s)^2 \xi (1+6 \xi) ,
\label{eq:zeroth}
\end{equation}
as also shown in Fig.~\ref{number1} for $\xi \gtrsim 10^{-2}$.
At linear order,
\begin{equation}
 \varphi_{*1} \simeq \frac{(l-4)^2 }{48} \varphi_{*0} \kappa_{*0} \cos{(n\theta_*+\delta)}, 
\quad 
\lambda_1 \simeq - \frac{ (l-2) (l-4)}{12} \lambda_0 \kappa_{*0} \cos{(n\theta_*+\delta)},
\label{varphilinear}
\end{equation}
where $\kappa_{*0}$ is as shown in the far right-hand side of (\ref{kappa}), but with the replacements $\varphi \to \varphi_{*0}$ and $\lambda \to \lambda_0$. 

Substituting (\ref{eq:zeroth}) and (\ref{varphilinear}) into (\ref{eq:starting}), we obtain up to linear order in $g$,
\begin{equation}
 N_* \simeq \frac{2}{1-n_s}
\left\{1 + \frac{(l-2)^2 (l-4)}{24 l}
\kappa_{*0} \cos{(n\theta_*+\delta)} \right\}.
\end{equation}
In the absence of higher-dimensional operators, the $e$-folding number is given by\footnote{The actual $e$-folding number is shifted to $N_* \approx 56.5$ as reported in Section \ref{Section2}, mainly due to deviations from slow-roll towards the end of inflation.} 
$N_* \simeq 2/(1-n_s) \approx 57$ for $n_s=0.965$. 
The leading correction by the higher-dimensional operator is controlled by the parameter $\kappa$;
this arises from both the first term of (\ref{eq:starting}) through the correction to~$\varphi_*$, as well as the second term.
For the $e$-folding number to lie within the range $50 < N_* < 60$, the higher-dimensional correction should be at most $ \sim 10\%$. This imposes 
\begin{equation}
 \frac{(l-2)^2 (l-4)}{24 l} \kappa_{*0}  \lesssim 10^{-1},
\label{eq:4.12}
\end{equation}
where we assumed the cosine factor to be of order unity. 
Rewriting $\kappa_{*0}$ in terms of observables using (\ref{eq:zeroth}), one obtains an upper bound on $\abs{g}$ as
\begin{equation}
\begin{split}
 \abs{g} &\lesssim  10^{-1} \frac{48 \pi^2 l A_s (1-n_s)}{(l-2)^2 (l-4)}
\left\{ \frac{(1+6 \xi ) (1-n_s)}{8}  \right\}^{l/2} , \\
& \sim 10^{-9}
\frac{ l  }{(l-2)^2 (l-4)}
\left( \frac{1+6 \xi }{230}  \right)^{l/2}.
\label{eq:linearprediction}
\end{split}
\end{equation}
Here, upon moving to the second line we used $A_s=2.1\times10^{-9}$ and $n_s=0.965$.

\begin{figure}[t!]
\centering
    \includegraphics[width=0.6\textwidth]{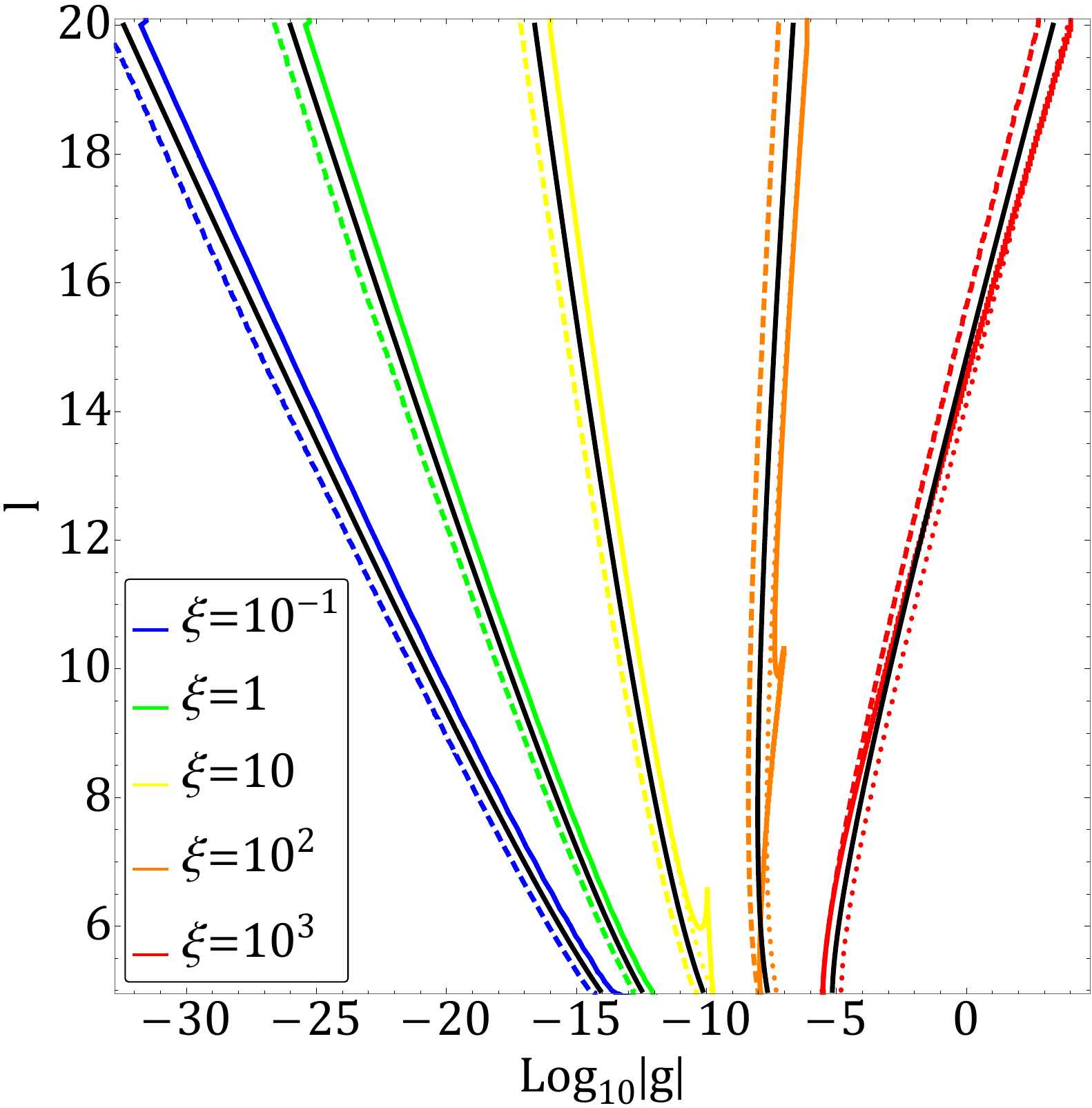}
    \captionof{figure}{Analytic bound (\ref{eq:linearprediction}) on higher-dimensional operators shown in black solid lines, compared with the numerical bounds from Fig.~\ref{fig:newquality}.}
      \label{fig:kappa}
\end{figure}

In Fig.~\ref{fig:kappa}, the analytic upper bound (\ref{eq:linearprediction}) is shown as black solid lines for different values of~$\xi$, overlaid with the numerical bounds from Fig.~\ref{fig:newquality}. 
The analytic estimate is seen to reproduce well the numerical results.
The tilt of the bound in the $l$ - $\log |g|$ plane is predominantly set by the sign of $\log [(1+6\xi) / 230]$; this factor derives from the ratio $\varphi_* / \sqrt{2} \Mp$ in (\ref{kappa}). 
For $\xi \lesssim 40$, the bound on $|g|$ becomes stronger with increasing~$l$. This reflects the fact that such values of $\xi$ require field excursions of $\varphi_* > \Mp$
(see also Fig.~\ref{number1}), rendering the system particularly sensitive to higher-dimensional operators.

Using the analytic bound on $\abs{g}$, we can also derive an upper bound for the axion mass during inflation.
Rewriting its ratio to the Hubble rate as $m_{\theta *}^2 / H_*^2 = n^2 \kappa_* \Mp^2 / \varphi_*^2$, approximating $\varphi_*$ and $\kappa_*$ by their zeroth order values, and using (\ref{eq:zeroth}), (\ref{eq:4.12}), and $n_s = 0.965$, one obtains
\begin{equation}
 \frac{m_{\theta *}^2}{H_*^2} \lesssim 10^{-2}
\frac{n^2 l (1+ 6 \xi)}{(l-2)^2 (l-4)} .
\label{eq:4.14}
\end{equation}
This expression explicitly shows that the maximum mass-Hubble ratio allowed within a consistent PQ inflation increases with $\xi$, as was seen in Fig.~\ref{fig:newqualityxi100}.

\section{Validity of single-field approximation}
\label{Section5}

In the previous sections we have treated PQ inflation as effectively a single-field model and computed the scalar perturbation, which was then compared to observational results to fix the model parameters. We now verify the validity of the single-field approximation by estimating the contribution of the axion field to the perturbation.

Let us evaluate scalar perturbations in the spatially flat gauge \citep{Bartolo:2004if},
\begin{align}\label{eq:Zeta}
\zeta = \frac{H}{\dot\rho} \delta\rho \simeq \frac{H}{\dot\rho} \Bigl( U_\varphi \delta\varphi + U_\theta \delta\theta \Bigr)  ,
\end{align}
where $\rho$ is the background energy density and 
$\delta \rho$ is the density perturbation.
In the far right-hand side we ignored terms containing spacetime derivatives of the fields and expanded the potential energy at linear order in the field fluctuations, with 
$U_\varphi = \partial U / \partial \varphi$ and 
$U_\theta = \partial U / \partial \theta$.
We can then go to Fourier space and write the scalar power spectrum as
\begin{equation}
    P_\zeta(k) \simeq \left(\frac{H}{\dot\rho} \right)^2 \Bigl( U_\varphi^2 P_{\delta\varphi}(k) + U_\theta^2 P_{\delta\theta}(k) \Bigr)   ,
    \label{eq:powerspectrum}
\end{equation}
where $P_{\delta\varphi}$ and $P_{\delta\theta}$ are the power spectra of the field fluctuations, and we ignored cross-correlations between 
$\delta \varphi$ and $\delta \theta$.\footnote{The interaction between $\theta$ and the almost canonical $\chi$ through the axion's kinetic term is suppressed at $\xi \varphi^2 \gg \Mp^2$ (cf. (\ref{eq:S-einstein})), while that through the higher-dimensional operator is suppressed by the coupling~$g$.}

The time derivative of the background density is evaluated using the continuity equation as,
\begin{equation}
 \frac{\dot{\rho}}{H} = -3 \left(
I^2 \dot{\varphi}^2 + \frac{\varphi^2}{\Omega^2}\dot{\theta}^2
\right)
\simeq -
\frac{1}{3 H^2}
 \left(
\frac{U_\varphi^2}{I^2} + 
\frac{\Omega^2 U_\theta^2}{\varphi^2}
\right).
\label{eq:continuity}
\end{equation}
The factors $I^2$ and $\varphi^2 / \Omega^2$ are respectively the coefficients of the kinetic terms of $\varphi$ and $\theta$ in the Einstein frame action~(\ref{eq:S-einstein}).
Upon moving to the far right-hand side we assumed both fields to slow-roll, namely, to follow
$3 H \dot{\varphi} \simeq -U_\varphi / I^2 $ and
$3 H  \dot{\theta} \simeq -U_\theta (\Omega^2 / \varphi^2)$.
We also evaluate the power spectra of the field fluctuations at the wave mode~$k_*$, when the mode exits the horizon, as
\begin{equation}
P_{\delta\varphi*}(k_*) \simeq 
\left. \frac{1}{I^2} \left(\frac{H}{2\pi}\right)^2 \right|_* , 
\quad 
P_{\delta\theta*} (k_*) \simeq 
\left. \frac{\Omega^2}{\varphi^2} \left(\frac{H}{2\pi}\right)^2 \right|_*.
\label{eq:5.4}
\end{equation}

Combining (\ref{eq:powerspectrum}), (\ref{eq:continuity}), and (\ref{eq:5.4}) with the slow-roll approximation for the Friedmann equation 
$3 \Mp^2 H^2 \simeq U$, we obtain the scalar power spectrum at the horizon exit of the wave mode~$k_*$ as
\begin{equation}
  P_{\zeta *} (k_*) \simeq \left.
\frac{I^2 U^3}{12 \pi^2 \Mp^6 U_\varphi^2}
\left( 1 + 
\frac{I^2 \Omega^2 }{\varphi^2 }
\frac{U_\theta^2}{U_\varphi^2}
\right)^{-1}
\right|_*.
\end{equation}
The expression in front of the parentheses is equivalent to the single-field result of (\ref{spectralindex2}).
The contribution from the axion kinetic term $\dot{\theta}^2$ through $\dot{\rho}$ (cf.~(\ref{eq:continuity})), and the axion fluctuation $P_{\delta \theta}$ (cf.~(\ref{eq:5.4})), 
both yield corrections to the scalar power spectrum of order:
\begin{equation}
\alpha= \left. \frac{I^2 \Omega^2 }{\varphi^2 }
\frac{U_\theta^2}{U_\varphi^2} \right|_* \, .
\label{eq:alpha}
\end{equation}

Focusing on the large-field regime $\xi \varphi_*^2 \gg \Mp^2$, and ignoring $g$ corrections in $U_\varphi$, the quantity~$\alpha$ is approximated by 
\begin{equation}
\alpha
\simeq
\frac{n^2 (1 + 6 \xi )}{144}\kappa_*^2
\sin^2 (n \theta_* + \delta),
\label{eq:alphalin}
\end{equation}
where $\kappa_*$ is as shown in the far right-hand side of (\ref{kappa}) but with the replacement $\varphi \to \varphi_*$. 
This shows that for instance with $n = 1$, $\kappa_* \lesssim 0.1$, and $\xi \lesssim 0.1$, the contribution by the axion to the scalar power is of $\alpha \lesssim 10^{-4}$.
One may expect from (\ref{eq:alphalin}) that the axion contribution increases with $\xi$, however a large $\xi$ also enhances 
$m_{\theta *}^2 / H_*^2$ and reduces the system to an effective single-field, as we see below. 

\begin{figure}[ht!]
\centering
\subfigure[$\xi=10$, $\theta_i+\delta=0.5$]{
\includegraphics[width=0.4\textwidth]
{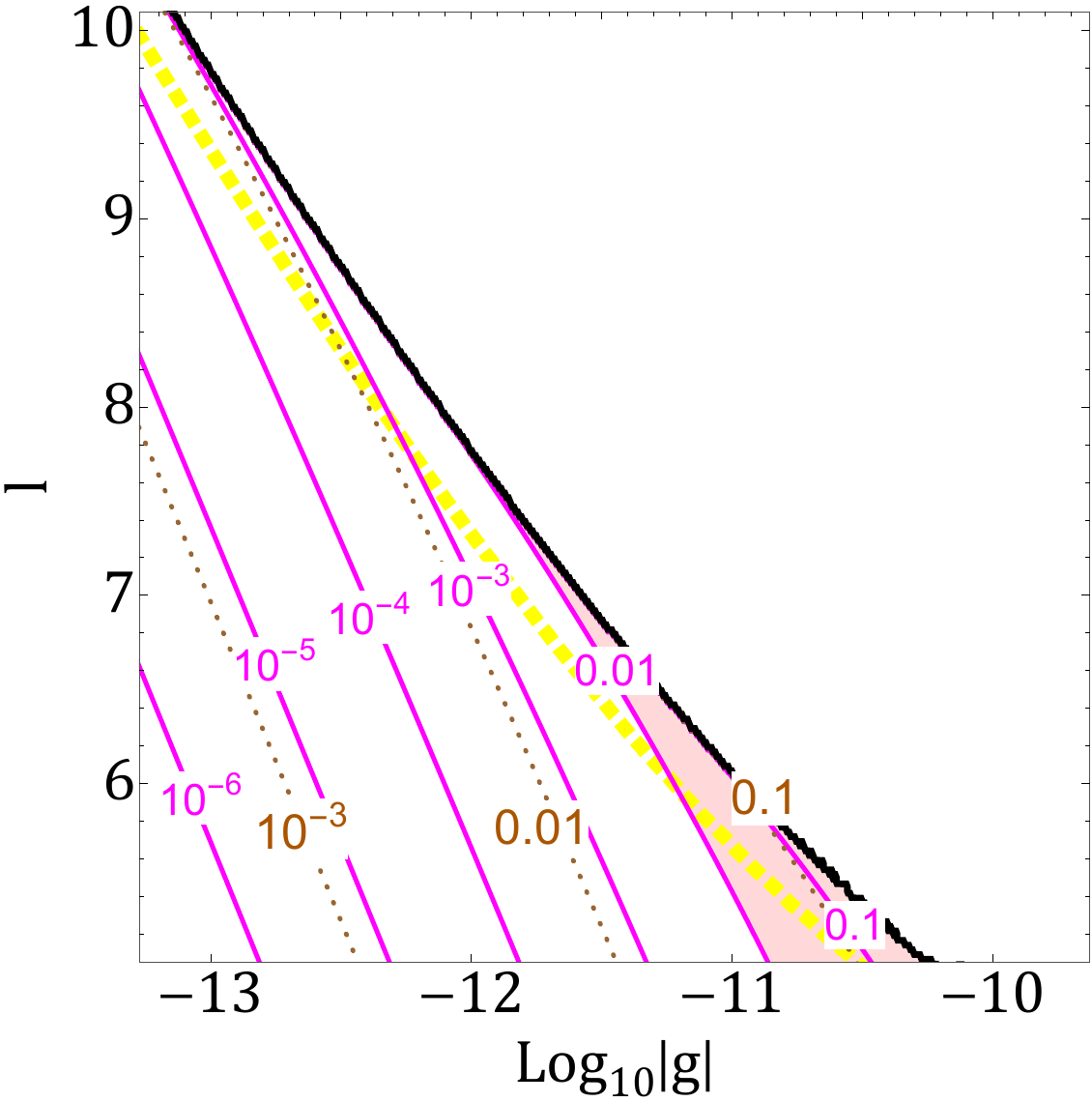}
\label{fig:10_0.5}
}
\subfigure[$\xi=10$, $\theta_i+\delta=\pi-0.5$]{
\includegraphics[width=0.4\textwidth]
{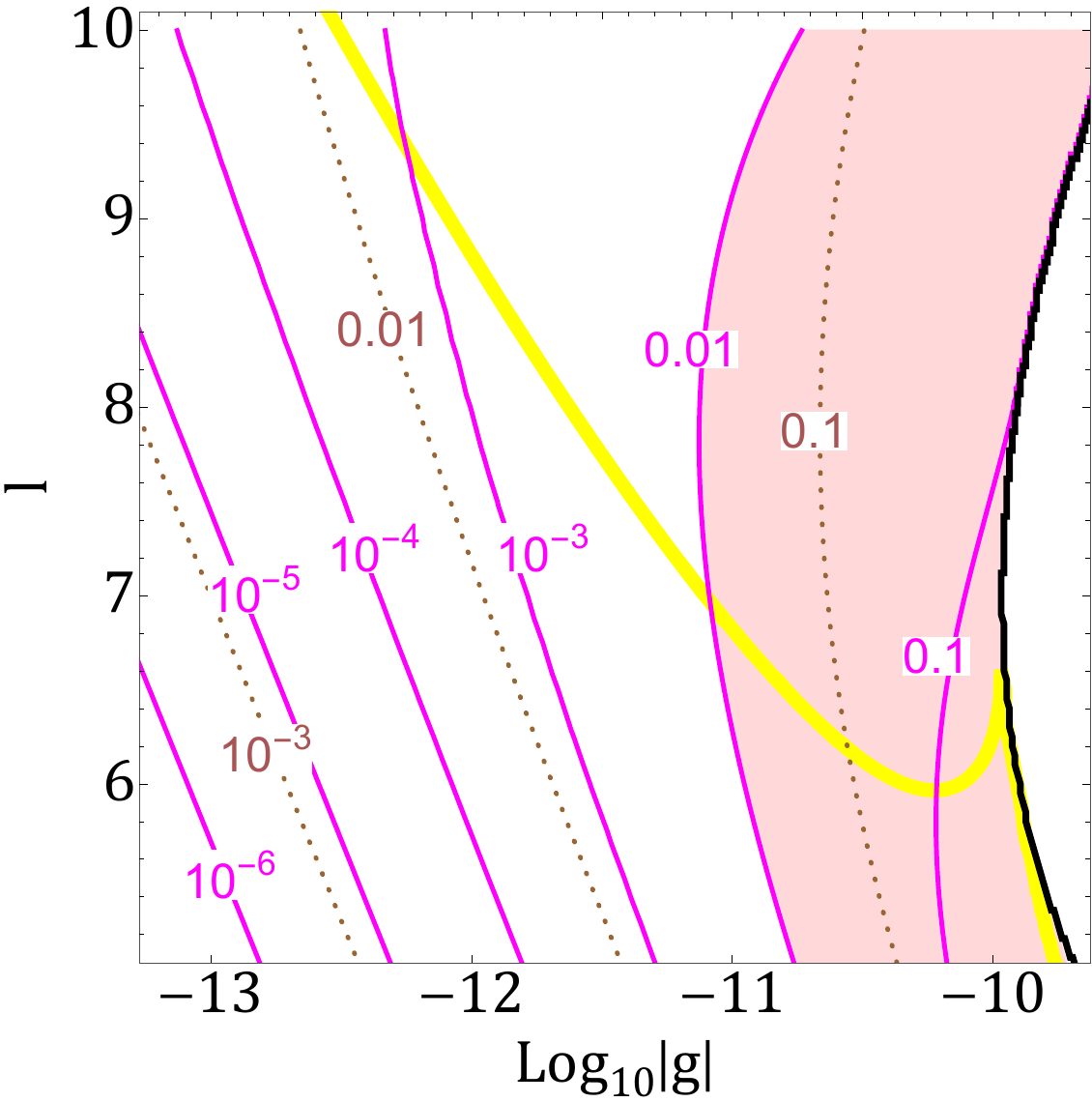}
\label{fig:10_pi-0.5}
}
\subfigure[$\xi=10^2$, $\theta_i+\delta=0.5$]{
\includegraphics[width=0.4\textwidth]
{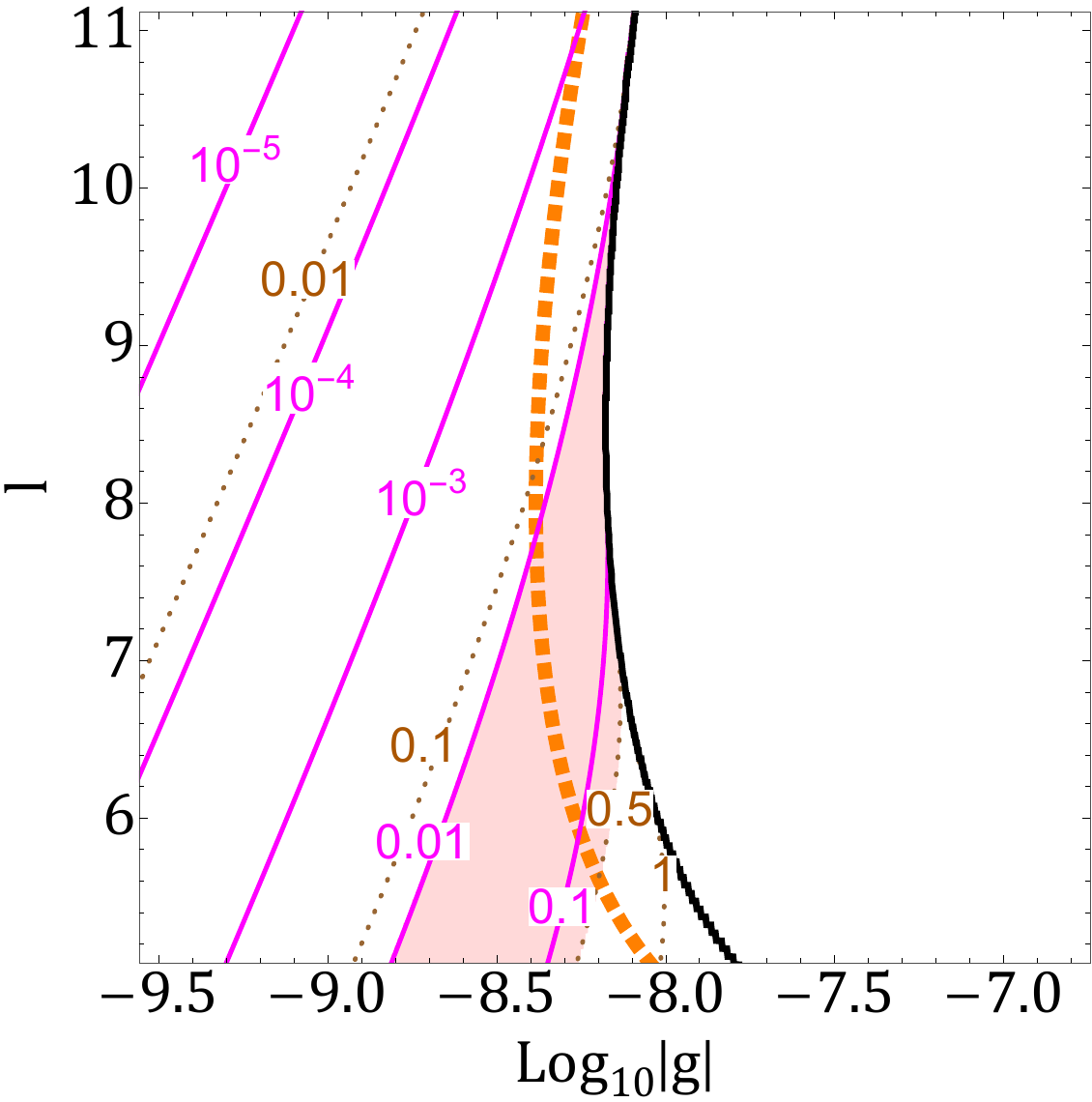}
\label{fig:100_0.5}
}
\subfigure[$\xi=10^2$, $\theta_i+\delta=\pi-0.5$]{
\includegraphics[width=0.4\textwidth]
{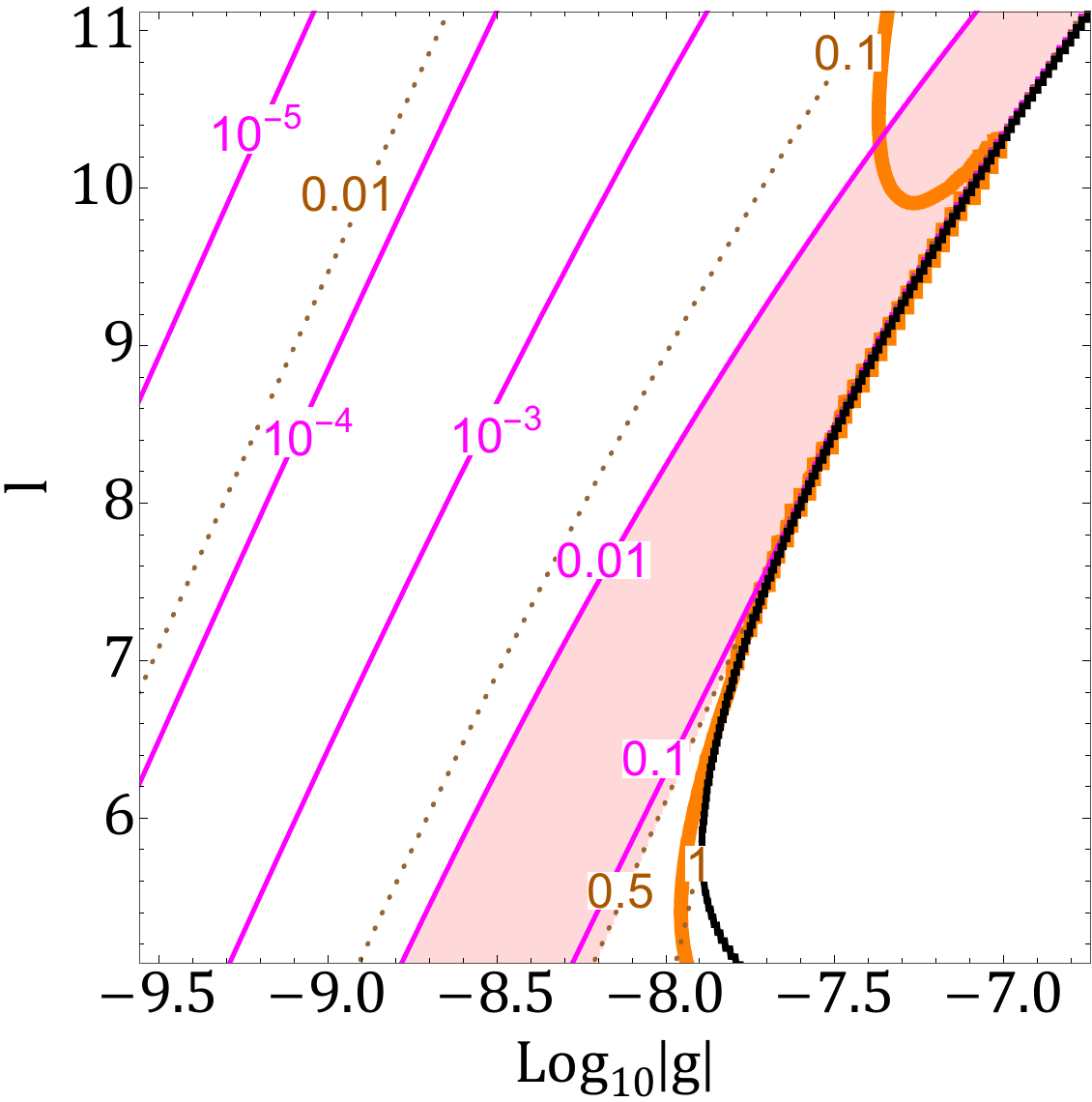}
\label{fig:100_pi-0.5}
}
\captionof{figure}{(Un)importance of multi-field effects in PQ inflation. The pink shaded regions show where the axion field affects the scalar power spectrum by more than~$1 \%$. The magenta solid contours show the axion's contribution to the scalar power, and the brown dotted show values of $m_{\theta *}^2 / H_*^2$. 
All the results are for $n = 1$, while $\xi$ and $\theta_i + \delta$ are varied in each panel.
Overlaid are constraints on higher-dimensional operators as shown in Fig.~\ref{fig:newquality}.
On the right of the black solid lines, $A_s$ and $n_s$ cannot simultaneously take the observed values.
}
\label{fig:single}
\end{figure}

In Fig.~\ref{fig:single} we show the values of $\alpha$ as defined in (\ref{eq:alpha}) by magenta contours in the $|g|$ - $l$ plane.
All the results are for $n = 1$, 
while the values of $\xi$ and $\theta_i + \delta$ are varied in each panel.
(Note that in the $\ro{U}(1)_{\ro{PQ}}$-symmetric case, i.e. $n = 0$, the axion field fluctuations do not source curvature perturbations.)
Upon plotting $\alpha$, we replaced $\theta_*$ in 
the definition~(\ref{eq:alpha}) with~$\theta_i$, which amounts to ignoring the axion rolling before the horizon exit of the pivot scale.
This is because if $m_{\theta *}^2 \ll H_*^2$, 
as assumed in the above discussions, 
then it follows from (\ref{eq:smothingtheta}) that $\theta_* \simeq \theta_i$.
On the other hand if $m_{\theta *}^2 \gtrsim H_*^2$, the analyses above break down; however in such cases the axion does not obtain super-horizon field fluctuations,
and thus it cannot source cosmological perturbations in the first place.
To also examine such cases, we show in the plots the values of $m_{\theta *}^2 / H_*^2$ as brown dotted contours. 
Both $\alpha$ and $m_{\theta *}^2 / H_*^2$ increase with~$\abs{g}$.
We assess the contribution from the axion to the scalar power spectrum to exceed~$\sim 1 \%$ when both $\alpha \gtrsim 0.01$ and $m_{\theta *}^2 / H_*^2 < 0.5$ are satisfied; these regions are shown in pink.

Overlaid in the plots are the bounds on higher-dimensional operators from Fig.~\ref{fig:newquality}, for 
$\xi = 10$ (yellow), $10^2$ (orange),
with dashed lines for $\theta_i + \delta = 0.5$ and 
solid lines for $\theta_i + \delta = \pi - 0.5$.
The black solid line denotes the left edge of the no-go region where the observed values of $A_s$ and $n_s$ cannot be simultaneously realized
(cf. Section~\ref{sec:no-go}).
Some parts of the bounds on higher-dimensional operators, which we have derived using the single-field approximation, are seen to run through the pink regions. 
Here the exact position of the bounds are subject to corrections from multi-field effects. However we remark that these correspond to very small parts in the displayed area of Fig.~\ref{fig:newquality}, 
and in particular that the ranges of $\abs{g}$ over which the bounds pass through the pink regions are less than about one order of magnitude.
Other parts of the bounds for $\xi = 10$ and $10^2$ in  Fig.~\ref{fig:newquality} do not cross the pink regions.
For $\xi=10^3$, the range of~$\abs{g}$ over which the bounds cross the pink regions are even smaller than the cases shown in Fig.~\ref{fig:single}.
The bounds for $\xi = 10^{-1}$ and $1$ do not enter the pink regions at all, as one expects from~(\ref{eq:alphalin}). 
Hence we conclude that PQ inflation is well approximated as single-field in most of the parameter space, 
and moreover, that multi-field effects alter our bounds on higher-dimensional operators by no more than an order of magnitude.

We should remark that in the above discussions we evaluated the curvature perturbation at the time when the mode~$k_*$ exits the horizon. However the curvature perturbation can evolve outside the horizon in the presence of isocurvature modes.
This effect is analyzed in Appendix~\ref{AppA} using the $\delta N$~formalism, where it is shown that the final contribution from the axion to the curvature spectrum is comparable to or smaller than $\alpha$ given in (\ref{eq:alpha}).
Hence the regions in parameter space where the axion's contribution exceeds $1 \%$ is actually even smaller than shown in Fig.~\ref{fig:single}.

Let us also comment that, when we consider the axion to slowly roll, we are implicitly assuming the quantum fluctuation (\ref{eq:5.4}) to be smaller than the classical rolling over a Hubble time, i.e. $P_{\delta \theta *}^{1/2} \ll \abs{\dot{\theta}_*}_{\ro{slow-roll}} / H_*$.
We have checked that this condition actually holds within the displayed regions of Fig.~\ref{fig:single}.
On the other hand, the quantum fluctuation can dominate as one moves to even smaller $\abs{g}$, larger $l$, or $n = 0$. However by combining 
(\ref{eq:5.4}) with (\ref{eq:large-field}), (\ref{eq:FR-SL}), and (\ref{eq:zeroth}), the quantum fluctuation is estimated to be of
\begin{equation}
  P_{\delta\theta*}^{1/2} \sim 10^{-7} \sqrt{1+ 6 \xi},
\end{equation}
which is much smaller than unity, unless $\xi$ is extremely large.
Hence we expect that our main conclusion that the axion sources negligible curvature perturbation remains valid even when the axion dynamics is governed by the quantum fluctuation. 
Likewise, we expect the main conclusions of the other sections that invoke axion slow-roll not to be altered by the domination of the quantum fluctuation.

\section{Parametric resonance}\label{Section6}

After inflation, the PQ field begins to oscillate around the vacuum. The oscillatory background can give rise to resonant amplifications of the field fluctuation, which would impact the post-inflation cosmology including reheating and axion dark matter production. 
One may expect that $\ro{U}(1)_{\ro{PQ}}$-breaking higher-dimensional operators can source an angular momentum to the PQ field and suppress resonant effects. However we now show that a resonant amplification is actually inevitably triggered after PQ inflation. 

Let us focus on the first oscillation of the radial field after the end of inflation. With the oscillation time scale being comparable to or shorter than the Hubble time, and the radial field bounded by $\varphi \leq \varphi_{\ro{end}}$, we ignore the expansion of the universe as well as higher dimensional operators.\footnote{It was claimed in \citep{Ema:2016dny} that a non-minimal gravitational coupling enhances a non-perturbative decay of the inflaton. Here we show that resonant effects are triggered even without the non-minimal coupling.}
Hence the $\ro{U}(1)_{\ro{PQ}}$ symmetry is unbroken and the PQ field's angular momentum is conserved, which we write as $L = \varphi^2 \dot{\theta}$. The equation of motion of the radial direction can be written as
\begin{equation}
 \ddot{\varphi} = -V_{\ro{eff}}' (\varphi), 
\end{equation}
where the effective potential is
\begin{equation}
 V_{\ro{eff}} (\varphi) = V(\varphi) + \frac{L^2}{2 \varphi^2},
\label{eq:Veff}
\end{equation}
with $V$ given in (\ref{eq:2.5}). This system has another conserved quantity, which is the energy density:
$\rho = \dot{\varphi}^2 / 2 + V_{\ro{eff}} (\varphi)$.
It is convenient to introduce dimensionless parameters as 
\begin{equation}
u = \frac{\varphi^2}{f^2}, \quad
Q = \sqrt{\frac{6}{\lambda }} \frac{\abs{L}}{ f^3}, \quad
E = \frac{4! \,  \rho}{\lambda f^4}.
\end{equation}
Then the field value that minimizes the effective potential is obtained by solving $V_{\ro{eff}}' \propto u^3 - u^2 - Q^2 = 0$ as
\begin{equation}
 u_{\ro{min}} = \frac{1}{3} 
\left\{ 1 + F(Q)^{1/3} + F(Q)^{-1/3} \right\},
\quad
 F(Q) = \frac{2 + 27 Q^2 + 3 Q \sqrt{12 + 81 Q^2}}{2} .
\label{eq:x_min}
\end{equation}

When the effective mass of the radial field, 
$m_{\varphi}^2 = V''(\varphi) = (\lambda /6)(3 \varphi^2 - f^2)$, 
varies with a time scale shorter than~$1/m_{\varphi}$, then adiabaticity is violated and the PQ field fluctuations are amplified (on wave modes typically of $\sim m_{\varphi}$) \citep{Kofman:1994rk,Kofman:1995fi,Tkachev:1995md,Kofman:1997yn}. As a measure of adiabaticity, let us evaluate the quantity $\dot{m}_{\varphi} / m_{\varphi}^2$, the square of which is written as
\begin{equation}
 \left(
\frac{\dot{m}_{\varphi}}{m_{\varphi}^2}
\right)^2
= \frac{E u - u (u - 1)^2 - 2 Q^2}{6 (u - \frac{1}{3})^3}.
\end{equation}
One necessary condition for the PQ mode functions to evolve adiabatically is that this quantity be smaller than unity at the minimum~(\ref{eq:x_min}) of the effective potential. 
If $Q \ll 1$, the adiabaticity parameter at $u = u_{\ro{min}}$ can be 
expanded in~$Q$ as
\begin{equation}
  \left(
\frac{\dot{m}_{\varphi}}{m_{\varphi}^2}
\right)^2_{\ro{min}} =
\left\{ \frac{9}{16} + \mathcal{O} (Q^2) \right\} E
-\frac{9}{8}Q^2 + \mathcal{O} (Q^4).
\label{eq:smallQ}
\end{equation}
This is smaller than unity for $E \lesssim 16/9$. 
On the other hand if $Q \gg 1$, then expanding in $1/Q$ yields
\begin{equation}
   \left(
\frac{\dot{m}_{\varphi}}{m_{\varphi}^2}
\right)^2_{\ro{min}} =
\left\{ \frac{1}{6 } Q^{-4/3} + \mathcal{O} (Q^{-2}) \right\} E
- \frac{1}{2} + \mathcal{O} (Q^{-2/3}),
\label{eq:largeQ}
\end{equation}
which is smaller than unity for 
\begin{equation}
Q \gtrsim \left( \frac{E}{9} \right)^{3/4}.
\label{eq:EQ9}
\end{equation}
The special case of $Q = (E / 3)^{3/4}$ leads to
$(\dot{m}_{\varphi} / m_{\varphi}^2)^2_{\ro{min}} = \mathcal{O} (Q^{-2/3}) \ll 1$; here the PQ field rotates in the complex plane with an almost circular orbit at $\varphi \simeq \varphi_{\ro{min}}$.

The potential and kinetic energies at the end of inflation are related by 
\begin{equation}
 V (\varphi_{\ro{end}}) = (\dot{\varphi}^2 + \varphi^2 \dot{\theta}^2)_{\ro{end}}.
\label{eq:PK_end}
\end{equation}
This yields $\rho = (3/2) V(\varphi_{\ro{end}})$, and by also using $\varphi_{\ro{end}}^2 \gg f^2$ we find
\begin{equation}
 E \simeq \frac{3}{2} \left( \frac{\varphi_{\ro{end}}}{f}\right)^4 \gg 1. 
\label{eq:6.10}
\end{equation}
Hence from the discussions around (\ref{eq:largeQ}), the adiabaticity condition 
$|\dot{m}_{\varphi} / m_{\varphi}^2|_{\ro{min}} < 1$ 
is satisfied only if $Q$ is as large as~(\ref{eq:EQ9}).
From (\ref{eq:PK_end}) it also follows that the ratio between the kinetic energy of the angular direction and the potential energy at the end of inflation is bounded as
\begin{equation}
 R = \left. \frac{\varphi^2 \dot{\theta}^2}{2 V (\varphi)}
\right|_{\ro{end}} \leq 0.5.
\label{eq:R}
\end{equation}
From (\ref{eq:6.10}), one finds that this energy ratio~$R$ is related to the dimensionless angular momentum as
\begin{equation}
 Q \simeq \left( \frac{2}{27} \right)^{1/4} E^{3/4} R^{1/2}
\end{equation}
The condition (\ref{eq:EQ9}) thus requires $R$ to lie within the range:
\begin{equation}
 0.1 \lesssim R \leq 0.5.
\label{eq:R0.1}
\end{equation}
We stress that this is a necessary but not a sufficient condition for adiabaticity during the oscillations.

Let us now estimate the size of $R$ by supposing that the angular field slow-rolls during inflation along the potential sourced by higher-dimensional operators as (\ref{eq:axion_slow-roll}).
Then the angular velocity at the end of inflation is estimated,
using also $\Omega_{\ro{end}}^2 \sim 1$, as
\begin{equation}
 3 H_{\ro{end}} \dot{\theta}_{\ro{end}} \sim - n \abs{g} \Mp^2 \left( \frac{\varphi_{\ro{end}} }{\sqrt{2} \Mp} \right)^{l-2} \sin(n \theta_{\ro{end}} + \delta),
\end{equation}
with the Hubble rate obtained using (\ref{eq:PK_end}) as
\begin{equation}
 H_{\ro{end}} \simeq \sqrt{\frac{\lambda }{48}} \frac{\varphi_{\ro{end}}^2}{\Mp}.
\end{equation}
\begin{figure}[t!]
\centering
\subfigure[ $\xi=10$, $\theta_i+\delta=0.5$]{
 \includegraphics[width=0.4\textwidth]{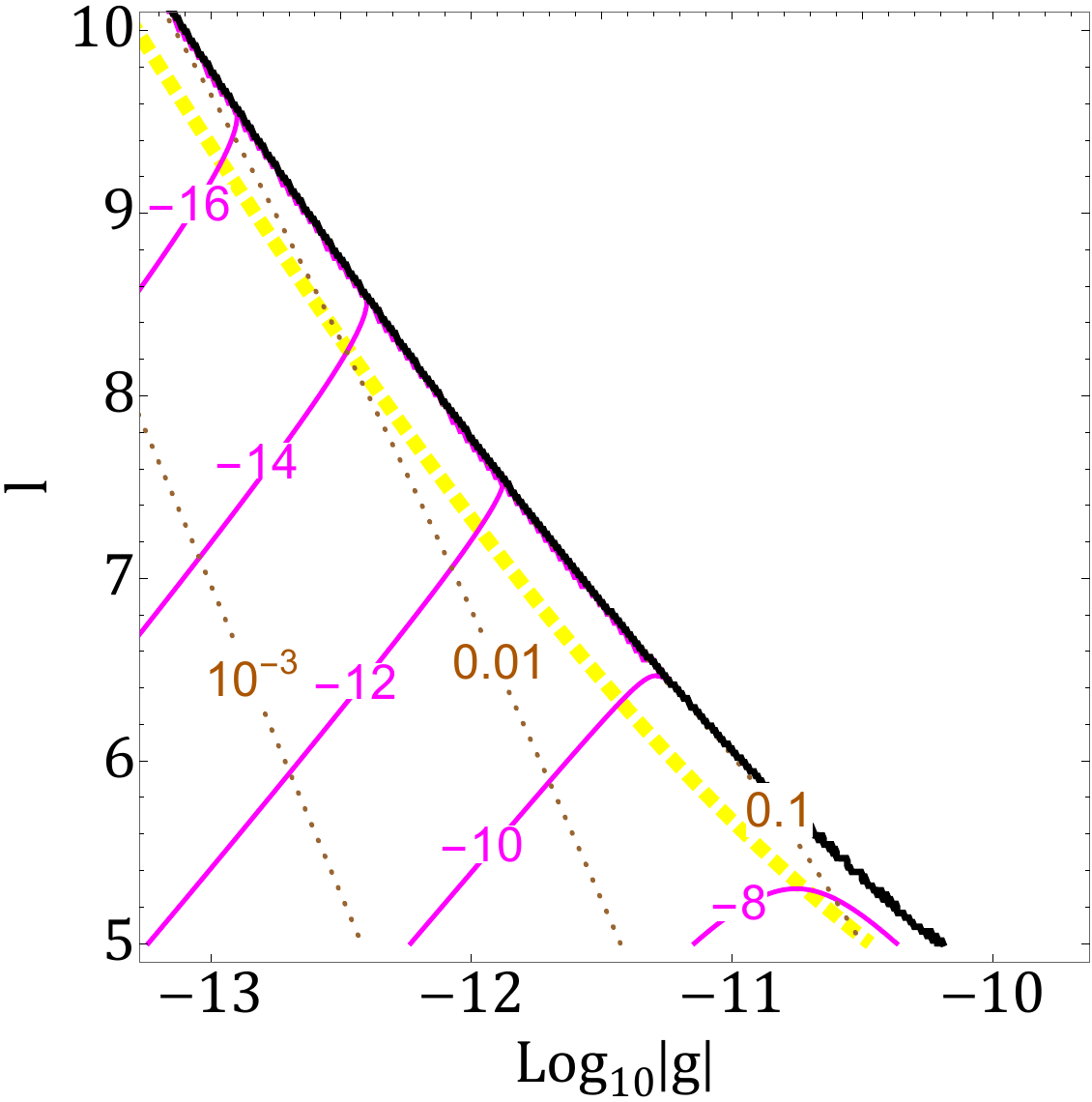} \label{fig:R10_0.5}
}
\subfigure[ $\xi=10$, $\theta_i+\delta=\pi-0.5$]{
\includegraphics[width=0.4\textwidth]{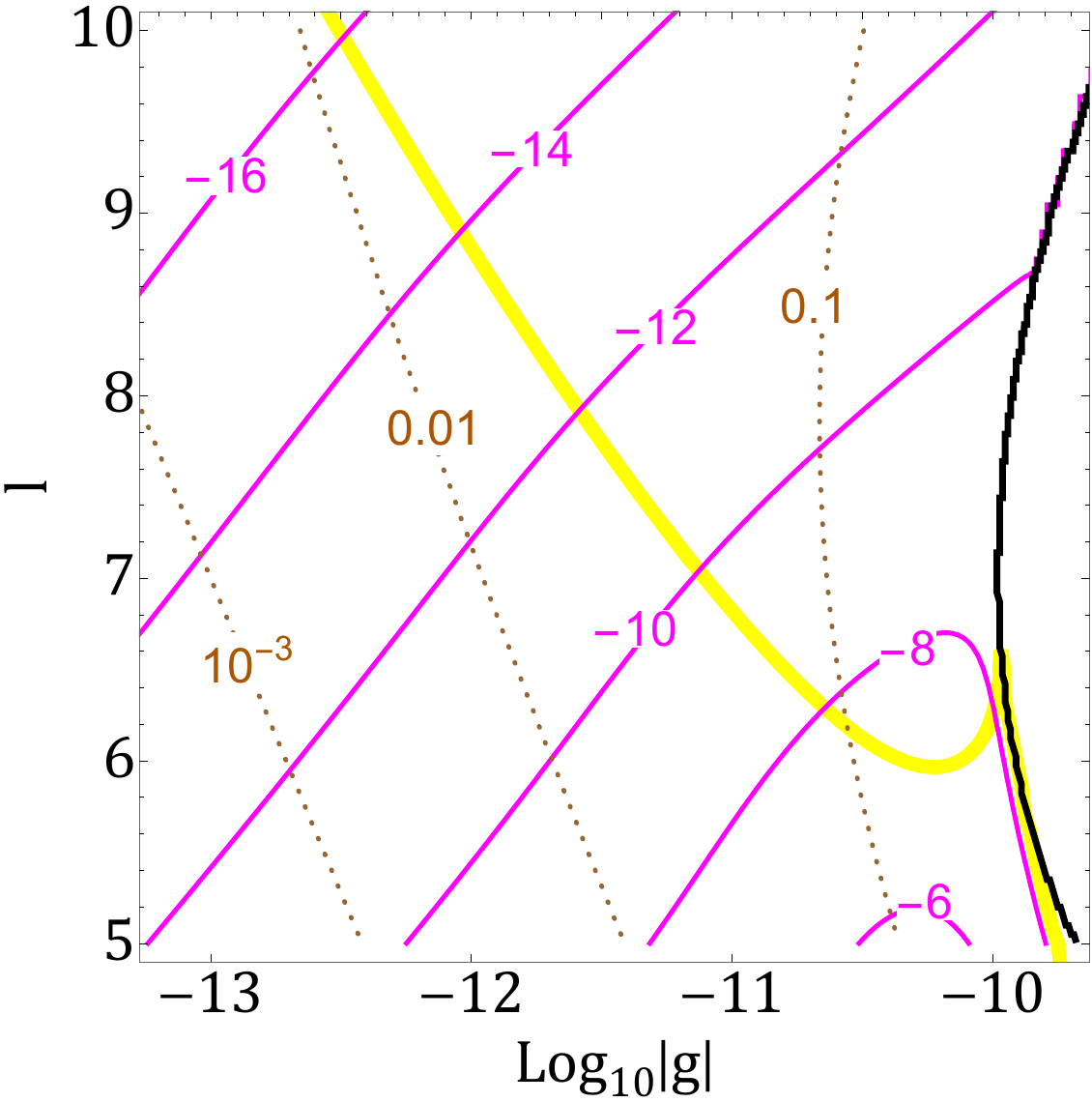}
\label{fig:R10_pi_0.5}
}
\subfigure[$\xi=10^2$, $\theta_i+\delta=0.5$]{
\includegraphics[width=0.4\textwidth]{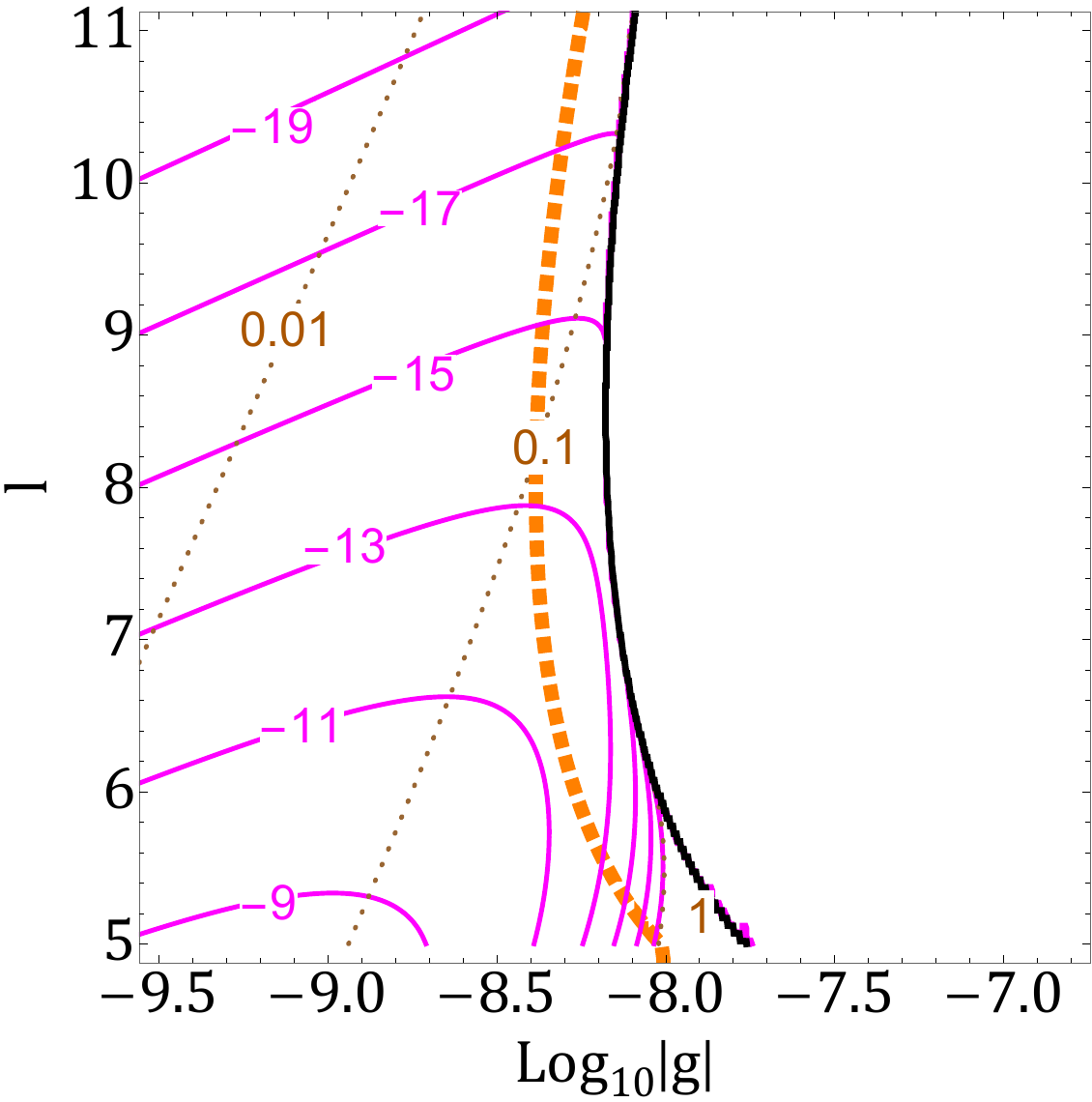}
\label{fig:R100_0.5}
}
\subfigure[$\xi=10^2$, $\theta_i+\delta=\pi-0.5$]{
\includegraphics[width=0.4\textwidth]{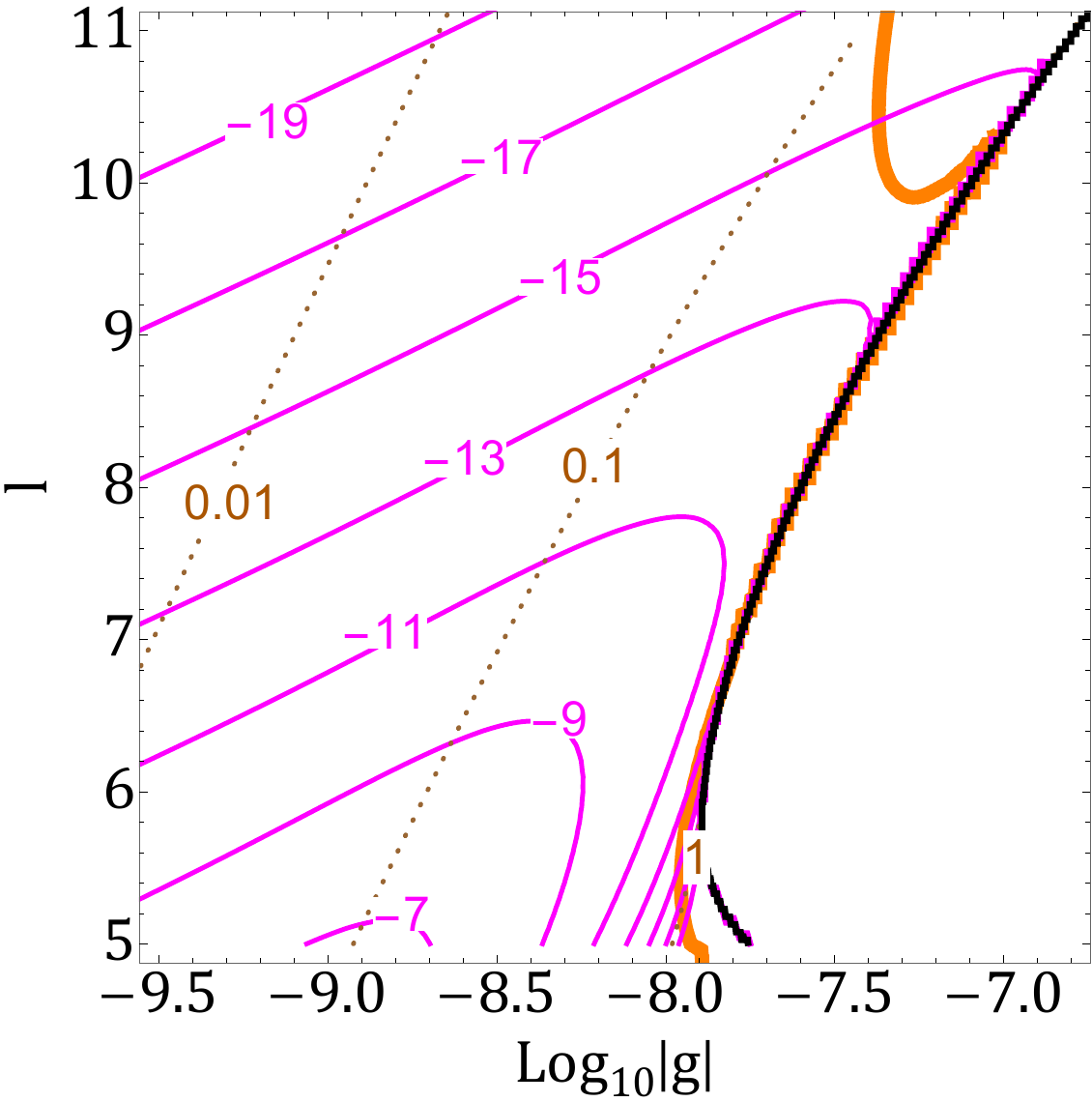}
\label{fig:R100_pi_0.5}
}
\captionof{figure}{Ratio~$R$ between the angular kinetic energy and potential energy at the end of PQ inflation.
The magenta contours show values of $\log_{10} R$, and the brown dotted contours show $m_{\theta *}^2 / H_*^2$. All the results are for $n=1$, while $\xi$ and $\theta_i+\delta$ are varied in each panel. Overlaid are constraints on higher-dimensional operators as shown in Fig.~\ref{fig:newquality}. On the right of the black solid lines, $A_s$ and $n_s$ cannot simultaneously take the observed values.}
\label{fig:angularenergy}
\end{figure}
Further plugging (\ref{eq:phiendPQ}) and (\ref{eq:zeroth}) respectively into $\varphi_{\ro{end}}$ and $\lambda$, 
as well as using the upper bound (\ref{eq:linearprediction}) on $|g|$ 
and $|\sin (n \theta_{\ro{end}} + \delta)| \leq 1$, one arrives at 
\begin{equation}
R \lesssim
  \begin{dcases}
 \frac{10^{-3}}{\xi^{2}}
\frac{n^2 l^2}{(l-2)^4 (l-4)^2}
\left(\frac{1-n_s}{2}\right)^{l-2}
\lesssim  10^{-9} \frac{n^2}{\xi^2}
 & \ro{for}\, \, \,  \xi \ll 10^{-1}, \\
10^{-1} \xi 
\frac{n^2 l^2}{(l-2)^4 (l-4)^2}
\left(\frac{\sqrt{3} (1-n_s)}{4}\right)^{l-2}
\lesssim 10^{-7} n^2 \xi 
  & \ro{for}\, \, \,  \xi \gg 10^{-1}.
 \end{dcases}
\label{eq:Rupper}
\end{equation}
Upon going to the far right-hand sides, we used $n_s = 0.965$,
and also that the expressions are maximized at $ l = 5$.
The upper limit in the first line increases with decreasing~$\xi$, however $\xi \gtrsim 10^{-2}$ is required for the model not to reduce to a $\varphi^4$~inflation (cf. Fig.~\ref{number1}); this constrains $R$ to be much smaller than~$10^{-1}$.
From the second line one may except $R$ to increase with~$\xi$, however this upper limit cannot always be saturated for large~$\xi$, as we will soon see.

We also numerically computed~$R$ as defined in (\ref{eq:R}), whose values are shown in the $|g|$ - $l$ plane in Fig.~\ref{fig:angularenergy}.
All the results are for $n = 1$, while $\xi$ and $\theta_i + \delta$ are varied in each panel.
The magenta contours show values of $\log_{10} R$. 
Also shown are brown dashed contours for the values of 
$m_{\theta *}^2/H_*^2$,
constraints on higher-dimensional operators from Fig.~\ref{fig:newquality}, 
and black lines denoting the left edge of the no-go region where $A_s$ and $n_s$ cannot simultaneously take the observed values.
For $\xi = 10$, the maximum value of $R$ is of $10^{-8}$ in Fig.~\ref{fig:R10_0.5} and $10^{-6}$ in Fig.~\ref{fig:R10_pi_0.5},
realized close to the lower right corner where the yellow line hits the lower edge of $l=5$.
These maximum values roughly match with that given in the second line of (\ref{eq:Rupper}).
On the other hand for $\xi = 10^2$, the maximum $R$ is of $10^{-9}$ in Fig.~\ref{fig:R100_0.5} and $10^{-7}$ in Fig.~\ref{fig:R100_pi_0.5}, which are significantly smaller than the upper limit of~(\ref{eq:Rupper}).
This is because in the lower right corners of these plots, $m_{\theta *}^2/H_*^2$ becomes of order unity and thus the axion rolls down to its potential minimum, suppressing the axion velocity~$\abs{\dot{\theta}}$ compared to the value $ m^2_\theta / 3 n H$ which was used for deriving~(\ref{eq:Rupper}).

We have thus seen that that, independently of $\xi$, the energy ratio~$R$ is generally much smaller than the minimum value of~$10^{-1}$ required in (\ref{eq:R0.1}) for adiabaticity. 
We therefore conclude that a non-adiabatic evolution of the field fluctuation is inevitably triggered after PQ inflation.

\section{Axion quality}\label{Section7}

After inflation, the higher-dimensional operators can also drive the axion field away from the $CP$-conserving vacuum and invalidate the PQ mechanism as a solution to the strong $CP$ problem~\citep{Kamionkowski:1992mf,Holman:1992us,Kallosh:1995hi}.
This is often referred to as the axion quality problem.
Let us compare the necessary suppression of higher-dimensional operators from this perspective, with the constraints derived in the previous sections.

With a coupling to gluons $(g_s^2 / 32 \pi^2) N_{\ro{DW}} \theta G \tilde{G}$ where $N_{\ro{DW}}$ is a positive integer (the so-called domain wall number), the axion obtains a potential at temperatures below the QCD scale of the form,
\begin{equation}\label{eq:Potential}
    V(\theta) = \frac{m_{\ro{QCD}}^2 f^2}{N_{\ro{DW}}^2} \left[1-\cos (N_{\ro{DW}} \theta) \right]- \frac{m_\theta^2 f^2}{n^2} \cos{\left( n\theta+\delta \right) }.
\end{equation}
Here we have taken the $CP$-conserving vacuum as $\theta = 0$ modulo $2 \pi/N_{\ro{DW}}$ without loss of generality, 
and $m_{\ro{QCD}}$ is the mass arising from QCD non-perturbative effects given by~\citep{GrillidiCortona:2015jxo,Borsanyi:2016ksw},
\begin{equation}
 m_{\ro{QCD}} \simeq 5.7 \mu \ro{eV} \left( \frac{10^{12} \ro{GeV} }{ f/N_{\ro{DW}} } \right).
\end{equation}
$m_\theta$ is the mass induced by the higher-dimensional operator after the radial field has stabilized to the minimum of the Mexican hat potential, given by (\ref{massoftheta}) with $\varphi = f$.

The axion at the minimum of the potential (\ref{eq:Potential}) needs to satisfy $\abs{N_{\ro{DW}} \theta_{\ro{min}}} \lesssim 10^{-10}$
(modulo $2 \pi$) as implied by experimental bounds on the neutron electric dipole moment~\citep{Baker:2006ts,Abel:2020pzs}.
This requirement imposes
\begin{equation}
\frac{N_{\ro{DW}} }{n} \frac{m^2_\theta }{m^2_{\ro{QCD}}}
\left| \sin \left( n \theta_{\ro{min}} + \delta \right) \right|
\lesssim  10^{-10}.
\end{equation}
Supposing $| \sin ( n \theta_{\ro{min}} + \delta ) | \sim 1$, then this translates into a bound on $\abs{g}$ as
\begin{equation}
 \abs{g} \lesssim 10^{-88} \, \frac{N_{\ro{DW}} }{n}
\left(\frac{\sqrt{2}M_{\ro{Pl}}}{f}\right)^{l}.
\label{eq:oldquality}
\end{equation}

\begin{figure}[t]
\centering
    \includegraphics[width=0.6\textwidth]{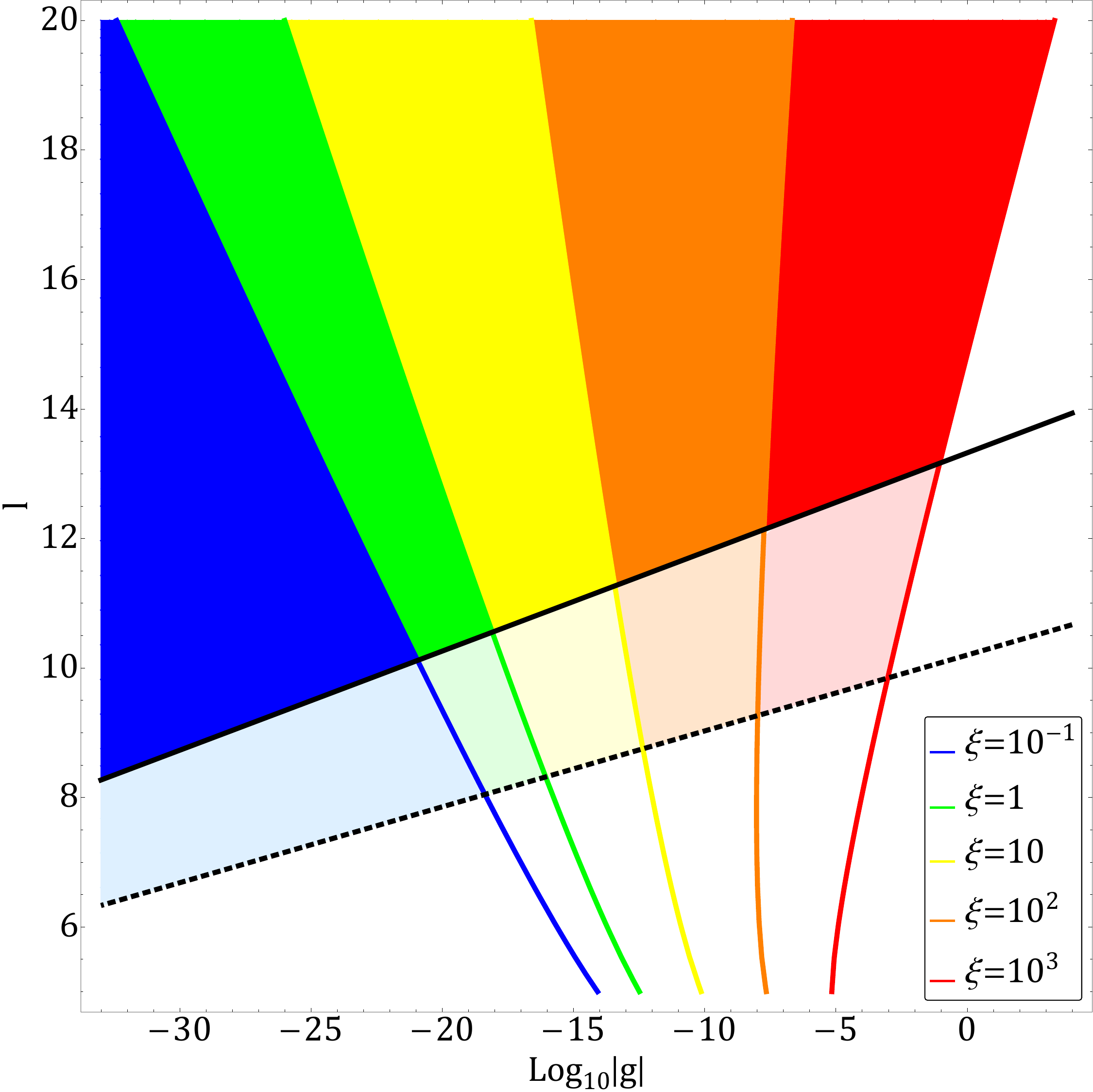}
    \captionof{figure}{Combined constraints on the coupling constant $|g|$ and dimension $l$ of higher-dimensional operators for the PQ field to drive inflation and solve the strong $CP$ problem. 
The black lines show the axion quality constraint on $\ro{U}(1)_{\ro{PQ}}$-breaking operators, for $f = 10^{12}\, \ro{GeV} $ (solid) and $f = 10^{10}\, \ro{GeV}$ (dashed).
The other lines show constraints on generic operators from the  consistency of PQ inflation, with the colors denoting different values of the non-minimal coupling $\xi$.
The colored regions show where both constraints are satisfied;
for $\xi = 10^{-1}$ this happens only in the blue region,
for $\xi = 1$ the region expands to include green, for $\xi = 10$ it further expands to yellow, etc.
The light-colored regions indicate the shift of the axion quality constraint as $f$ decreases from $10^{12}\, \ro{GeV} $ to $10^{10}\, \ro{GeV}$.}
    \label{fig:Axion quality}
\end{figure}

In Fig.~\ref{fig:Axion quality} the axion quality constraint  (\ref{eq:oldquality}) is shown in the $|g|$ - $l$ plane as the black lines, for the parameter choice of $N_{\ro{DW}} =  n$ with $f = 10^{12}\, \ro{GeV}$ (solid) and  $f = 10^{10}\, \ro{GeV}$ (dashed). 
The constraint (\ref{eq:linearprediction}) from the consistency of PQ inflation is overlaid, with the same color scheme based on the value of $\xi$ as in Fig.~\ref{fig:newquality}.
The axion quality constraint becomes weaker for smaller~$f$, while the inflationary constraint becomes weaker for larger~$\xi$ (although a larger
$\xi$ on the other hand lowers the cutoff of the effective field theory). 
The PQ field can drive inflation and solve the strong $CP$ problem
in the colored regions where both constraints are satisfied:
With $\xi = 10^{-1}$ this happens only in the blue region,
for $\xi = 1$ the region expands to include green, for $\xi = 10$ it further expands to yellow, etc.
The light-colored regions indicate the shift of the axion quality constraint as $f$ is varied between $10^{12}\, \ro{GeV} $ and $10^{10}\, \ro{GeV}$.

In the plane of $\log |g|$ and $l$, the tilt of the axion quality bound is given by $\log (\sqrt{2} \Mp / f)$, while that of the inflationary bound is predominantly given by $\log (\sqrt{2} \Mp / \varphi_* )$ (see discussions below (\ref{eq:linearprediction})). The latter is smaller since $\varphi_* \gg f$. As a consequence the axion quality bound is more sensitive to $l$, while the inflationary bound is more sensitive to $\abs{g}$. The two bounds are thus complementary to each other, and exclude a large parameter region of higher-dimension operators when combined.
It should also be noted that, while the axion quality bound only constrains $\ro{U}(1)_{\ro{PQ}}$-breaking operators ($n \neq 0$), the inflationary bound applies to general higher-dimensional operators independently of whether they break the~$\ro{U}(1)_{\ro{PQ}}$.

\section{Conclusions}
\label{Section8}

The PQ inflation scenario is sensitive to the ultraviolet completion of the theory for a wide range of values for the non-minimal gravitational coupling~$\xi$, since a large~$\xi$ entails a low cutoff of the effective field theory, while a small~$\xi$ requires a large inflaton field excursion.
In this paper we studied the latter effect and demonstrated that the predictions of the scenario is very sensitive to Planck-suppressed higher-dimensional operators. We showed in particular that for PQ inflation to produce the appropriate number of $e$-folds and curvature perturbations that match with observations, the coupling constant of an operator of dimension~$l$ suppressed by $\Mp$ in the Jordan frame is bounded as
\begin{equation}
 \abs{g} \lesssim 10^{-9} \frac{l}{(l-4) (l-2)^2}
\left( \frac{1+6 \xi }{230}  \right)^{l/2}.
\end{equation}
For instance with $\xi = 1$, a dimension-five Planck-suppressed operator needs to be further suppressed by a coupling as small as $\abs{g} \lesssim 10^{-13}$, with the constraint becoming more severe with increasing dimension.

Our constraint based on the consistency of PQ inflation is similar in spirit to the axion quality argument which constrains $\ro{U}(1)_{\ro{PQ}}$-breaking higher-dimensional operators by requiring the axion to be able to solve the strong $CP$ problem. 
However the resulting constraints are quite different, since the former is a high-energy/large-field effect, while the latter is low-energy/small-field.
Moreover, the constraint from PQ inflation applies to generic higher-dimensional operators, independently of whether they break the $\ro{U}(1)_{\ro{PQ}}$ symmetry.
In other words, even if the $\ro{U}(1)_{\ro{PQ}}$ symmetry is protected for some reason, PQ inflation is vulnerable to higher-dimensional corrections.
The inflationary and axion quality constraints are complementary to each other, and when combined exclude a large parameter space of higher-dimensional operators, as depicted in Fig.~\ref{fig:Axion quality}. 

Based on the constraint on higher-dimensional operators, we further showed that the oscillation of the PQ field after inflation inevitably triggers resonant amplifications of the field fluctuations.
It remains to be analyzed whether 
this necessarily leads to a restoration of the PQ symmetry,
in which case topological defects will later form when the symmetry becomes broken again.
Such studies require numerical simulations,
which we leave for the future.
Here we simply note that in order to fully understand the cosmology after PQ inflation including the reheating process, possible formation of axion strings and domain walls, as well as the production of axion dark matter, it is crucial to ascertain the impact of parametric resonance.
Our analyses can also be extended to other axion scenarios that make use of higher-dimensional operators, such as~\citep{Co:2019jts,Co:2021lkc,Gouttenoire:2021jhk}.

We should remark that most of the analyses in this paper neglect interactions between the PQ field and other matter fields, whose inclusion will be necessary for understanding the reheating process.
Matter couplings may also modify the PQ field dynamics during/after inflation, and alter our results. 
Another important direction for further study is to seek for ultraviolet completions that control higher-dimensional operators.
See e.g. \citep{Silverstein:2008sg,Kaloper:2014zba} for attempts for realizing high-scale inflation.
It would be interesting to study if such constructions exist also for the PQ inflation scenario.

\section*{Acknowledgments}

We thank Sabino Matarrese, Marco Peloso, Roberto Percacci, and Christophe Ringeval for helpful discussions. 
D.D.C. and T.K. acknowledge support from the INFN program on Theoretical Astroparticle Physics (TAsP). T.K. also acknowledges support from JSPS KAKENHI (Grant No. JP22K03595).

\appendix

\section{Effective-Planck-suppressed operators}
\label{sec:Mp_eff}

In this appendix we suppose that quantum gravity corrections give rise to higher-dimensional operators suppressed by the effective Planck scale in the Jordan frame, namely, operators of the form~(\ref{eq:HDO}) but with $\Mp$ replaced by $(\Mp^2 + \xi \varphi^2)^{1/2} $.
This scale at large fields becomes linear in the field as $\simeq \sqrt{\xi} \varphi $, and thus the operators reduce to the form
\begin{equation}
 - \frac{\abs{g} }{2 (2 \xi )^{\frac{l}{2} - 2}}
\varphi^4 
\cos( n\theta + \delta) .
\end{equation}
This merely has the effect of shifting the PQ self-coupling by 
\begin{equation}
\Delta \lambda = 
 - \frac{12 \abs{g} }{(2 \xi )^{\frac{l}{2} - 2}}
\cos( n\theta + \delta) .
\label{eq:Delta_lambda}
\end{equation}
PQ inflation would hence match with observations in the presence of such operators, given that the effective coupling $\lambda_{\ro{eff}} = \lambda + \Delta \lambda$ takes an appropriate value.

On the other hand, if one wishes to make reliable predictions of PQ inflation without detailed knowledge of quantum gravity, then $\Delta \lambda$ needs to be small enough such that it has little effect on the cosmological observables.
When fixing the other parameters, the self-coupling is related to the spectral index as 
$\lambda \propto (n_s - 1)^2$, cf. (\ref{eq:zeroth}).
Hence a shift in $\lambda$ affects $n_s$ by 
\begin{equation}
 \frac{\Delta n_s}{n_s - 1}\simeq
\frac{\Delta \lambda }{2 \lambda }.
\end{equation}
Combining with (\ref{eq:Delta_lambda}), approximating the cosine to unity, and substituting (\ref{eq:zeroth}) into $\lambda$, we obtain
\begin{equation}
\begin{split}
 \abs{g} 
&\sim \frac{3 \pi^2}{2} A_s (1-n_s) \abs{\Delta n_s}
(1 + 6 \xi) (2 \xi)^{\frac{l}{2} - 1} \\
&\lesssim 10^{-12} (1 + 6 \xi) (2 \xi )^{\frac{l}{2} - 1}.
\end{split}
\end{equation}
Upon moving to the second line we substituted
$A_s=2.1\times 10^{-9} $, $n_s = 0.9649$, 
and also required the shift in $n_s$ to be smaller than the
Planck $1\sigma$ uncertainty~\citep{Planck:2018jri}, 
i.e., $\abs{\Delta n_s} < 0.0042$.
For $\xi \lesssim 0.1$, this bound indicates that effective-Planck-suppressed operators should be further suppressed by couplings as small as $\abs{g}\lesssim 10^{-12}$.
On the other hand at $\xi \gtrsim 10^4$, the second line is of order unity or larger for $l \geq 5$ and hence the suppression by the effective Planck scale alone renders the operators negligible.

\section{Detailed computation of multi-field effects}
\label{AppA} 

We compute the contribution from the axion field to the curvature perturbation by using the $ \delta N$ formalism~\citep{Starobinsky:1985ibc,Sasaki:1995aw,Wands:2000dp,Lyth:2004gb} to take into account the evolution of the perturbation outside the horizon.
The analysis in this appendix is close to that presented in~\citep{Kobayashi:2010fm}.

The power spectrum of the curvature perturbation can be written in terms of field derivatives of the $e$-folding number as
\begin{equation}
 P_{\zeta} (k_*) \simeq 
\left( \frac{\partial N_*}{\partial \varphi_*} \right)^2
P_{\delta\varphi *} (k_*)
+ \left(
\frac{\partial N_*}{\partial \theta_*} 
\right)^2
P_{\delta\theta *} (k_*),
\end{equation}
where we expanded the perturbation in terms of field fluctuations up to linear order, and ignored cross-correlations between the radial and angular fluctuations.
$P_{\delta \varphi *}$ and $P_{\delta \theta *}$ denote the power spectra of the field fluctuations on the initial flat slice when the wave mode~$k_*$ exits the horizon.

Let us focus on PQ inflation in the large-field regime.
In particular, we assume the conditions laid out in (\ref{eq:SLcond}), plus $m_\theta^2 \ll H^2 $ to hold throughout inflation.
Then the equations of motion (\ref{eq:chi-EoM}), (\ref{eq:theta-EoM}), and the Friedmann equation (\ref{eq:Friedmann}) are approximated by the slow-roll expressions (\ref{eq:varphi-SL}), (\ref{eq:axion_slow-roll}), and (\ref{eq:FR-SL}), respectively.
Moreover, the power of the field fluctuations at horizon exit are (cf. (\ref{eq:5.4})),
\begin{equation}
 P_{\delta\varphi *} (k_*) \simeq
\frac{\xi }{1 + 6 \xi } 
\left( \frac{\varphi_*}{\Mp} \right)^2
\left(\frac{H_*}{2 \pi }  \right)^2,
 \quad
 P_{\delta\theta *} (k_*) \simeq
\frac{\xi }{\Mp^2} \left(\frac{H_*}{2 \pi }  \right)^2.
\label{eq:field_fluc}
\end{equation}
The infinitesimal variation of the $e$-folding number is written using 
(\ref{eq:varphi-SL}) and (\ref{eq:FR-SL}) as
\begin{equation}
  d N \simeq - \frac{1 + 6 \xi }{4} \frac{\varphi \, d\varphi }{\Mp^2},
\label{eq:dN-leading}
\end{equation}
hence the radial contribution to the curvature power spectrum is obtained as
\begin{equation}
\left( \frac{\partial N_*}{\partial \varphi_*} \right)^2
P_{\delta\varphi *}
\simeq
\frac{\xi (1 + 6 \xi) }{16 }
 \frac{\varphi_*^4}{\Mp^6}
\left( \frac{H_*}{2 \pi} \right)^2.
\label{eq:A.4}
\end{equation}
Rewriting $H_*$ using (\ref{eq:FR-SL}), this reproduces the expression for $A_s$ in (\ref{eq:2.16}).

Considering inflation to end when the radial field approaches some field value $\varphi_{\ro{end}}$ which is independent of $\theta_*$, 
then it is convenient to express the evolution of the angular field in terms of the radial field. Combining (\ref{eq:varphi-SL}) and (\ref{eq:axion_slow-roll}) gives
\begin{equation}
 \frac{d \theta }{d \varphi } \simeq
\frac{6 \xi^2 (1 + 6 \xi )}{n \lambda }
\frac{\varphi \, m_\theta^2}{\Mp^4}
 \sin (n \theta + \delta),
\end{equation}
which can be integrated to yield,
\begin{equation}
 \tan \left( \frac{n \theta + \delta }{2} \right)
\simeq
 \tan \left( \frac{n \theta_* + \delta }{2} \right)
\exp \left[
\frac{\ca{G}}{2}
(x^{l-2} - x_*^{l-2})
\right].
\label{eq:theta-varphi}
\end{equation}
Here we have introduced 
\begin{equation}
 x = \frac{\varphi}{\sqrt{2} \Mp},
\quad
 \ca{G} = \frac{12 n ^2}{l-2} \frac{\xi (1 + 6 \xi ) \abs{g}}{\lambda }.
\end{equation}

Following similar prescriptions as in Section~\ref{Section4}, one finds that under the conditions~(\ref{eq:SLcond}) and $m_\theta^2 \ll H^2 $, the leading correction to the expression~(\ref{eq:dN-leading}) for $dN$ from terms that explicitly\footnote{In principle the axion can also affect $N$ through terms that only depend on $\varphi$ by modulating the radial field dynamics; however we consider such effects to be negligible.} depend on $\theta$ yields~(\ref{eq:3.4}), but with the replacement $\theta_* \to \theta$.
(Unlike in Section~\ref{Section4}, here we are not limiting ourselves to corrections linear in $g$, and hence the axion is allowed to roll.)
Then $N_*$ can be obtained by integrating this expression
over $\varphi_* \geq \varphi \geq \varphi_{\ro{end}}$, noting that $\theta$ is a function of $\varphi$ as given in (\ref{eq:theta-varphi}). 
In particular, its $\theta_*$-derivative is written as
\begin{equation}
 \frac{\partial N_*}{\partial \theta_*} \simeq
\frac{(l-2) (l-4)}{2 n } \ca{G}
\int_{x_*}^{x_{\ro{end}}} dx\, x^{l-1}
\frac{\tan \left( \frac{n \theta_* + \delta }{2} \right) 
\left\{1 +\tan^2 \left( \frac{n \theta_* + \delta }{2} \right) \right\}
e^{\ca{G} (x^{l-2} - x_*^{l-2})} 
}{
\left\{ 1 + \tan^2 \left( \frac{n \theta_* + \delta }{2} \right) 
e^{\ca{G} (x^{l-2} - x_*^{l-2})}
\right\}^2
}.
\label{eq:dNdtheta}
\end{equation}
In terms of this quantity, the ratio of the contributions to the curvature power spectrum from the angular and radial fields is written as
\begin{equation}
 \beta =  \frac{(\partial \mathcal{N}_* / \partial \theta_*)^2 P_{\delta\theta *}}
{(\partial \mathcal{N}_* / \partial \varphi_*)^2 P_{\delta\varphi *} }
\simeq \frac{16}{1 + 6 \xi } \left(\frac{\Mp}{\varphi_*}\right)^4
\left( \frac{\partial N_*}{\partial \theta_*} \right)^2.
\label{eq:beta}
\end{equation}
Here, upon moving to the far right-hand side we used (\ref{eq:field_fluc}) and (\ref{eq:A.4}).

In order to analytically compute the integral in (\ref{eq:dNdtheta}), let us for the moment focus on the case of
\begin{equation}
 \abs{n \theta_* + \delta } \ll 1.
\end{equation}
Then by expanding the integrand up to linear
order in $(n \theta_* + \delta )$, one can perform the integral and obtain
\begin{equation}
 \frac{\partial N_*}{\partial \theta_*}
\simeq
\frac{l-4}{4 n } \ca{G} e^{- \ca{G} x_*^{l-2} }
\biggl[
x^{l}
E_{- \frac{2}{l-2 }} (-\ca{G} x^{l-2})
\biggr]_{x_{\ro{end}}}^{x_*}
(n \theta_* + \delta) ,
\label{eq:3.6}
\end{equation}
where 
\begin{equation}
 E_p (z) = z^{p-1}
\int^\infty_z \frac{e^{-y}}{y^p} dy
\end{equation}
is the generalized exponential integral.
The argument $\ca{G} x^{l-2}$ in (\ref{eq:3.6}) is, under the
conditions~(\ref{eq:SLcond}), approximated by
\begin{equation}
 \ca{G} x^{l-2} \simeq
\frac{1 + 6 \xi }{6 (l-2)}\frac{m_\theta^2}{H^2}
\frac{ \varphi^2}{\Mp^2}
= \frac{n^2 (1 + 6 \xi)}{6 (l-2)} \kappa ,
\end{equation}
where $\kappa$ is defined in~(\ref{kappa}). 
Let us further assume this to be sufficiently small such that
$\ca{G} x^{l-2} \ll 1$, and expand the generalized exponential integral using~\citep{Olver:2010:NHMF}
\begin{equation}
 E_p (z) = z^{p-1} \Gamma (1-p) - 
\sum^\infty_{s = 0} \frac{(-z)^s}{s ! (1-p+s)},
\end{equation}
which is valid for $\abs{\ro{ph}\, z} \leq \pi $.
Keeping terms up to $s = 0$ order, and further using $e^{-\ca{G} x_*^{l-2}} \simeq 1$ and $x_*^l \gg x_{\ro{end}}^l$, we arrive at
\begin{equation}
\frac{\partial  N_*}{\partial \theta_*} \simeq
- \frac{(l-2) (l-4)}{4 n l  }\ca{G}
 x_*^{l} (n \theta_* + \delta ).
\end{equation}
This approximate expression starts at linear order in~$g$, and ignores the axion rolling; the same result can also be obtained by taking a $\theta_*$-derivative of the $e$-folding number derived in~(\ref{eq:starting}). 
We thus obtain
\begin{equation}
\beta \simeq
 \frac{n^2 (l-4)^2 (1 + 6 \xi)}{144 \, l^2} 
\kappa_*^2 (n \theta_* + \delta)^2
\simeq
\frac{(l-4)^2}{l^2} \alpha.
\label{eq:betaapp}
\end{equation}
Here $\alpha$ is the relative contribution to the curvature perturbation from the angular field which was derived in (\ref{eq:alphalin})
by ignoring the evolution of the perturbations outside the horizon.

\begin{figure}[t!]
\centering
\subfigure[$\xi=10$, $\theta_i+\delta=0.5$]{
\includegraphics[width=0.4\textwidth]
{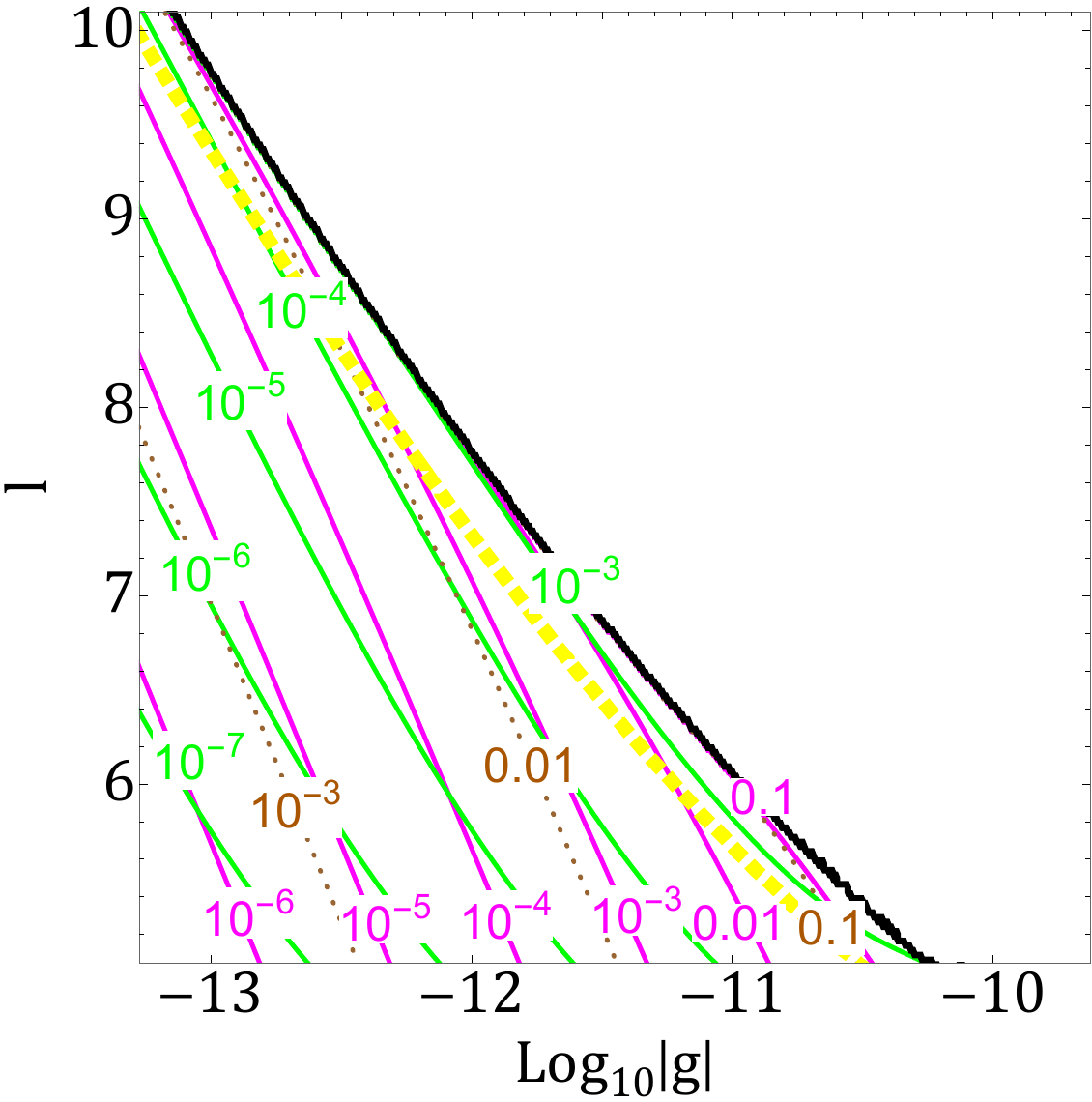}
\label{fig:App10_0.5}
}
\subfigure[$\xi=10$, $\theta_i+\delta=\pi-0.5$]{
\includegraphics[width=0.4\textwidth]
{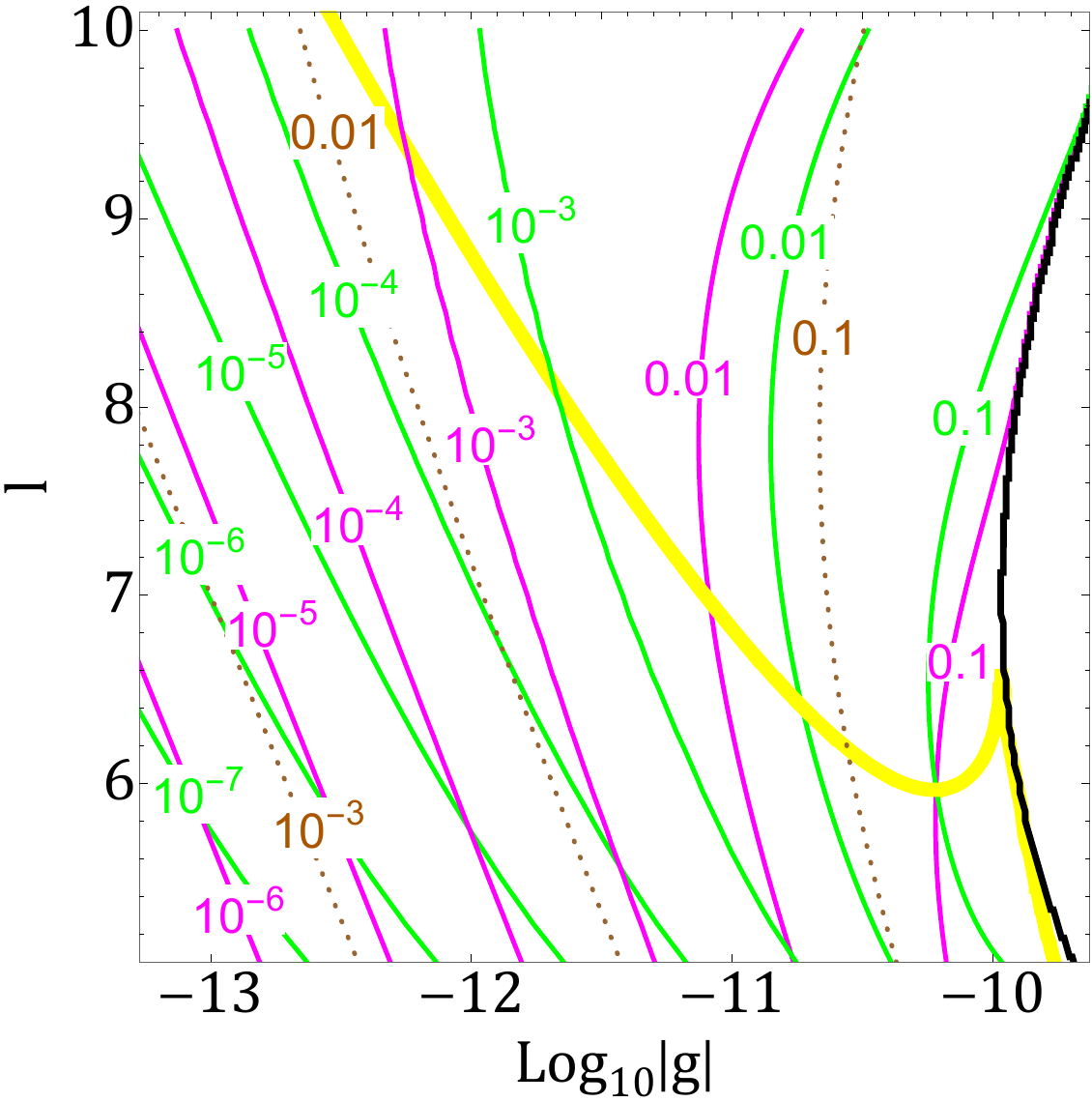}
\label{fig:App10_pi-0.5}
}
\subfigure[$\xi=10^2$, $\theta_i+\delta=0.5$]{
\includegraphics[width=0.4\textwidth]
{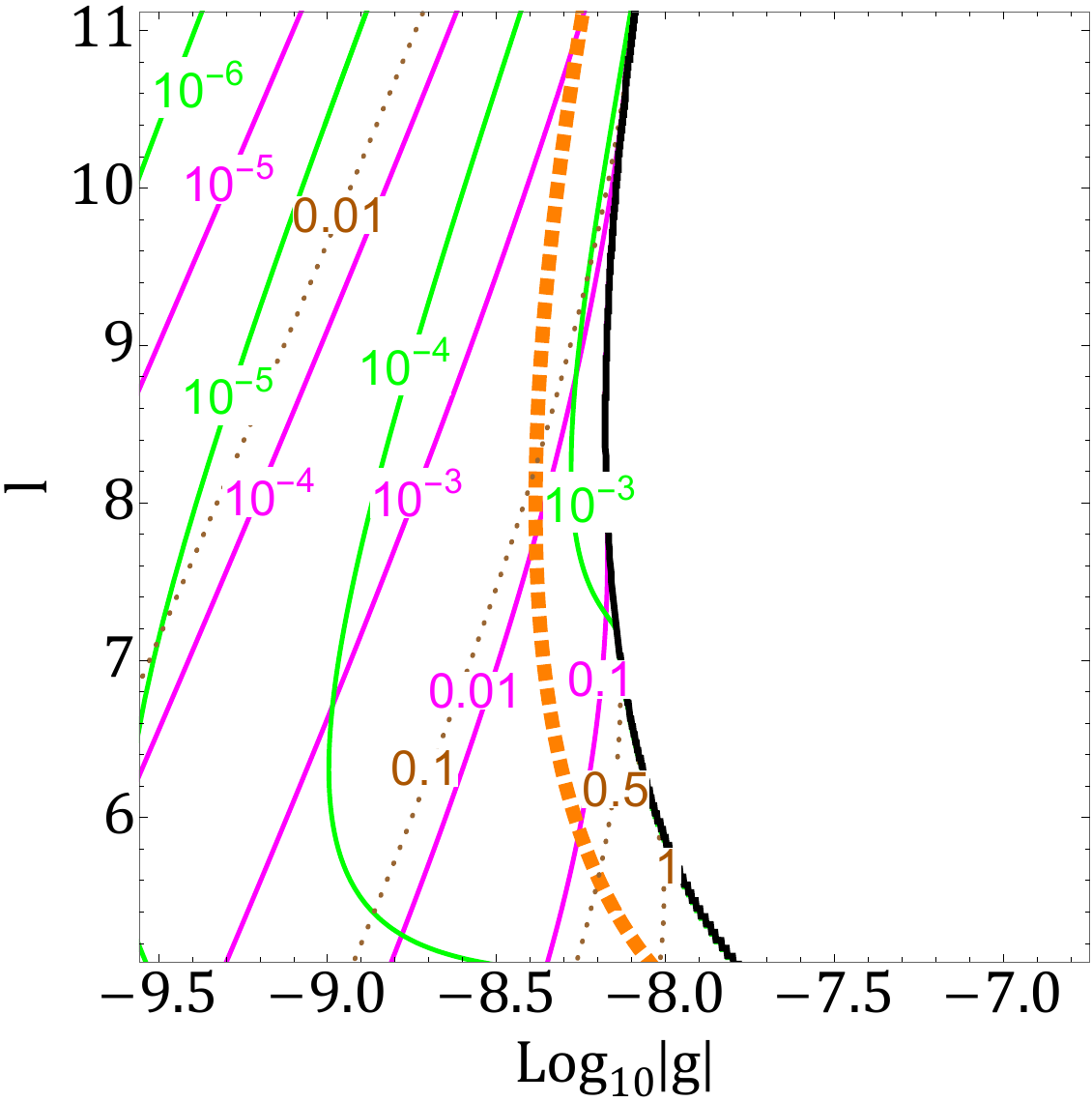}
\label{fig:App100_0.5}
}
\subfigure[$\xi=10^2$, $\theta_i+\delta=\pi-0.5$]{
\includegraphics[width=0.4\textwidth]
{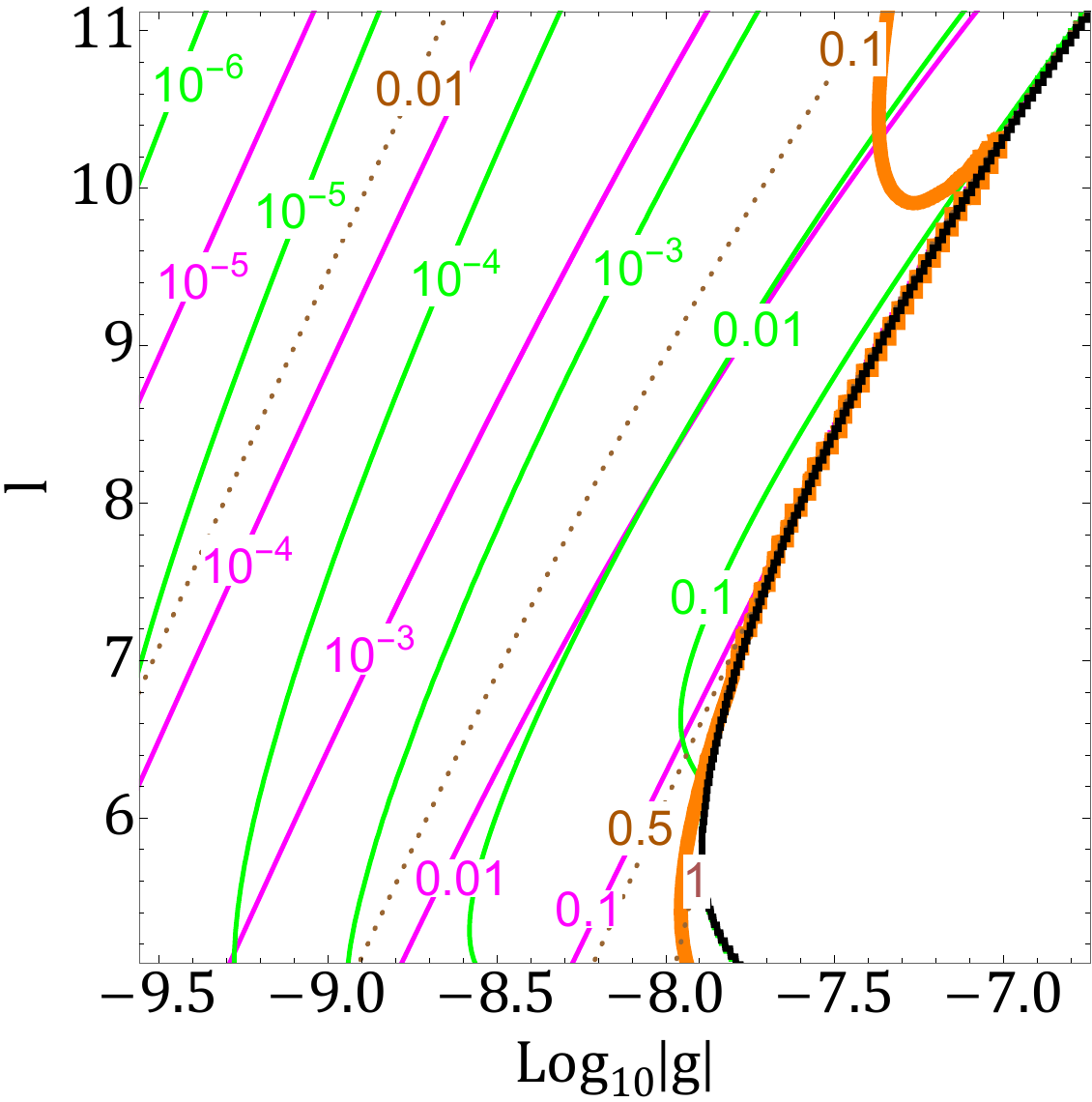}
\label{fig:App100_pi-0.5}
}
\captionof{figure}{Contribution of the axion to the curvature perturbation amplitude evaluated from $\beta$ (green contours) which takes into account the evolution of the perturbation outside the horizon, 
and $\alpha$ (magenta contours) which ignores the evolution.
The brown dotted contours show values of $m_{\theta *}^2 / H_*^2$.
Constraints on higher-dimensional operators are taken from Fig.~\ref{fig:newquality}.
On the right of the black solid lines, $A_s$ and $n_s$ cannot simultaneously take the observed values.
All the results are for $n = 1$.}
\label{Fig:single}
\end{figure}

We show in Fig.~\ref{Fig:single} the values of $\beta$ as given in (\ref{eq:beta}), by numerically computing the integral~(\ref{eq:dNdtheta}) to evaluate $\partial N_* / \partial \theta_*$. 
However, here we have replaced $\theta_*$ with $\theta_i$, for the same reason as described below~(\ref{eq:alphalin}). The contours for~$\beta$ are shown in green. 
The other lines are the same as in Fig.~\ref{fig:single}:
contours of $\alpha$ in magenta, contours of $m_{\theta*}^2/H^2_*$ in brown dotted, the edge of the no-go region in black, and the constraints on higher-dimensional operators with the same color scheme in Fig.~\ref{fig:newquality}. 
All the results are for $n = 1$, while $\xi$ and $\theta_i + \delta$ are varied in each panel.
One sees that $\beta < \alpha $ generically holds 
in regions where $m_{\theta*}^2/H^2_* \ll 1$.
In particular, as expected from the relation~(\ref{eq:betaapp}), 
$\beta$ is smaller than $\alpha$ by about two orders of magnitude at $l = 5$, while their difference shrinks towards larger~$l$.
On the other hand in regions where $m_{\theta*}^2/H^2_* \sim 1$, 
the angular field's super-horizon fluctuations get suppressed and their actual contribution to the curvature perturbation becomes smaller than $\alpha$ or $\beta$.
Hence we conclude that the regions in parameter space where the angular contribution exceeds~$1 \%$, namely, where
both $\beta \geq 10^{-2}$ and $m_{\theta*}^2/H^2_* \ll 1$ are satisfied, 
are actually even smaller than shown in Fig.~\ref{fig:single}. 

Before ending this appendix, we remark that upon using the $\delta N$ formalism, one should consider the $e$-folding number up to a final uniform-density slice when the curvature perturbation has approached a constant. 
In the above discussions we have simply computed the $e$-folding until the end of inflation, and also assumed the radial field value~$\varphi_{\ro{end}}$ when inflation ends to be independent of the initial conditions $\varphi_*$ and $\theta_*$. 
If the end of inflation and/or the post-inflation expansion history are also affected by $\varphi_*$ and $\theta_*$, then the curvature perturbation would continue to evolve after inflation~\citep{Lyth:2005qk,Sasaki:2008uc}.

\section{Axion isocurvature in vacuum misalignment scenario}
\label{app:iso}

We showed in Section~\ref{Section6} that a resonant amplification of 
the PQ field fluctuation is triggered after inflation, which is likely to render the axion field highly inhomogeneous. 
However in this appendix, we assume that somehow the amplification is suppressed, and moreover that the PQ symmetry is not restored after inflation, such that axion dark matter is produced through the conventional vacuum misalignment scenario. 
We compute the axion isocurvature perturbation in this hypothetical situation and derive constraints on the model parameters.

The relic abundance of axions produced via vacuum misalignment   is~\citep{Turner:1985si}
\begin{equation}
 \Omega_\theta h^2 \approx 0.1 (N_{\ro{DW}} \theta_*)^2
\left( \frac{ f/N_{\ro{DW}} }{10^{12} \ro{GeV} } \right)^{7/6},
\label{eq:C.1}
\end{equation}
where the $CP$-conserving vacuum is taken as $\theta = 0$,
and $N_{\ro{DW}}$ is the domain wall number.
We have assumed $m_\theta^2 \ll H^2 $ to hold throughout inflation, 
and also that the initial misalignment angle satisfies 
$\abs{N_{\ro{DW}} \theta_*} < 1$ such that the QCD potential in~(\ref{eq:Potential}) is approximated by a quadratic.\footnote{It is also assumed that the axion begins to oscillate during radiation domination, at temperatures above the QCD scale. This amounts to requiring $f/N_{\ro{DW}} \lesssim 10^{17} \, \ro{GeV}$.}
We suppose that the axions make up the entire dark matter, i.e. $\Omega_\theta h^2 \approx 0.1$. This sets a direct relation between $\theta_*$ and $f$ through the above equation.

Since the angle~$\theta_*$ acquires super-horizon fluctuations during inflation, 
axions obtain isocurvature perturbations as
$\delta \Omega_\theta / \Omega_\theta \simeq 2 \delta \theta_* / \theta_*$,
up to linear order in the angular fluctuations.
The isocurvature power spectrum is thus written, using (\ref{eq:5.4}), as
\begin{equation}
 P_{\ro{iso}} (k_*) \simeq 
\frac{1}{\theta_*^2}
\frac{\Omega_*^2}{\varphi_*^2} \frac{H_*^2}{\pi^2}.
\label{eq:C.2}
\end{equation}
Here $\Omega_*^2$ is the conformal factor at $\varphi = \varphi_*$ and should not be confused with the axion relic abundance $\Omega_\theta$.
Using (\ref{eq:C.1}) to rewrite $\theta_*$ in terms of~$f$, one obtains
\begin{equation}
 P_{\ro{iso}} (k_*) \approx
N_{\ro{DW}}^2
\left( \frac{ f/N_{\ro{DW}} }{10^{12} \ro{GeV} } \right)^{7/6}
\frac{\Omega_*^2}{\varphi_*^2} \frac{H_*^2}{\pi^2}.
\label{eq:C.3}
\end{equation}
The last factor is determined by the inflationary observables. In particular for large fields, $\xi \varphi_*^2 \gg \Mp^2$, 
one can plug (\ref{eq:FR-SL}) and (\ref{eq:zeroth}) respectively into $H_*$ and $\lambda$ to obtain 
\begin{equation}
\frac{\Omega_*^2}{\varphi_*^2} \frac{H_*^2}{\pi^2}
\simeq \frac{P_{\zeta} (k_*) (1-n_s)^2 (1 + 6 \xi)}{4}.
\label{eq:C.4}
\end{equation}

\begin{figure}[t]
\centering
    \includegraphics[width=0.5\textwidth]{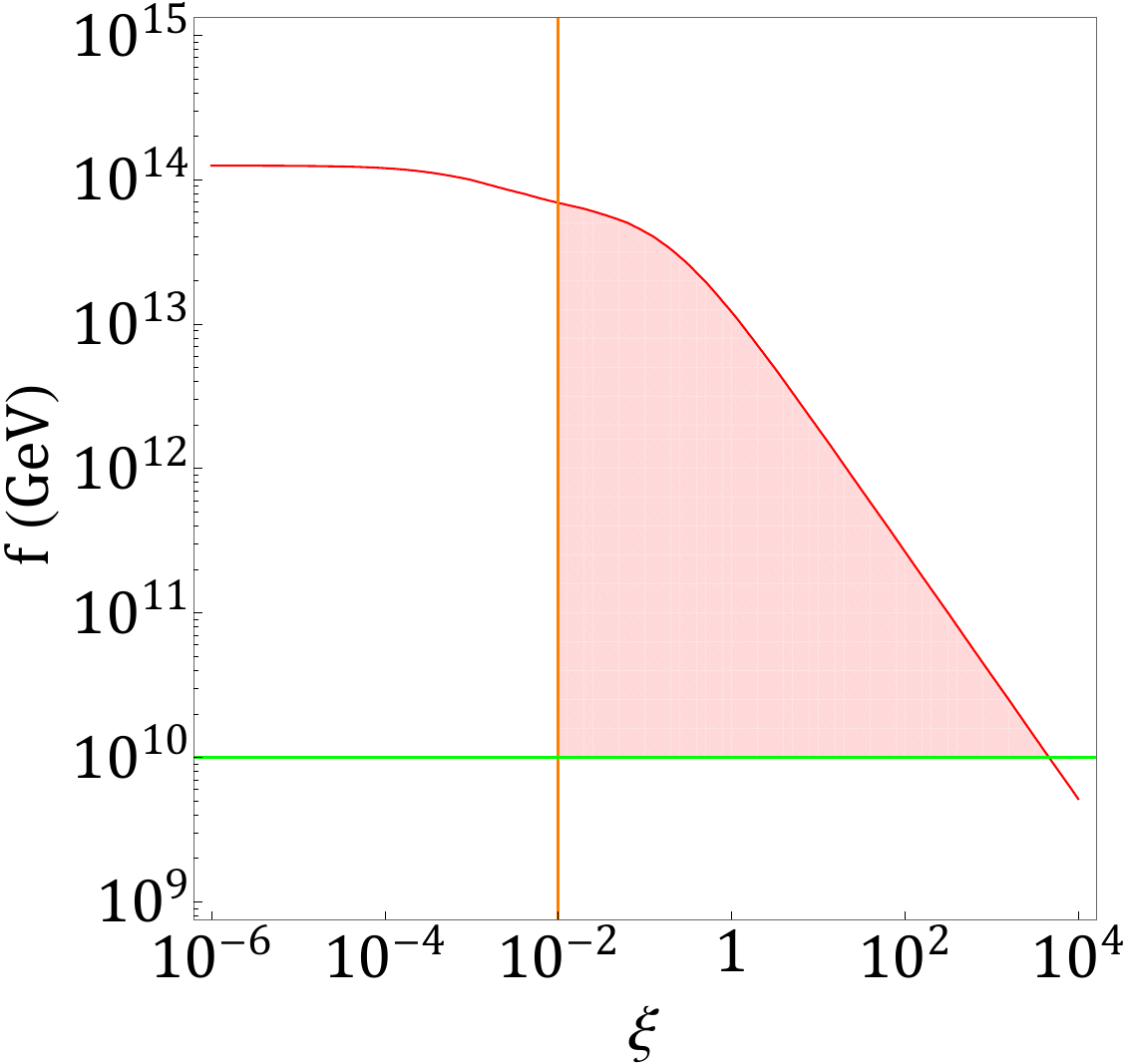}
    \caption{Constraints on the decay constant and gravitational coupling for the hypothetical scenario where axions are produced by vacuum misalignment after PQ inflation. 
The region where axions can make up dark matter is shown in red.
Axion isocurvature exceeds the observational limit in regions above the red line (in a quadratic axion potential), as well as below the green line (anharmonic potential). The tensor-to-scalar ratio exceeds the observational limit on the left of the orange line.
Here we have set $N_{\ro{DW}} = 1$.
}
    \label{initialHubble}
\end{figure}

Considering the $k_*$~dependence of (\ref{eq:C.2}) to be small and also ignoring cross-correlations between $\delta \theta$ and $\delta \varphi$, the axion isocurvature is almost scale-invariant and uncorrelated with curvature perturbations. 
Parametrizing the power spectrum as
$P_{\ro{iso}}(k) / P_\zeta (k) = \beta_{\ro{iso}}(k) / 
\{1- \beta_{\ro{iso}} (k) \}$,
this kind of cold dark matter isocurvature is constrained by Planck~\citep{Planck:2018jri} as $\beta_{\ro{iso}} (k_*) < 0.038$ ($95\%$ C.L.)
at $k_*=0.05 \, \mathrm{Mpc}^{-1} $.
Imposing this constraint on (\ref{eq:C.3}), and substituting (\ref{eq:C.4}) and $n_s = 0.965$, we obtain an upper bound on the axion decay constant,
\begin{equation}
 f \lesssim 
\frac{10^{14}\, \ro{GeV}}{N_{\ro{DW}}^{5/7} (1 + 6 \xi)^{6/7}}.
\label{eq:C.5}
\end{equation}
We derived this result assuming $\xi \varphi_*^2 \gg \Mp^2 $; 
however can check that the small $\xi$ limit of this expression matches with the bound in the minimal $\varphi^4$-inflation regime at small~$\xi$, at the order-of-magnitude level.
The upper bound originates from the fact that a larger~$f$ requires 
a smaller~$\theta_*$ for a fixed relic abundance, which in turn leads to a larger isocurvature.\footnote{If the PQ field is not the inflaton and its radial component is fixed to~$f$, the axion isocurvature is given by (\ref{eq:C.2}) with the factor $\Omega_*^2 / \varphi_*^2 $ replaced by $1/f^2$. The isocurvature then comes with a negative power of~$f$, and the isocurvature constraint yields a \textit{lower} limit on~$f$ in terms of an undetermined inflation scale.}
We also remark that for arbitrary~$\xi$, the bound (\ref{eq:C.5}) is stronger than the condition (\ref{eq:xi-f}) we have been assuming throughout this paper. 
In Fig.~\ref{initialHubble} we show the upper bound on~$f$ as a function of~$\xi$ by the red line. Here the factor $\Omega_*^2 H_*^2 / \varphi_*^2$ is numerically computed as was done for Fig.~\ref{number1}. In addition to using the aforementioned values for the observables, in the plot we further set $N_{\ro{DW}} = 1$.

It should be noted that the above discussions break down 
at $f/N_{\ro{DW}} < 10^{12}\, \ro{GeV}$, for which 
$\abs{N_{\ro{DW}} \theta_*} > 1$ is required for the axions to make up the entire dark matter. 
In this regime the full axion potential needs to be taken into account, and its departure from a quadratic increases the relic abundance compared to (\ref{eq:C.1})~\citep{Turner:1985si,Bae:2008ue}, and moreover significantly enhances the axion isocurvature compared to (\ref{eq:C.2})~\citep{Lyth:1991ub,Strobl:1994wk,Kobayashi:2013nva}. 
Each of these effects makes the upper bound on~$f$ more stringent than (\ref{eq:C.5}).
The isocurvature hence becomes large at both large~$f$ in a quadratic axion potential, 
as well as at small~$f$ in an anharmonic potential.\footnote{The work \citep{Fairbairn:2014zta} also evaluated axion isocurvature assuming vacuum misalignment after PQ inflation. 
However the conformal factor~$\Omega_*^2$ in (\ref{eq:C.2}) was overlooked, and moreover the isocurvature power was underestimated by~$8 \pi$. It also did not take into account anharmonic effects, and consequently obtained a much larger parameter window than in this appendix.}
Since the anharmonic enhancement of the isocurvature becomes particularly strong at 
$f/N_{\ro{DW}} \lesssim 10^{10}\, \ro{GeV}$~\citep{Kobayashi:2013nva},
in Fig.~\ref{initialHubble} we show $ f = 10^{10}\, \ro{GeV}$ by the green line as a rough lower bound below which anharmonic effects produce isocurvature beyond the observational limit.
Note that the upper bound (\ref{eq:C.5}) with $N_{\ro{DW}} = 1$ becomes of 
$10^{10}\, \ro{GeV}$ at $\xi \sim 10^3$; this sets an absolute upper limit for~$\xi$. 

In the plot we also show in orange the lower limit $\xi \gtrsim 10^{-2}$, which derives from the observational limit on the tensor-to-scalar ratio, cf.~Fig.~\ref{fig:r}. This, together with the upper and lower limits for~$f$, leave us with the red region as the allowed parameter space for the scenario where PQ inflation is followed by a vacuum misalignment.

\bibliography{bibliography}

\providecommand{\href}[2]{#2}\begingroup\raggedright\begin{thebibliography}{10}

\bibitem{Peccei:1977hh}
R.~D. Peccei and H.~R. Quinn, {\it Cp conservation in the presence of
  instantons},  {\em Phys. Rev. Lett.} {\bf 38} (1977) 1440--1443.

\bibitem{Weinberg:1977ma}
S.~Weinberg, {\it A new light boson?},  {\em Phys. Rev. Lett.} {\bf 40} (1978)
  223--226.

\bibitem{Wilczek:1977pj}
F.~Wilczek, {\it Problem of strong $p$ and $t$ invariance in the presence of
  instantons},  {\em Phys. Rev. Lett.} {\bf 40} (1978) 279--282.

\bibitem{Preskill:1982cy}
J.~Preskill, M.~B. Wise, and F.~Wilczek, {\it Cosmology of the invisible
  axion},  {\em Phys. Lett. B} {\bf 120} (1983) 127--132.

\bibitem{Abbott:1982af}
L.~F. Abbott and P.~Sikivie, {\it A cosmological bound on the invisible axion},
   {\em Phys. Lett. B} {\bf 120} (1983) 133--136.

\bibitem{Dine:1982ah}
M.~Dine and W.~Fischler, {\it The not so harmless axion},  {\em Phys. Lett. B}
  {\bf 120} (1983) 137--141.

\bibitem{Starobinsky:1979ty}
A.~A. Starobinsky, {\it Spectrum of relict gravitational radiation and the
  early state of the universe},  {\em JETP Lett.} {\bf 30} (1979) 682--685.

\bibitem{Starobinsky:1980te}
A.~A. Starobinsky, {\it A new type of isotropic cosmological models without
  singularity},  {\em Phys. Lett. B} {\bf 91} (1980) 99--102.

\bibitem{Sato:1980yn}
K.~Sato, {\it First-order phase transition of a vacuum and the expansion of the
  universe},  {\em Monthly Notices of the Royal Astronomical Society} {\bf 195}
  (1981), no.~3 467--479.

\bibitem{Guth:1980zm}
A.~H. Guth, {\it The inflationary universe: A possible solution to the horizon
  and flatness problems},  {\em Phys. Rev. D} {\bf 23} (1981) 347--356.

\bibitem{Mukhanov:1981xt}
V.~F. Mukhanov and G.~V. Chibisov, {\it Quantum fluctuations and a nonsingular
  universe},  {\em JETP Lett.} {\bf 33} (1981) 532--535.

\bibitem{Linde:1981mu}
A.~D. Linde, {\it A new inflationary universe scenario: A possible solution of
  the horizon, flatness, homogeneity, isotropy and primordial monopole
  problems},  {\em Phys. Lett. B} {\bf 108} (1982) 389--393.

\bibitem{Fairbairn:2014zta}
M.~Fairbairn, R.~Hogan, and D.~J.~E. Marsh, {\it Unifying inflation and dark
  matter with the peccei-quinn field: observable axions and observable
  tensors},  {\em Phys. Rev. D} {\bf 91} (2015)
  [\href{http://arxiv.org/abs/1410.1752}{{\tt arXiv:1410.1752}}].

\bibitem{Kearney:2016vqw}
J.~Kearney, N.~Orlofsky, and A.~Pierce, {\it High-scale axions without
  isocurvature from inflationary dynamics},  {\em Phys. Rev. D} {\bf 93} (2016)
  [\href{http://arxiv.org/abs/1601.03049}{{\tt arXiv:1601.03049}}].

\bibitem{Ballesteros:2016xej}
G.~Ballesteros, J.~Redondo, A.~Ringwald, and C.~Tamarit, {\it {Standard
  Model\textemdash{}axion\textemdash{}seesaw\textemdash{}Higgs portal
  inflation. Five problems of particle physics and cosmology solved in one
  stroke}},  {\em JCAP} {\bf 08} (2017) 001,
  [\href{http://arxiv.org/abs/1610.01639}{{\tt arXiv:1610.01639}}].

\bibitem{Boucenna:2017fna}
S.~M. Boucenna and Q.~Shafi, {\it Axion inflation, proton decay, and
  leptogenesis in $su(5)\times u(1)_{PQ}$},  {\em Phys. Rev. D} {\bf 97} (2018)
  [\href{http://arxiv.org/abs/1712.06526}{{\tt arXiv:1712.06526}}].

\bibitem{Hamaguchi:2021mmt}
K.~Hamaguchi, Y.~Kanazawa, and N.~Nagata, {\it Axion quality problem alleviated
  by nonminimal coupling to gravity},  {\em Phys. Rev. D} {\bf 105} (2022)
  [\href{http://arxiv.org/abs/2108.13245}{{\tt arXiv:2108.13245}}].

\bibitem{LiteBIRD:2022cnt}
{\bf LiteBIRD} Collaboration, E.~Allys et~al., {\it {Probing Cosmic Inflation
  with the LiteBIRD Cosmic Microwave Background Polarization Survey}},  {\em
  PTEP} {\bf 2023} (2023), no.~4 042F01,
  [\href{http://arxiv.org/abs/2202.02773}{{\tt arXiv:2202.02773}}].

\bibitem{Parker:2009uva}
L.~E. Parker and D.~Toms, {\em {Quantum Field Theory in Curved Spacetime}:
  {Quantized Field and Gravity}}.
\newblock Cambridge Monographs on Mathematical Physics. Cambridge University
  Press, 8, 2009.

\bibitem{Salopek}
D.~S. Salopek, J.~R. Bond, and J.~M. Bardeen, {\it Designing density
  fluctuation spectra in inflation},  {\em Phys. Rev. D} {\bf 40} (Sep, 1989)
  1753--1788.

\bibitem{Kaiser:1994vs}
D.~I. Kaiser, {\it Primordial spectral indices from generalized einstein
  theories},  {\em Phys. Rev. D} {\bf 52} (1995) 4295--4306,
  [\href{http://arxiv.org/abs/astro-ph/9408044}{{\tt astro-ph/9408044}}].

\bibitem{Okada:2010jf}
N.~Okada, M.~U. Rehman, and Q.~Shafi, {\it Tensor to scalar ratio in
  non-minimal $\phi^4$ inflation},  {\em Phys. Rev. D} {\bf 82} (2010)
  [\href{http://arxiv.org/abs/1005.5161}{{\tt arXiv:1005.5161}}].

\bibitem{Linde:2011nh}
A.~Linde, M.~Noorbala, and A.~Westphal, {\it Observational consequences of
  chaotic inflation with nonminimal coupling to gravity},  {\em JCAP} {\bf 03}
  (2011) [\href{http://arxiv.org/abs/1101.2652}{{\tt arXiv:1101.2652}}].

\bibitem{Bezrukov:2007ep}
F.~L. Bezrukov and M.~Shaposhnikov, {\it The standard model higgs boson as the
  inflaton},  {\em Phys. Lett. B} {\bf 659} (2008) 703--706,
  [\href{http://arxiv.org/abs/0710.3755}{{\tt arXiv:0710.3755}}].

\bibitem{Barbon:2009ya}
J.~L.~F. Barbon and J.~R. Espinosa, {\it On the naturalness of higgs
  inflation},  {\em Phys. Rev. D} {\bf 79} (2009)
  [\href{http://arxiv.org/abs/0903.0355}{{\tt arXiv:0903.0355}}].

\bibitem{Burgess:2009ea}
C.~P. Burgess, H.~M. Lee, and M.~Trott, {\it Power-counting and the validity of
  the classical approximation during inflation},  {\em JHEP} {\bf 09} (2009)
  [\href{http://arxiv.org/abs/0902.4465}{{\tt arXiv:0902.4465}}].

\bibitem{Hertzberg:2010dc}
M.~P. Hertzberg, {\it On inflation with non-minimal coupling},  {\em JHEP} {\bf
  11} (2010) [\href{http://arxiv.org/abs/1002.2995}{{\tt arXiv:1002.2995}}].

\bibitem{Burgess:2014lza}
C.~P. Burgess, S.~P. Patil, and M.~Trott, {\it On the predictiveness of
  single-field inflationary models},  {\em JHEP} {\bf 06} (2014)
  [\href{http://arxiv.org/abs/1402.1476}{{\tt arXiv:1402.1476}}].

\bibitem{Bezrukov:2010jz}
F.~Bezrukov, A.~Magnin, M.~Shaposhnikov, and S.~Sibiryakov, {\it Higgs
  inflation: consistency and generalisations},  {\em JHEP} {\bf 01} (2011)
  [\href{http://arxiv.org/abs/1008.5157}{{\tt arXiv:1008.5157}}].

\bibitem{Jinno:2019und}
R.~Jinno, M.~Kubota, K.-y. Oda, and S.~C. Park, {\it {Higgs inflation in metric
  and Palatini formalisms: Required suppression of higher dimensional
  operators}},  {\em JCAP} {\bf 03} (2020) 063,
  [\href{http://arxiv.org/abs/1904.05699}{{\tt arXiv:1904.05699}}].

\bibitem{Kamionkowski:1992mf}
M.~Kamionkowski and J.~March-Russell, {\it Planck scale physics and the
  peccei-quinn mechanism},  {\em Phys. Lett. B} {\bf 282} (1992) 137--141,
  [\href{http://arxiv.org/abs/hep-th/9202003}{{\tt hep-th/9202003}}].

\bibitem{Holman:1992us}
R.~Holman, S.~D.~H. Hsu, T.~W. Kephart, E.~W. Kolb, R.~Watkins, and L.~M.
  Widrow, {\it Solutions to the strong cp problem in a world with gravity},
  {\em Phys. Lett. B} {\bf 282} (1992) 132--136,
  [\href{http://arxiv.org/abs/hep-ph/9203206}{{\tt hep-ph/9203206}}].

\bibitem{Kallosh:1995hi}
R.~Kallosh, A.~D. Linde, D.~A. Linde, and L.~Susskind, {\it Gravity and global
  symmetries},  {\em Phys. Rev. D} {\bf 52} (1995) 912--935,
  [\href{http://arxiv.org/abs/hep-th/9502069}{{\tt hep-th/9502069}}].

\bibitem{Linde:1990yj}
A.~D. Linde and D.~H. Lyth, {\it Axionic domain wall production during
  inflation},  {\em Phys. Lett. B} {\bf 246} (1990) 353--358.

\bibitem{Linde:1991km}
A.~D. Linde, {\it Axions in inflationary cosmology},  {\em Phys. Lett. B} {\bf
  259} (1991) 38--47.

\bibitem{Tkachev:1995md}
I.~I. Tkachev, {\it Phase transitions at preheating},  {\em Phys. Lett. B} {\bf
  376} (1996) 35--40, [\href{http://arxiv.org/abs/hep-th/9510146}{{\tt
  hep-th/9510146}}].

\bibitem{Kasuya:1996ns}
S.~Kasuya, M.~Kawasaki, and T.~Yanagida, {\it Cosmological axion problem in
  chaotic inflationary universe},  {\em Phys. Lett. B} {\bf 409} (1997)
  94--100, [\href{http://arxiv.org/abs/hep-ph/9608405}{{\tt hep-ph/9608405}}].

\bibitem{Kasuya:1998td}
S.~Kasuya and M.~Kawasaki, {\it Topological defects formation after inflation
  on lattice simulation},  {\em Phys. Rev. D} {\bf 58} (1998)
  [\href{http://arxiv.org/abs/hep-ph/9804429}{{\tt hep-ph/9804429}}].

\bibitem{Tkachev:1998dc}
I.~Tkachev, S.~Khlebnikov, L.~Kofman, and A.~D. Linde, {\it Cosmic strings from
  preheating},  {\em Phys. Lett. B} {\bf 440} (1998) 262--268,
  [\href{http://arxiv.org/abs/hep-ph/9805209}{{\tt hep-ph/9805209}}].

\bibitem{Harigaya:2015hha}
K.~Harigaya, M.~Ibe, M.~Kawasaki, and T.~T. Yanagida, {\it Dynamics of
  peccei-quinn breaking field after inflation and axion isocurvature
  perturbations},  {\em JCAP} {\bf 11} (2015)
  [\href{http://arxiv.org/abs/1507.00119}{{\tt arXiv:1507.00119}}].

\bibitem{Kobayashi:2016qld}
T.~Kobayashi and F.~Takahashi, {\it Cosmological perturbations of axion with a
  dynamical decay constant},  {\em JCAP} {\bf 08} (2016)
  [\href{http://arxiv.org/abs/1607.04294}{{\tt arXiv:1607.04294}}].

\bibitem{Co:2020dya}
R.~T. Co, L.~J. Hall, K.~Harigaya, K.~A. Olive, and S.~Verner, {\it Axion
  kinetic misalignment and parametric resonance from inflation},  {\em JCAP}
  {\bf 08} (2020) [\href{http://arxiv.org/abs/2004.00629}{{\tt
  arXiv:2004.00629}}].

\bibitem{Ballesteros:2021bee}
G.~Ballesteros, A.~Ringwald, C.~Tamarit, and Y.~Welling, {\it Revisiting
  isocurvature bounds in models unifying the axion with the inflaton},  {\em
  JCAP} {\bf 09} (2021) [\href{http://arxiv.org/abs/2104.13847}{{\tt
  arXiv:2104.13847}}].

\bibitem{Planck:2018jri}
{\bf Planck} Collaboration, Y.~Akrami et~al., {\it {Planck 2018 results. X.
  Constraints on inflation}},  {\em Astron. Astrophys.} {\bf 641} (2020) A10,
  [\href{http://arxiv.org/abs/1807.06211}{{\tt arXiv:1807.06211}}].

\bibitem{BICEP:2021xfz}
{\bf BICEP, Keck} Collaboration, P.~A.~R. Ade et~al., {\it {Improved
  Constraints on Primordial Gravitational Waves using Planck, WMAP, and
  BICEP/Keck Observations through the 2018 Observing Season}},  {\em Phys. Rev.
  Lett.} {\bf 127} (2021), no.~15 151301,
  [\href{http://arxiv.org/abs/2110.00483}{{\tt arXiv:2110.00483}}].

\bibitem{Lyth:1996im}
D.~H. Lyth, {\it What would we learn by detecting a gravitational wave signal
  in the cosmic microwave background anisotropy?},  {\em Phys. Rev. Lett.} {\bf
  78} (1997) 1861--1863, [\href{http://arxiv.org/abs/hep-ph/9606387}{{\tt
  hep-ph/9606387}}].

\bibitem{Abbott}
L.~F. Abbott and M.~B. Wise, {\it Wormholes and global symmetries.},  {\em
  Nuclear Physics B} {\bf 325} (1989), no.~3 687--704.

\bibitem{Banks:2010zn}
T.~Banks and N.~Seiberg, {\it Symmetries and strings in field theory and
  gravity},  {\em Phys. Rev. D} {\bf 83} (2011)
  [\href{http://arxiv.org/abs/1011.5120}{{\tt arXiv:1011.5120}}].

\bibitem{Harlow:2018tng}
D.~Harlow and H.~Ooguri, {\it Symmetries in quantum field theory and quantum
  gravity},  {\em Commun. Math} {\bf 383} (2021) 1669--1804,
  [\href{http://arxiv.org/abs/1810.05338}{{\tt arXiv:1810.05338}}].

\bibitem{Bartolo:2004if}
N.~Bartolo, E.~Komatsu, S.~Matarrese, and A.~Riotto, {\it Non-gaussianity from
  inflation: Theory and observations},  {\em Phys. Rept.} {\bf 402} (2004)
  103--266, [\href{http://arxiv.org/abs/astro-ph/0406398}{{\tt
  astro-ph/0406398}}].

\bibitem{Ema:2016dny}
Y.~Ema, R.~Jinno, K.~Mukaida, and K.~Nakayama, {\it Violent preheating in
  inflation with nonminimal coupling},  {\em JCAP} {\bf 02} (2017)
  [\href{http://arxiv.org/abs/1609.05209}{{\tt arXiv:1609.05209}}].

\bibitem{Kofman:1994rk}
L.~Kofman, A.~D. Linde, and A.~A. Starobinsky, {\it Reheating after inflation},
   {\em Phys. Rev. Lett.} {\bf 73} (1994) 3195--3198,
  [\href{http://arxiv.org/abs/hep-th/9405187}{{\tt hep-th/9405187}}].

\bibitem{Kofman:1995fi}
L.~Kofman, A.~D. Linde, and A.~A. Starobinsky, {\it Nonthermal phase
  transitions after inflation},  {\em Phys. Rev. Lett.} {\bf 76} (1996)
  1011--1014, [\href{http://arxiv.org/abs/hep-th/9510119}{{\tt
  hep-th/9510119}}].

\bibitem{Kofman:1997yn}
L.~Kofman, A.~D. Linde, and A.~A. Starobinsky, {\it Towards the theory of
  reheating after inflation},  {\em Phys. Rev. D} {\bf 56} (1997) 3258--3295,
  [\href{http://arxiv.org/abs/hep-ph/9704452}{{\tt hep-ph/9704452}}].

\bibitem{GrillidiCortona:2015jxo}
G.~G. di~Cortona, E.~Hardy, J.~P. Vega, and G.~Villadoro, {\it The qcd axion,
  precisely},  {\em JHEP} {\bf 01} (2016)
  [\href{http://arxiv.org/abs/1511.02867}{{\tt arXiv:1511.02867}}].

\bibitem{Borsanyi:2016ksw}
S.~Borsanyi, Z.~Fodor, J.~Guenther, K.~H. Kampert, S.~D. Katz, T.~Kawanai,
  T.~G. Kovacs, S.~W. Mages, A.~Pasztor, and F.~Pittler, {\it et al},  {\em
  ``Calculation of the axion mass based on high-temperature lattice quantum
  chromodynamics,'' Nature} {\bf 539} (2016) 69--71,
  [\href{http://arxiv.org/abs/1606.07494}{{\tt arXiv:1606.07494}}].

\bibitem{Baker:2006ts}
C.~A. Baker, D.~D. Doyle, P.~Geltenbort, K.~Green, M.~G.~D. van~der Grinten,
  P.~G. Harris, P.~Iaydjiev, S.~N. Ivanov, D.~J.~R. May, and J.~M. Pendlebury,
  {\it et al},  {\em ``An Improved experimental limit on the electric dipole
  moment of the neutron,'' Phys. Rev. Lett.} {\bf 97} (2006)
  [\href{http://arxiv.org/abs/hep-ex/0602020}{{\tt hep-ex/0602020}}].

\bibitem{Abel:2020pzs}
C.~Abel et~al., {\it {Measurement of the Permanent Electric Dipole Moment of
  the Neutron}},  {\em Phys. Rev. Lett.} {\bf 124} (2020), no.~8 081803,
  [\href{http://arxiv.org/abs/2001.11966}{{\tt arXiv:2001.11966}}].

\bibitem{Co:2019jts}
R.~T. Co, L.~J. Hall, and K.~Harigaya, {\it Axion kinetic misalignment
  mechanism},  {\em Phys. Rev. Lett.} {\bf 124} (2020)
  [\href{http://arxiv.org/abs/1910.14152}{{\tt arXiv:1910.14152}}].

\bibitem{Co:2021lkc}
R.~T. Co, D.~Dunsky, N.~Fernandez, A.~Ghalsasi, L.~J. Hall, K.~Harigaya, and
  J.~Shelton, {\it Gravitational wave and cmb probes of axion kination},  {\em
  JHEP} {\bf 09} (2022) [\href{http://arxiv.org/abs/2108.09299}{{\tt
  arXiv:2108.09299}}].

\bibitem{Gouttenoire:2021jhk}
Y.~Gouttenoire, G.~Servant, and P.~Simakachorn, {\it {Kination cosmology from
  scalar fields and gravitational-wave signatures}},
  \href{http://arxiv.org/abs/2111.01150}{{\tt arXiv:2111.01150}}.

\bibitem{Silverstein:2008sg}
E.~Silverstein and A.~Westphal, {\it Monodromy in the cmb: Gravity waves and
  string inflation},  {\em Phys. Rev. D} {\bf 78} (2008)
  [\href{http://arxiv.org/abs/0803.3085}{{\tt arXiv:0803.3085}}].

\bibitem{Kaloper:2014zba}
N.~Kaloper and A.~Lawrence, {\it Natural chaotic inflation and ultraviolet
  sensitivity},  {\em Phys. Rev. D} {\bf 90} (2014)
  [\href{http://arxiv.org/abs/1404.2912}{{\tt arXiv:1404.2912}}].

\bibitem{Starobinsky:1985ibc}
A.~A. Starobinsky, {\it Multicomponent de sitter (inflationary) stages and the
  generation of perturbations},  {\em JETP Lett.} {\bf 42} (1985) 152--155.

\bibitem{Sasaki:1995aw}
M.~Sasaki and E.~D. Stewart, {\it A general analytic formula for the spectral
  index of the density perturbations produced during inflation},  {\em Prog.
  Theor. Phys.} {\bf 95} (1996) 71--78,
  [\href{http://arxiv.org/abs/astro-ph/9507001}{{\tt astro-ph/9507001}}].

\bibitem{Wands:2000dp}
D.~Wands, K.~A. Malik, D.~H. Lyth, and A.~R. Liddle, {\it A new approach to the
  evolution of cosmological perturbations on large scales},  {\em Phys. Rev. D}
  {\bf 62} (2000) [\href{http://arxiv.org/abs/astro-ph/0003278}{{\tt
  astro-ph/0003278}}].

\bibitem{Lyth:2004gb}
D.~H. Lyth, K.~A. Malik, and M.~Sasaki, {\it A general proof of the
  conservation of the curvature perturbation},  {\em JCAP} {\bf 05} (2005)
  [\href{http://arxiv.org/abs/astro-ph/0411220}{{\tt astro-ph/0411220}}].

\bibitem{Kobayashi:2010fm}
T.~Kobayashi and S.~Mukohyama, {\it Effects of light fields during inflation},
  {\em Phys. Rev. D} {\bf 81} (2010)
  [\href{http://arxiv.org/abs/1003.0076}{{\tt arXiv:1003.0076}}].

\bibitem{Olver:2010:NHMF}
F.~W.~J. Olver, D.~W. Lozier, R.~F. Boisvert, and C.~W. Clark, {\em NIST
  Handbook of Mathematical Functions}.
\newblock Cambridge University Press, 2010.

\bibitem{Lyth:2005qk}
D.~H. Lyth, {\it Generating the curvature perturbation at the end of
  inflation},  {\em JCAP} {\bf 11} (2005)
  [\href{http://arxiv.org/abs/astro-ph/0510443}{{\tt astro-ph/0510443}}].

\bibitem{Sasaki:2008uc}
M.~Sasaki, {\it Multi-brid inflation and non-gaussianity},  {\em Prog. Theor.
  Phys.} {\bf 120} (2008) 159--174, [\href{http://arxiv.org/abs/0805.0974}{{\tt
  arXiv:0805.0974}}].

\bibitem{Turner:1985si}
M.~S. Turner, {\it {Cosmic and Local Mass Density of Invisible Axions}},  {\em
  Phys. Rev. D} {\bf 33} (1986) 889--896.

\bibitem{Bae:2008ue}
K.~J. Bae, J.-H. Huh, and J.~E. Kim, {\it {Update of axion CDM energy}},  {\em
  JCAP} {\bf 09} (2008) 005, [\href{http://arxiv.org/abs/0806.0497}{{\tt
  arXiv:0806.0497}}].

\bibitem{Lyth:1991ub}
D.~H. Lyth, {\it {Axions and inflation: Sitting in the vacuum}},  {\em Phys.
  Rev. D} {\bf 45} (1992) 3394--3404.

\bibitem{Strobl:1994wk}
K.~Strobl and T.~J. Weiler, {\it {Anharmonic evolution of the cosmic axion
  density spectrum}},  {\em Phys. Rev. D} {\bf 50} (1994) 7690--7702,
  [\href{http://arxiv.org/abs/astro-ph/9405028}{{\tt astro-ph/9405028}}].

\bibitem{Kobayashi:2013nva}
T.~Kobayashi, R.~Kurematsu, and F.~Takahashi, {\it {Isocurvature Constraints
  and Anharmonic Effects on QCD Axion Dark Matter}},  {\em JCAP} {\bf 09}
  (2013) 032, [\href{http://arxiv.org/abs/1304.0922}{{\tt arXiv:1304.0922}}].

\end{thebibliography}\endgroup

\end{document}